\documentclass[11pt,a4paper]{article}

\usepackage{hyperref}

\usepackage[fleqn]{amsmath}
\usepackage{amsthm,amssymb,epsfig,latexsym,euscript}
\usepackage[T1]{fontenc} 
\usepackage{lmodern}
\usepackage{times}
\usepackage{pgf}
\usepackage{tikz}
\usepackage{bbold}

\usepackage{graphicx}

\newtheorem{thm}{Theorem}[section]
\newtheorem{prop}{Proposition}[section]

{\theoremstyle{remark}
\newtheorem{rem}{Remark}[section]}
{\theoremstyle{remark}
}
\def\Proof{\medskip\noindent {\it Proof --- \ }}

\let\qed=\cqfd

\makeatletter
\@addtoreset{equation}{section}
\makeatother

\newcommand{\tr}{\operatorname{tr}}

\newcommand{\pl}{\prod\limits}

\newcommand{\bra}[1]{\langle\,#1\,|}
\newcommand{\ket}[1]{|\,#1\,\rangle}
\newcommand{\braket}[2]{\ensuremath{\langle\, #1 \mid  #2\, \rangle }}
\newcommand{\moy}[1]{\langle\,#1\,\rangle}

\newcommand{\zero}{\mathbb{0}}

\def\eps{\epsilon}

\begin{document}

\begin{flushright}
LPENSL-TH-11/13
\end{flushright}

\vspace{24pt}

\begin{center}
\begin{LARGE}
{\bf Multi-point local height probabilities of the CSOS model 
within the algebraic Bethe Ansatz framework}
\end{LARGE}

\vspace{50pt}

\begin{large}
{\bf D.~Levy-Bencheton}\footnote{damien.levybencheton@ens-lyon.fr}
{\bf and V.~Terras}\footnote{veronique.terras@ens-lyon.fr}
\end{large}

\vspace{.5cm}
 
{Laboratoire de Physique, ENS Lyon \& CNRS UMR 5672,\\
Universit\'e de Lyon, France}
\vspace{2cm}

\today

\end{center}

\vspace{1cm}

\begin{abstract}
We study the local height probabilities of the exactly solvable cyclic solid-on-solid model within the algebraic Bethe Ansatz framework.
We more specifically consider multi-point local height probabilities at adjacent sites on the lattice.
We derive multiple integral representations for these quantities at the thermodynamic limit,
starting from finite-size expressions for the corresponding multi-point matrix elements in the Bethe basis as sums of determinants of elliptic functions.

\end{abstract}

\vspace{1cm}

\section{Introduction}
\label{sec-intro}


The exactly solvable solid-on-solid (SOS) model is a two-dimensional model of statistical mechanics describing the atomic structure of a crystal-vapor interface.
The surface of the crystal is modeled by a two-dimensional square lattice with, attached to each lattice site, a fluctuation variable labeling the `height' of the crystal relatively to a flat reference surface. The statistics of the system is then governed by local Boltzmann weights, associated to all possible configurations of interface heights around a given face of the lattice, and satisfying the star-triangle relation \cite{Bax82L}.
This model plays an important role in the context of two-dimensional exactly solvable models of statistical mechanics, since it is  the archetype of the class of so-called {\em interaction-round-faces} (IRF) models, by opposition to vertex models that describe instead interactions of spin variables around vertices of a lattice. It is also famous for its crucial role in Baxter's solution \cite{Bax73a} of the eight-vertex model: using the fact that the local Boltzmann weights of these two models are related by the so-called vertex-IRF transformation, Baxter managed to construct the eigenstates of the eight-vertex transfer matrix by means of the eigenstates of the (cyclic) SOS transfer matrix.


The underlying algebraic structure of the SOS model is the elliptic quantum group $E_{\tau,\eta}(sl_2)$ \cite{Fel95,FelV96a}, associated to an elliptic $R$-matrix satisfying the dynamical (or modified) Yang-Baxter equation \cite{GerN84,Fel95}.
This model is solvable by algebraic Bethe Ansatz (ABA) \cite{FelV96b}. However,
compared to usual vertex models such as the six-vertex model, the corresponding Yang-Baxter algebra has a slightly more complicated structure due to the shifts undergone by the dynamical (`height') parameter. This has for a long time been problematic for the computation of correlation functions within the ABA framework. In fact, the ABA approach to correlation functions \cite{KitMT99} usually relies on the existence of a compact and manageable expression, preferably in the form of a single determinant of usual functions, for the scalar products of Bethe states \cite{Sla89}. The latter representation happens to be strongly related, in the six-vertex case, to Izergin's determinant representation for the partition function of the model with domain wall boundary conditions \cite{Kor82,Ize87}. In the SOS case, however, it seems that the corresponding partition function does not admit a so simple representation \cite{PakRS08,Ros09}.
In our previous article \cite{LevT13a}, we have nevertheless managed to obtain, at least in the cyclic (CSOS) case, i.e. when the crossing parameter of the model is a rational number, a single determinant representation for the scalar products of Bethe states and, through the solution of the inverse problem, for the finite-size form factors associated to eigenstates of the transfer matrix. The latter representation was used in \cite{LevT13b} to explicitly compute the spontaneous staggered polarizations of the CSOS model at the thermodynamic limit, in a way quite similar as what had been done in \cite{IzeKMT99} in the six-vertex case.

The study of explicitly height-dependent quantities such as local height probabilities is however more complicated in this context. Indeed, the simplest of these quantities, namely the one-point local height probabilities (i.e. the probabilities to have a specified height at a given site of the model), are related to what we called in \cite{LevT13a} the `partial scalar products'. The latter do not admit, to the best of our knowledge, a representation as simple as the true scalar products of Bethe states: in the cyclic (CSOS) case, they can at best be represented as a finite sum of determinants. 
This is the purpose of the present paper to show that such a representation can nevertheless be used to compute the one-point local height probabilities, and more generally any multi-point local height probability, in the CSOS model at the thermodynamic limit. 
As a result, we obtain multiple integral representations for multi-point local height probabilities at adjacent sites. When these adjacent sites belong to a same vertical line of the lattice, these representation have a quite similar form as the multiple integral representations obtained in the simpler six-vertex case \cite{JimM95L,KitMT00}, as well as those obtained in the restricted (RSOS) case by means of the $q$-vertex operator approach \cite{LukP96}. We moreover recover the expression obtained in \cite{PeaS89} for the one-point local height probabilities of the CSOS model.

This paper is organized as follows. In Section~\ref{sec-ABA}, we define the SOS model, recall the ABA construction of the space of states and the characterization of the degenerate ground states at the thermodynamic limit in the cyclic case. 
In Section~\ref{sec-ABA-corr}, we formulate our problem, namely express the local heights probabilities of the model within the ABA framework.
We notably express the multi-point local height probabilities at adjacent sites in terms of matrix elements, between Bethe ground states of the model, of particular combinations of entries of the monodromy matrix. 
In Section~\ref{sec-MPME}, we compute the corresponding finite-size multi-point matrix elements in the Bethe basis as sums of determinants of elliptic functions. In Section~\ref{sec-therm}, we explain how to take the thermodynamic limit of these expressions.
Finally, in Section~\ref{sec-MPLHP}, we obtain multiple integral representations for the multi-point local height probabilities at adjacent sites in the basis of ground states in which local operators are diagonal.

\section{The SOS model in the ABA framework}
\label{sec-ABA}


Let us consider a two-dimensional lattice of size $M\times N$, with $M$ elementary square faces in a line and $N$ in a column ($N$ even).
A fluctuation variable (the `height') $s$ is attached to each vertex of the lattice, so that heights on adjacent sites differ by $\pm 1$. The difference of heights between two adjacent sites is hence described by a variable (`spin' variable) $\alpha=\pm 1$ attached to the corresponding bond.
There are six different allowed configurations of the height variables around a face,

\medskip

\begin{pgfpicture}{0cm}{0cm}{2cm}{2cm}

\pgfnodecircle{Node1}[fill]{\pgfxy(2.5,1.5)}{0.05cm}
\pgfnodecircle{Node2}[fill]{\pgfxy(3.5,1.5)}{0.05cm}
\pgfnodecircle{Node4}[fill]{\pgfxy(2.5,0.5)}{0.05cm}
\pgfnodecircle{Node3}[fill]{\pgfxy(3.5,0.5)}{0.05cm}
\pgfnodeconnline{Node1}{Node2}
\pgfnodeconnline{Node2}{Node3}
\pgfnodeconnline{Node3}{Node4}
\pgfnodeconnline{Node1}{Node4}
 

 \pgfputat{\pgfxy(2.4,1.6)}{\pgfbox[right,center]{$s$}}
 \pgfputat{\pgfxy(3.6,1.6)}{\pgfbox[left,center]{$s+\alpha'_i$}}
 \pgfputat{\pgfxy(2.45,0.1)}{\pgfbox[right,bottom]{$s+\alpha_j$}}
 \pgfputat{\pgfxy(3.6,0.2)}{\pgfbox[left,bottom]{$s+\alpha_i+\alpha_j$}}
 \pgfputat{\pgfxy(3.6,-0.2)}{\pgfbox[left,bottom]{$=s+\alpha'_i+\alpha'_j$}}

\pgfputat{\pgfxy(3,1.6)}{\pgfbox[center,bottom]{$\alpha'_i$}}
\pgfputat{\pgfxy(3,0.4)}{\pgfbox[center,top]{$\alpha_i$}}
\pgfputat{\pgfxy(2.4,0.95)}{\pgfbox[right,center]{$\alpha_j$}}
\pgfputat{\pgfxy(3.6,0.95)}{\pgfbox[left,center]{$\alpha'_j$}}

\pgfputat{\pgfxy(9.12,1.2)}{\pgfbox[center,center]{with $\alpha_i,\alpha'_i,\alpha_j,\alpha'_j\in\{+1,-1\}$}}

\pgfputat{\pgfxy(9,0.6)}{\pgfbox[center,center]{such that $\alpha_i+\alpha_j=\alpha'_i+\alpha'_j$,}}
 
\end{pgfpicture}

\medskip
\noindent
and the corresponding statistical weights ${W}\binom{s\qquad s+\alpha'_i}{s+\alpha_j\ s+\alpha_i+\alpha_j}$ can be understood as the six non-zero elements ${R}(w_i-\xi_j; s)^{\alpha_i,\alpha_j}_{\alpha'_i,\alpha'_j}$ of the following $R$-matrix (see Fig.~\ref{6faces}):
\begin{equation}\label{R-mat}
  R(w_i-\xi_j;s)=
  \begin{pmatrix} 1 & 0 & 0 & 0 \\
                              0 & \mathsf{b}(w_i-\xi_j;s) & \mathsf{c}(w_i-\xi_j;s) & 0 \\
                              0 & \mathsf{c}(w_i-\xi_j;-s) & \mathsf{b}(w_i-\xi_j;-s) & 0 \\
                              0 & 0 & 0 & 1 
  \end{pmatrix}
  \in\mathrm{End}(\mathbb{C}^2\otimes \mathbb{C}^2).
\end{equation}
Here $w_i$ (respectively $\xi_j$) is an inhomogeneity parameter attached to the column $i$ (resp. row $j$) of cells of the lattice, labelled from left to right (resp. from top to bottom). The functions $\mathsf{b}(u;s)$ and $\mathsf{c}(u;s)$ are given as
\begin{equation}
 \mathsf{b}(u;s)=\frac{[s+1] \,  [u]}{[s] \, [u+1]},           \qquad   \mathsf{c}(u;s)=\frac{[s+u] \,  [1]}{[s] \, [u+1]} , \label{bc1}
 \qquad \text{with} \quad [u]=\theta_1(\eta u;\tau),
\end{equation} 
where $\theta_1$ denotes the usual theta function \eqref{theta1} with quasi-periods 1 and $\tau$ ($\Im\tau>0$), and $\eta$ is the crossing parameter of the model.
The height $s$ in \eqref{R-mat} is called dynamical parameter. 
\begin{figure}[h]
\centering
\begin{tikzpicture}
    \draw(0,5.5) ;
    \draw(0,-2) ;   

    \draw (0,3.5) -- (0,4.5) node[above left]{$s$} ; 
    \draw (0,4.5) -- (1,4.5) node[above right]{$s+1$} ;
    \draw (1,4.5) -- (1,3.5) node[below right]{$s+2$} ;
    \draw (1,3.5) -- (0,3.5) node[below left]{$s+1$} ;
    \draw (0.5,4.5) node[above]{$+$} ;
    \draw (0.5,3.5) node[below]{$+$} ;
    \draw (0,4) node[left]{$+$} ;
    \draw (1,4) node[right]{$+$} ;
    
    \draw (0.5,2) node[above]{$1$} ;

    \draw (0,0) -- (0,1) node[above left]{$s$} ; 
    \draw (0,1) -- (1,1) node[above right]{$s-1$} ;
    \draw (1,1) -- (1,0) node[below right]{$s-2$} ;
    \draw (1,0) -- (0,0) node[below left]{$s-1$} ;
    \draw (0.5,1) node[above]{$-$} ;
    \draw (0.5,0) node[below]{$-$} ;
    \draw (0,0.5) node[left]{$-$} ;
    \draw (1,0.5) node[right]{$-$} ;  
    
    \draw (0.5,-1.5) node[above]{$1$} ;

    \draw (4,3.5) -- (4,4.5) node[above left]{$s$} ; 
    \draw (4,4.5) -- (5,4.5) node[above right]{$s+1$} ;
    \draw (5,4.5) -- (5,3.5) node[below right]{$s$} ;
    \draw (5,3.5) -- (4,3.5) node[below left]{$s-1$} ;
    \draw (4.5,4.5) node[above]{$+$} ;
    \draw (4.5,3.5) node[below]{$+$} ;
    \draw (4,4) node[left]{$-$} ;
    \draw (5,4) node[right]{$-$} ;
    
    \draw (4.5,2) node[above]{$\mathsf{b}(u;s)$} ;

    \draw (4,0) -- (4,1) node[above left]{$s$} ; 
    \draw (4,1) -- (5,1) node[above right]{$s-1$} ;
    \draw (5,1) -- (5,0) node[below right]{$s$} ;
    \draw (5,0) -- (4,0) node[below left]{$s+1$} ;
    \draw (4.5,1) node[above]{$-$} ;
    \draw (4.5,0) node[below]{$-$} ;
    \draw (4,0.5) node[left]{$+$} ;
    \draw (5,0.5) node[right]{$+$} ;  
    
    \draw (4.5,-1.5) node[above]{$\bar{\mathsf{b}}(u;s)$} ;

    \draw (8,3.5) -- (8,4.5) node[above left]{$s$} ; 
    \draw (8,4.5) -- (9,4.5) node[above right]{$s-1$} ;
    \draw (9,4.5) -- (9,3.5) node[below right]{$s$} ;
    \draw (9,3.5) -- (8,3.5) node[below left]{$s-1$} ;
    \draw (8.5,4.5) node[above]{$-$} ;
    \draw (8.5,3.5) node[below]{$+$} ;
    \draw (8,4) node[left]{$-$} ;
    \draw (9,4) node[right]{$+$} ;
    
    \draw (8.5,2) node[above]{${\mathsf{c}}(u;s)$} ;

    \draw (8,0) -- (8,1) node[above left]{$s$} ; 
    \draw (8,1) -- (9,1) node[above right]{$s+1$} ;
    \draw (9,1) -- (9,0) node[below right]{$s$} ;
    \draw (9,0) -- (8,0) node[below left]{$s+1$} ;
    \draw (8.5,1) node[above]{$+$} ;
    \draw (8.5,0) node[below]{$-$} ;
    \draw (8,0.5) node[left]{$+$} ;
    \draw (9,0.5) node[right]{$-$} ;  
    
    \draw (8.5,-1.5) node[above]{$\bar{\mathsf{c}}(u;s)$} ;     
\end{tikzpicture}\vspace{-5mm}
\caption{\label{6faces}The 6 different local configurations around a face and their associated local statistical weights.}
\end{figure}

The $R$-matrix \eqref{R-mat} with dynamical parameter $s$  satisfies the dynamical (or modified) quantum Yang-Baxter equation on $\mathbb{C}^2\otimes \mathbb{C}^2\otimes \mathbb{C}^2$ \cite{GerN84,Fel95},
\begin{multline}\label{YB}
  R_{12}(u_1-u_2; s+h_3) \; R_{13}(u_1-u_3 ; s) \; R_{23}( u_2-u_3 ; s+h_1) \\
  =
  R_{23}(u_2-u_3; s) \; R_{13}(u_1-u_3 ; s+h_2) \; R_{12}( u_1-u_2 ; s) , \quad
  \text{with} \quad
  h=\begin{pmatrix} 1 & 0 \\ 0 & -1 \end{pmatrix},
\end{multline}
where the indices label as usual the space of the tensor product on which the corresponding operator acts. 
This equation is equivalent to Baxter's star-triangle relation for the local Boltzmann weights $W$.

In the general (unrestricted) SOS model, the crossing parameter $\eta$ is arbitrary and the dynamical parameter $s$ belongs to the infinite set $s_0+\mathbb{Z}$ for some given parameter $s_0$, so that the space of states of the model is infinite dimensional even for a finite lattice. In the cyclic case that we consider in this paper\footnote{We do not consider here the restricted SOS (or ABF) model \cite{AndBF84}, for which $\eta=1/L$, $s_0=0$, and $s\in\{1,2,\ldots,L-1\}$. The construction of the eigenvectors of the transfer matrix is indeed slightly more subtle in that case due to possible poles in \eqref{bc1} (see \cite{FelV99}).}, the parameter $\eta$ is rational ($\eta=r/L$, $r$ and $L$ being relatively prime integers), so that heights are periodic in $L$. In other words, it means that the dynamical parameter takes its values in the finite set $s_0+\mathbb{Z}/L\mathbb{Z}$.

The thermodynamic properties of the cyclic SOS model with periodic boundary conditions have been studied in a series of papers \cite{KunY88,PeaS89,PeaB90,KimP89,DatJKM90,RieN91,SeaN92}.
In particular, in  \cite{PeaS89}, the $2(L-r)$ ground states of the (infinite-size) CSOS model with crossing parameter $\eta=r/L$ and real phase angle $\eta\tilde{s}_0\equiv\eta s_0-\frac{\tau}{2}$ have been identified in the completely ordered low-temperature limit $\tau\to 0$: they correspond, in this limit, to the flat configurations of the type $(\mathsf{s},\mathsf{s}+1,\mathsf{s},\mathsf{s}+1,\ldots)$ or $(\mathsf{s}+1,\mathsf{s},\mathsf{s}+1,\mathsf{s},\ldots)$, i.e. to the two possible assignments of the heights $\mathsf{s}$ and $\mathsf{s}+1$ to the two sublattices of the square lattice, for the $L-r$ values of $\mathsf{s}\in s_0+\mathbb{Z}/L\mathbb{Z}$ satisfying $\lfloor \eta \tilde{\mathsf{s}}\rfloor$=$\lfloor \eta(\tilde{\mathsf{s}}+1) \rfloor$  (here $\lfloor \cdot \rfloor$ denotes the floor function, and $\tilde{\mathsf{s}}=\mathsf{s}-\frac{\tau}{2\eta}$)\footnote{Our parameter $\eta s_0$, where $s_0$ is a fixed global shift of the dynamical parameter (such that $s-s_0$ is an integer) introduced so as to avoid the singularities in \eqref{bc1}, is related to the phase angle $w_0/\pi$ (that we denote here by $\eta\tilde{s}_0$) of the physical model considered in \cite{PeaS89,PeaB90} by a shift of $\tau/2$. Note also that the (positive) statistical weights of  \cite{PeaS89,PeaB90} correspond to a diagonal dynamical gauge transformation of the $R$-matrix \eqref{R-mat} which leaves the local height probabilities invariant.}.
In the same paper, one-point local height probabilities $\mathbf{P}(s)$, i.e. probabilities that a given site of the lattice corresponds to a given value $s$ of the height in one of the previously identified ground state configurations, have been explicitly computed by means of Baxter's corner transfer matrix method \cite{Bax82L}.  

\begin{figure}[h]\centering
\includegraphics[height=4.5cm]{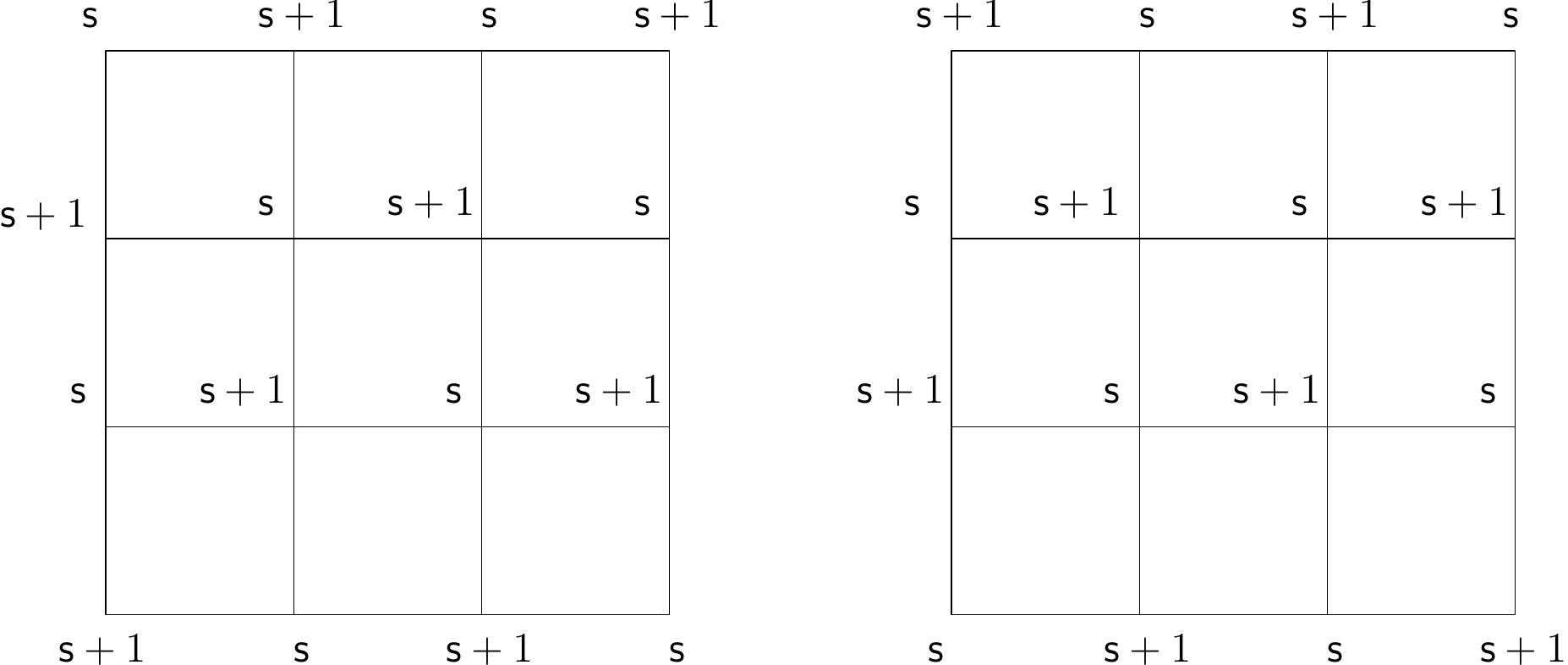}
\caption{\label{flat-gs}The flat ground state configurations in the low temperature limit.}
\end{figure}

In the present paper, we explain how to compute more general multi-point local height probabilities.
For the model with periodic boundary conditions, such quantities are commonly expressed in terms of  the transfer matrix of the model such that, in the thermodynamic limit, their computation can be reduced to the computation of matrix elements of appropriate combinations of local operators between the eigenstates of the transfer matrix corresponding to the maximal eigenvalue (see Section~\ref{sec-ABA-corr}).
We recall in this section the construction of these eigenstates in the algebraic Bethe Ansatz framework, as well as the characterization of the $2(L-r)$ ground states of the model.

\subsection{Construction of the space of states in the ABA framework}
\label{subsec-space-ABA}

Eigenstates of the transfer matrix of the SOS model with periodic boundary conditions
can be obtained by means of algebraic Bethe Ansatz (see \cite{FelV96a,FelV96b}).
The associated space of states corresponds to the space of functions $\mathbb{H}[0]=\mathrm{Fun}(\mathcal{H}[0])$ of the dynamical parameter $s$ with values in the zero-weight space $\mathcal{H}[0]=\{\,\ket{\!\mathbf{v} \! }\in\mathcal{H}\mid  h_{1\ldots N}\,\ket{\! \mathbf{v} \! }=0\}$, with $\mathcal{H}\sim (\mathbb{C}^2)^{\otimes N}$ and $h_{1\ldots N} = h_1 + \ldots + h_N$.
We recall that, in the unrestricted case, the dynamical parameter belongs to the set $s_0+\mathbb{Z}$, whereas in the cyclic case it belongs to $s_0+\mathbb{Z}/L\mathbb{Z}$. In the latter case, the space of states $\mathbb{H}[0]$ of the finite-size model is finite dimensional and the zero-weight criterion has to be understood modulo $L$.
A basis of $\mathbb{H}=\mathrm{Fun}(\mathcal{H})$ is given by elementary $\delta$-function states of the form
\begin{equation}\label{delta-state}
   \ket{s_1 ;\varepsilon_1,\varepsilon_2,\ldots,\varepsilon_N} =\delta_{s_1}\
    \mathsf{e}_{\varepsilon_1}\otimes\mathsf{e}_{\varepsilon_2}\otimes\cdots\otimes\mathsf{e}_{\varepsilon_N},
    \qquad s_1 \in s_0+\mathbb{Z}/L\mathbb{Z},\quad \varepsilon_j\in\{+,-\},
\end{equation}
where $\mathsf{e}_+=\binom{1}{0}$, $\mathsf{e}_-=\binom{0}{1}$, $\delta_s$ being the numerical function of the dynamical parameter defined as
\begin{equation}\label{delta-s}
  \delta_s: \tilde{s} \mapsto \delta_s(\tilde{s})
  =\begin{cases} 1 &\text{if } \tilde{s}=s,\\ 0 &\text{otherwise}.\end{cases} 
\end{equation}
A natural basis of $\mathbb{H}[0]$ then corresponds to the elementary $\delta$-function states \eqref{delta-state} such that $\varepsilon_1+\varepsilon_2+\cdots +\varepsilon_N=0\ (\text{mod } L)$.

\begin{rem}
The elementary $\delta$-function state \eqref{delta-state} can be understood as the particular configuration of heights $s_1,s_2\equiv s_1+\varepsilon_1, s_3\equiv s_1+\varepsilon_1+\varepsilon_2,\ldots,s_{N+1}\equiv s_1+\varepsilon_1+\varepsilon_2+\cdots+\varepsilon_{N}$ (from top to bottom) along a vertical line of vertices of the lattice.
\end{rem}

In the ABA framework, the central object is the monodromy matrix. It is defined here as the following ordered product of $R$-matrices along a column of elementary cells of the lattice:
\begin{align}\label{monodromy}
 T_{a, 1\ldots N}(u ; \xi_1,\ldots, \xi_N ; s)
  &= R_{a N} (u - \xi_N ; s + h_1 +\cdots + h_{N-1} ) \ldots    R_{a 1} (u - \xi_1 ; s  ) \nonumber\\
  &= \begin{pmatrix} A(u ; s) & B(u ; s) \\
                                    C(u ; s) & D(u ; s) \end{pmatrix}_{\!\! [a]}.   
\end{align}
It acts on $V_a\otimes \mathcal{H}$, where $V_a\sim\mathbb{C}^2$ is usually called auxiliary space.
Hence, the entries $A,B,C,D$ of the monodromy matrix are linear operators acting on $\mathcal{H}$. Their commutation relations are given in terms of the $R$-matrix \eqref{R-mat} by the following quadratic equation on $V_{a_1} \otimes V_{a_2} \otimes \mathcal{H}$,
\begin{multline}\label{RTT}
R_{a_1 a_2} (u_1-u_2 ; s + h_{1\ldots N}) \;
        T_{a_1, 1\ldots N}(u_1 ; s) \;  T_{a_2, 1\ldots N}(u_2 ; s+h_{a_1}) \\
= T_{a_2, 1\ldots N}(u_2 ; s)\;  T_{a_1, 1\ldots N}(u_1 ; s+h_{a_2}) \;
   R_{a_1 a_2} (u_1-u_2 ; s),
\end{multline}
which is a consequence of the dynamical Yang-Baxter relation \eqref{YB}.

To deal with the shifts of the dynamical parameter, it is convenient to introduce two operators  $\widehat{s}$ and $\widehat{\tau}_s$, acting on functions $f$ of the dynamical parameter as
\begin{equation}
    (\widehat{s}\, f)(s)= s\, f(s), \qquad (\widehat{\tau}_s \, f)(s)= f(s+1),
\end{equation}
and such that  $\widehat{\tau}_s\, \widehat{s}=(\widehat{s}+1)\, \widehat{\tau}_s$.
An action on the left can similarly be defined as
\begin{equation}
    (f\, \widehat{s})(s)=  f(s)\, s, \qquad (f \,\widehat{\tau}_s)(s)= f(s-1).
\end{equation}
This enables us to define an operator algebra  (see \cite{FelV96a,FelV96b}) generated by the operator entries  $\widehat{A},\ \widehat{B},\ \widehat{C},\ \widehat{D}$ of the matrix $\widehat{T}$ constructed in terms of the monodromy matrix \eqref{monodromy} as
\begin{equation}\label{mon-op}
   \widehat{T}(u)= \begin{pmatrix} \widehat{A}(u) & \widehat{B}(u) \\
                                    \widehat{C}(u) & \widehat{D}(u) \end{pmatrix}_{\!\! [a]} 
                            = T(u; \widehat{s}) \, \begin{pmatrix} \widehat{\tau}_s & 0 \\ 0 & \widehat{\tau}_s^{-1}\end{pmatrix}_{\!\! [a]}       
           \quad \in\mathrm{End}(V_a\otimes\mathbb{H}) ,            
\end{equation}
and satisfying the commutation relations given in terms of the $R$-matrix \eqref{R-mat} as
\begin{multline}\label{RTTop}
  R_{a_1 a_2} (u_1-u_2 ; \widehat{s} + h_{1\ldots N}) \;
        \widehat{T}_{a_1, 1\ldots N}(u_1) \;  \widehat{T}_{a_2, 1\ldots N}(u_2 ) \\
= \widehat{T}_{a_2, 1\ldots N}(u_2)\;  \widehat{T}_{a_1, 1\ldots N}(u_1) \;
   R_{a_1 a_2} (u_1-u_2 ; \widehat{s}) .
\end{multline}
The transfer matrix $\widehat{t}(u)\in \mathrm{End}(\mathbb{H})$ of the model with periodic boundary conditions along the vertical direction is then defined as the trace, over the auxiliary space $V_a$, of the monodromy matrix \eqref{mon-op}:
\begin{equation}\label{transfer}
    \widehat{t}(u)\equiv\widehat{t}(u;\xi_1,\ldots,\xi_N)=\widehat{A}(u)+\widehat{D}(u).
\end{equation} 
These operators preserve the zero-weight space $\mathbb{H}[0]$ and mutually commute on  $\mathbb{H}[0]$.

In this picture, we define Bethe states to be the following elements of $\mathbb{H}[0]$, depending on a set of $n$ spectral parameter $\{v\}\equiv\{v_1,\ldots, v_n\}$  (which are supposed to be such that $\eta v_i\neq\eta v_j\mod\mathbb{Z}+\tau\mathbb{Z}$) and a root of unity $\omega$ satisfying $(-1)^{rn} \omega^L=1$:
%
\begin{equation}\label{state}
  \ket{ \{v\}, \omega }
 = \varphi_\omega \prod_{j=1}^n\widehat{B}(v_j) \, \ket{\zero},
 \qquad
 \text{with}\quad
 \varphi_\omega(s)=\frac{\omega^s}{\sqrt{L}}\prod_{j=1}^n\frac{[1]}{[s-j]}.
\end{equation}
In \eqref{state}, $\ket{\zero}$ is a constant function of $\mathbb{H}$ (the reference state) given as $ \ket{\zero}:s\mapsto \ket{0}\equiv(\mathsf{e}_+)^{\otimes N}$, and $n$ is such that $N=2n+\aleph L$ for some integer $\aleph$, so that \eqref{state} effectively belongs to $\mathbb{H}[0]$.
Similarly, Bethe states in the dual space of states are constructed as a multiple action of $n$ $\widehat{C}$ operators on the left reference state $\bra{\zero}:s\mapsto \bra{0}\equiv(\,1\; 0\,)^{\otimes N}$:
\begin{equation}\label{dual}
   \bra{\{ v\} ,\omega}= \bra{\zero}\prod_{j=1}^n \widehat{C}(v_j) \; \widetilde{\varphi}_\omega,
   \qquad \text{with}\quad 
   \widetilde{\varphi}_\omega(s)=  \frac{{\omega}^{- s}}{\sqrt{L}}  \prod_{j=0}^{n-1} \frac{[s+j]}{[1]}.
\end{equation}
%
It is easy to see that, when $(\{v\},\omega)$ satisfies the system of Bethe equations
\begin{equation}\label{Bethe}
    \mathsf{a}(v_j)  \prod_{l\ne j} \frac{[v_l-v_j+1]}{[v_l-v_j]} 
    =  (-1)^{r\aleph} \omega^{-2} \; \mathsf{d}(v_j)  \prod_{l\ne j} \frac{[v_j-v_l+1]}{[v_j-v_l]} ,
    \quad j=1,\ldots n,
\end{equation}
with
\begin{equation}\label{a-d}
  \mathsf{a}(u)=1, \qquad \mathsf{d}(u)=\prod_{j=1}^N \frac{[u-\xi_j]}{[u-\xi_j+1]},
\end{equation}
the states \eqref{state} and \eqref{dual} are respectively right and left eigenstates of the transfer matrix \eqref{transfer}:
\begin{equation}\label{act-transfer}
   \widehat{t}(u)\, \ket{\{ v\},\omega }= \tau (u; \{ v\},\omega )\, \ket{\{ v\},\omega },
   \qquad
   \bra{ \{v\}, \omega }\,\widehat{t} (u) = \tau(u; \{v\},\omega )\,  \bra{ \{v\} ,\omega },
\end{equation}
with eigenvalues
\begin{equation}
   \tau (u; \{v\},\omega)
    = \omega \; \mathsf{a}(u) \prod_{l=1}^n \frac{[v_l-u+1]}{[v_l-u]}
        +  (-1)^{r\aleph}  \omega^{-1} \; \mathsf{d}(u) \prod_{l=1}^n \frac{[u-v_l+1]}{[u-v_l]}.
    \label{tau}
\end{equation}

The scalar product of a left Bethe state $\bra{\{u\},\omega_u}$ \eqref{dual} with a right Bethe state $\ket{\{v\},\omega_v } $ \eqref{state} is defined as\footnote{Compared to \cite{LevT13a}, we have included the normalization factor $1/L$ into the definition of the states \eqref{state}, \eqref{dual}, and the scalar product is normalized in such a way that elementary $\delta$-function states \eqref{delta-state} are orthonormal.}
\begin{align}
 \moy{ \{u\},\omega_u \mid \{v\},\omega_v } 
  &= 
         \sum_{s\in s_0+\mathbb{Z}/L\mathbb{Z} } 
         \widetilde\varphi_{\omega_u}(s)\,\varphi_{\omega_v}(s) \, S_n ( \{u\}; \{v\} ; s),\label{scalarproduct}
\end{align}
in term of the quantity
\begin{align}
S_n ( \{u\}; \{v\} ; s)
  &= \bra{0} C(u_{n};s-n)\ldots  C(u_{1};s-1)
                  B(v_{1};s) \ldots B(v_{n};s-n+1) \ket{0},
                  \nonumber\\
  &=   \bra{\zero}\prod_{j=1}^n \widehat{C}(v_j) \; \delta_s \,       \prod_{j=1}^n\widehat{B}(v_j) \, \ket{\zero},    \label{Sn}
\end{align}
that we called in \cite{LevT13a} {\em partial scalar product}\footnote{As explained in \cite{LevT13a}, the partial scalar product \eqref{Sn} is not a scalar product of Bethe states but of $\delta$-function states of the form
$\delta_s \,       \prod_{j=1}^n\widehat{B}(v_j) \, \ket{\zero}.$}.
The scalar product of Bethe states \eqref{scalarproduct} and the partial scalar product \eqref{Sn} have been computed in \cite{LevT13a} in the case where $\{u\}\equiv\{u_1,\ldots,u_n\}$ is a solution of the set of Bethe equations \eqref{Bethe} associated to the complex root of unity $\omega_u$ ($\{v\}\equiv\{v_1,\ldots,v_n\}$ being a set of arbitrary parameters). In particular, it has been shown in \cite{LevT13a} that scalar products of Bethe states \eqref{scalarproduct} could, in that case, be expressed as a unique determinant as in the case of the XXZ chain (see Theorem~3.1 of \cite{LevT13a}). In the same way, form factors of (1-point) local operators can be expressed as a single determinant, and such a representation was used in \cite{LevT13b} to compute spontaneous staggered polarizations of the CSOS model. In the present paper, we however do not need these formulas, and therefore we do not recall them. The only crucial point that we will use here is the fact that the scalar product of a Bethe eigenstate \eqref{state} with the corresponding left Bethe eigenstate \eqref{dual} (`square of the norm') can be represented as a unique determinant of the form:
\begin{align}
   \moy{ \{u\},\omega_u \mid \{u\},\omega_u } 
        &=        
     \frac{ (-1)^{nr\aleph} }{(-[0]')^n}
     \frac{ \prod_{t=1}^{n}  \mathsf{a}(u_{t})  \mathsf{d}(u_{t}) \,
                    \prod_{j,k=1}^n [u_{j}-u_{k}+1] }
             { \prod_{j\not= k} [u_{j}-u_{k}]  }   
     \det_n \big[ \Phi (\{u\} ) \big]   ,\label{gaudin}
\end{align}  
with
\begin{multline}\label{mat-Phi}
   \big[ \Phi (\{u\} ) \big]_{jk}
   = \delta_{jk}\Bigg\{ \log'\frac{\mathsf{a}}{\mathsf{d}}(u_{j})
                                    +\sum_{t=1}^n \left(\frac{[u_j-u_t-1]'}{[u_j-u_t-1]}-\frac{[u_j-u_t+1]'}{[u_j-u_t+1]}\right)\Bigg\}\\
      -  \left(\frac{[u_j-u_k-1]'}{[u_j-u_k-1]}-\frac{[u_j-u_k+1]'}{[u_j-u_k+1]}\right).
\end{multline}   

The situation is slightly more complicated for the partial scalar product \eqref{Sn}. This partial scalar product is directly related to the partition function of the model with domain wall boundary conditions computed in \cite{PakRS08,Ros09}. In fact, up to a normalization factor, the latter coincides with a particular case of the former. Hence, as for the partition function, it seems  that the partial scalar product \eqref{Sn} cannot be expressed in the simple form of a single determinant as the scalar product of Bethe state \eqref{scalarproduct} or the form factors computed in \cite{LevT13a}. It can nevertheless be expressed, still as for the partition function \cite{Ros09}, in the form of a sum of determinants (see \cite{LevT13a}). In fact, such a representation exists even in the case of a general (unrestricted) SOS model but, in the cyclic case $\eta=r/L$, it can be shown that the formula simplifies as a sum of only $L$ terms:
\begin{multline}\label{result-spL}
   S_n ( \{u\}; \{v\} ; s)
  = \frac{[\gamma]\, [s]}{ [0]'\,[ |u| -|v|+\gamma+s] }
      \prod_{j=1}^{n}\frac{[s-j]}{[s+j-1]}   \,   
     \frac{ \prod_{t=1}^{n}  \mathsf{d}(u_{t})   }{ \prod_{j<k} [u_{j}-u_{k}] [v_{k}-v_{j}] }     
  \\
 \times \sum_{ \nu=0  }^{L-1} q^{\nu s}\, 
     a_\gamma^{(\nu)}(s_0)\,
         \det_n \big[ \Omega_\gamma^{(\nu)} (\{u\},\omega_u; \{ v\}  ) \big]  ,
\end{multline}
with $q=e^{2\pi i\eta}$ and
\begin{equation}\label{a-nu}
   a_\gamma^{(\nu)}(s_0)= \eta\,
   \frac{\theta_1(r s_0+\eta\gamma+\nu\tau;L\tau)\, \theta_1'(0;L\tau)}
           {\theta_1(r s_0;L\tau)\, \theta_1(\eta\gamma+\nu\tau;L\tau)},
\end{equation}
\begin{multline}\label{def-OmegaL}
   [ \Omega_\gamma^{(\nu)} (\{u \},\omega_u; \{ v \}  ) \big]_{ij} 
   = \frac{(-1)^{r\aleph} }{[\gamma]}\Bigg\{\! \frac{[u_{i}-v_{j}+\gamma]}{[u_{i}-v_{j}]}
                  -q^{-\nu}\frac{[u_{i}-v_{j}+\gamma+1]}{[u_{i}-v_{j}+1]} \! \Bigg\} \,
       \mathsf{a}(v_{j}) \! \prod_{t=1}^{n} [u_{t}-v_{j}+1]
       \\
     + \frac{1}{[\gamma]}
     \Bigg\{\! \frac{[u_{i}-v_{j}+\gamma]}{[u_{i}-v_{j}]}
                  -q^\nu\frac{[u_{i} -v_{j}+\gamma-1]}{[u_{i}-v_{j}-1]} \!  \Bigg\}    \,
        \omega_u^{-2} \mathsf{d}(v_{j}) \! \prod_{t=1}^{n} [u_{t}-v_{j}-1]. 
\end{multline}
Here $\gamma$ is an arbitrary complex parameter (the expression \eqref{result-spL} does not depend on the value of $\gamma$), and we have set $|u|\equiv u_1+u_2+\dots + u_n$, $|v|\equiv v_1+v_2+\dots +v_n$.
We recall that, in \eqref{result-spL}, $(\{u\},\omega_u)$ is solution of the Bethe equations, whereas $\{v\}$ is a set of arbitrary parameters.

\subsection{The degenerate ground states in the thermodynamic limit}
\label{sec-gs-th}

It has been shown in \cite{PeaS89,PeaB90} that the transfer matrix of the CSOS model possesses $2(L-r)$ largest (in magnitude) eigenvalues which are asymptotically degenerate at the thermodynamic limit.
The corresponding $2(L-r)$ degenerate ground states correspond to particular solutions of the Bethe equations \eqref{Bethe} in the $n=N/2$ sector. As shown in \cite{LevT13b}, these Bethe ground states are completely determined by two quantum numbers $\mathsf{k}\in\mathbb{Z}/2\mathbb{Z}$ and $\ell\in  \mathbb{Z}/(L-r)\mathbb{Z}$, so that, as in \cite{LevT13b}, it will be convenient to denote the corresponding Bethe vectors by $\ket{\mathsf{k},\ell}$.

The inhomogeneous Bethe equations \eqref{Bethe} in the $n=N/2$ sector for $\omega=e^{i\pi\frac{rn+2\ell}{L}}$ can be rewritten in the logarithmic form by means of Jacobi's imaginary transformation \eqref{jacobi} as
\begin{equation}\label{Bethe-log}
  N {p_0}_{\text{tot}}(z_j) -\sum_{l=1}^n \vartheta(z_j-z_l)
  =2\pi \bigg(n_j -\frac{n+1}{2}+\frac{rn+2\ell}{L}+2\eta\sum_{l=1}^nz_l+\eta\bar\xi\bigg),\quad j=1,\ldots , n,
\end{equation}
in which we have set $z_j=\tilde\eta v_j$, $j=1,\ldots, n$, with $\tilde\eta=-\eta/\tau$. Here and in the following, unless explicitly specified, the considered theta functions are of imaginary quasi-period $\tilde\tau=-1/\tau$, i.e. $\theta_1(z)\equiv \theta_1(z;\tilde\tau)$.
In \eqref{Bethe-log}, $n_j$ are integers, ${p_0}_{\text{tot}}$ and $\vartheta$ are the bare momentum and bare phase
\begin{align}
  &{p_0}_{\text{tot}}(z) =\frac{1}{N}\sum_{k=1}^N p_0\Big(z-\tilde\xi_k+\frac{\tilde\eta}{2}\Big)
  \qquad \text{with}\quad
  p_0(z)=i\log\frac{\theta_1(\tilde\eta/2+z)}{\theta_1(\tilde\eta/2-z)}, \label{bare-mom}
  \\
  &\vartheta(z)=i\log\frac{\theta_1(\tilde\eta+z)}{\theta_1(\tilde\eta-z)}, \label{bare-ph}
\end{align}
and
\begin{equation}
  \bar\xi=\sum_{l=1}^N \Big(\frac{\tilde\eta}{2}-\tilde\xi_l\Big),
\end{equation}
with inhomogeneity parameters $\tilde\xi_l=\tilde\eta\xi_l,\ l=1,\ldots,N$.
For simplicity, we suppose in this paper that  these inhomogeneity parameters are such that $\Im\tilde\xi_l=\Im\frac{\tilde\eta}{2}$, $l=1,\ldots,N$.

A Bethe ground state $\ket{\mathsf{k},\ell}$ corresponds to a set of real solutions of \eqref{Bethe-log} for a given choice of $\ell\in\mathbb{Z}/(L-r)\mathbb{Z}$ and a successive set of $n$ integers $n_j =  j+\mathsf{k}$,  $j =1,\ldots, n$,  for a given choice of $\mathsf{k}\in\mathbb{Z}/2\mathbb{Z}$.
In the thermodynamic limit $N\to\infty$ (with $n=N/2$), the distribution of the Bethe roots corresponding to any of such ground states tends to a function $\rho_{\text{tot}}(z)$ on the interval $[-1/2,1/2]$, solution of the following integral equation:
\begin{equation}\label{liebeq}
   \rho_{\text{tot}}(z)+\int_{-1/2}^{1/2} K(z-w)\, \rho_{\text{tot}}(w)\, dw =\frac{{p'_0}_{\text{tot}}(z)}{2\pi},
\end{equation} 
where
\begin{align}
  &{p_0'}_{\text{tot}}(z) =\frac{1}{N}\sum_{k=1}^N p'_0\Big(z-\tilde\xi_k+\frac{\tilde\eta}{2}\Big)
  \quad \text{with}\quad
    p_0'(z)=i \left\{ \frac{\theta_1'(z+\tilde\eta/2)}{\theta_1(z+\tilde\eta/2)}   -\frac{\theta_1'(z-\tilde\eta/2)}{\theta_1(z-\tilde\eta/2)}\right\},\label{p-prime}\\
  &K(z)=\frac{1}{2\pi}\vartheta'(z)
         = \frac{i}{2\pi} \left\{ \frac{\theta_1'(z+\tilde\eta)}{\theta_1(z+\tilde\eta)}   -\frac{\theta_1'(z-\tilde\eta)}{\theta_1(z-\tilde\eta)}\right\}. \label{K}
\end{align}
The solution of the inhomogeneous integral equation \eqref{liebeq} is
\begin{equation}
   \rho_{\text{tot}}(z)=\frac{1}{N}\sum_{k=1}^N \rho\Big(z-\tilde\xi_k+\frac{\tilde\eta}{2}\Big),
\end{equation}
where $\rho$ is the density function solution of \eqref{liebeq} for the homogeneous model corresponding to $\tilde\xi_k=\tilde\eta/2$, $k=1,\ldots,N$:
\begin{equation}\label{rho}
   \rho(z)=\sum_{m=-\infty}^\infty \frac{e^{2\pi i m z}}{2\cosh(i\pi m\tilde\eta)}
  = \frac{1}{2\pi}
     \frac{\theta_1'(0;\tilde\eta)\,\theta_3(z;\tilde\eta)}{\theta_2(0;\tilde\eta)\,\theta_4(z;\tilde\eta)}.
\end{equation}
%

In \cite{LevT13b} we have studied the behavior of the ground states roots with respect to finite size corrections.
We have notably obtained the following result, which can easily be generalized to the inhomogeneous case:

\begin{prop}\label{prop-sum-int}
Let $f$ be a $\mathcal{C}^\infty$ 1-periodic function on $\mathbb{R}$. Then, the sum of all the values $f(x_j)$, where the set of spectral parameters $\{x_j\}_{1\leq j \leq n}$ parametrizes one of the quasi-ground states solution to \eqref{Bethe-log}, can be replaced by an integral in the thermodynamic limit according to the following rule:
\begin{equation}\label{sum-int}
\frac{1}{N} \sum_{j=1}^n f(x_j) = \int_{-1/2}^{1/2}  f(z)\, \rho_{\text{\rm tot}}(z) \, dz+ O(N^{-\infty}).
\end{equation}
Similarly, if $g$ is a $\mathcal{C}^\infty$ function such that $g'$ is 1-periodic, then
\begin{equation}\label{sum-int2}
\frac{1}{N} \sum_{j=1}^n g(x_j) = \int_{-1/2}^{1/2} g(z)\, \rho_{\text{\rm tot}}(z)\, dz +\frac{c_g}{N}\sum_{j=1}^n x_j+ O(N^{-\infty}),
\end{equation}
where $c_g=\int_{-1/2}^{1/2} g'(z)\, dz = g(1/2)-g(-1/2)$.
\end{prop}

Proposition~\ref{prop-sum-int} was used in \cite{LevT13b} to obtain a sum rule for the corresponding ground state roots, and 
to study the infinitesimal shift behavior of the ground states roots with respect to the size of the system. These results can easily be extended to the inhomogeneous case.

In particular, let us consider two different Bethe ground states $\ket{\mathsf{k}_x,\ell_x}$ and $\ket{\mathsf{k}_y,\ell_y}$, parameterized by two sets of Bethe roots  $\{x_j\}_{j=1,\ldots,n}$ and $\{y_j\}_{j=1,\ldots,n}$ for the system of Bethe equations \eqref{Bethe-log} with respective quantum numbers $\mathsf{k}_x,\ell_x$ and $\mathsf{k}_y,\ell_y$. Then we have the following sum rule:
\begin{equation}
  |x|-|y|\equiv\sum_l (x_l-y_l) 
           =\frac{L(\mathsf{k}_x-\mathsf{k}_y)+2(\ell_x-\ell_y)}{2(L-r)}+O(N^{-\infty}),\label{dif-sum}
\end{equation}
which also means that, setting $\omega_x=e^{i\pi\frac{rn+2\ell_x}{L}}$ and $\omega_y=e^{i\pi\frac{rn+2\ell_y}{L}}$,
\begin{equation}\label{id-om}
  e^{2\pi i(1-\eta)(|x|-|y|)}
                             =e^{i\pi(\mathsf{k}_x -\mathsf{k}_y)}\, \frac{\omega_x}{\omega_y}+O(N^{-\infty}).
\end{equation}
%
Still applying Proposition~\ref{prop-sum-int}, one can also evaluate several useful quantities depending on these sets of roots.
For instance, one obtains that 
\begin{equation} \label{phi-t}
  \phi(t;\{x\},\{y\})\equiv\prod_{l=1}^n \frac{\theta_1(x_l+t)}{\theta_1(y_l+t)}
    = e^{ i \pi (2k_t-1)(|x| - |y|)}+O(N^{-\infty}),
\end{equation}
where $k_t\in\mathbb{Z}$ is such that $0<\Im(t+k_t\tilde\tau)<\Im\tilde\tau$. One also obtains that
(see \cite{LevT13b} for details)
\begin{equation}
\phi_j(\{x\},\{y\})\equiv
\frac{\prod_{l=1}^n\theta_1(y_j-x_l)}{\prod_{l\not= j} \theta_1(y_j-y_l)}
     =-\frac{\theta'_1(0) }{N\pi\rho_{\text{tot}}(y_j)} \, \sin\pi(|x|-|y|)+O(N^{-\infty}) .
   \label{phi-zero}
\end{equation}
The evaluation of the products \eqref{phi-t}-\eqref{phi-zero} in terms of the difference of roots \eqref{dif-sum} will be used in Section~\ref{sec-therm} for taking the thermodynamic limit of the finite-size representations obtained in Section~\ref{sec-MPME} for the multi-point matrix elements.

To conclude this brief description of the degenerate grounds states of the model, let us finally mention that the $2(L-r)$ Bethe ground states that we have characterized above are in fact linear combinations of the $2(L-r)$ states corresponding to the flat ground state configurations of the type $(\mathsf{s},\mathsf{s}+1,\mathsf{s},\mathsf{s}+1,\ldots)$ or $(\mathsf{s}+1,\mathsf{s},\mathsf{s}+1,\mathsf{s},\ldots)$ identified in \cite{PeaS89} in the low temperature limit $\tilde\tau\to +\infty$ in which the system becomes completely ordered (see Fig.~\ref{flat-gs}).
To come back to such configurations, one has therefore to perform  the appropriate change of basis in the  $2(L-r)$-dimensional subspace $\mathrm{Fun}(\mathcal{H}_g[0])$ of the space of states  $\mathrm{Fun}(\mathcal{H}[0])$ generated by all the ground states, i.e. associated to the $2(L-r)$ largest (in magnitude) eigenvalues of the transfer matrix in the thermodynamic limit. This change of basis is defined as follows: if
\begin{equation}\label{basis-Bethe}
   \ket{\psi_g^{(\mathsf{k}_\alpha,\ell_\alpha)}}\equiv\frac{\ket{\mathsf{k}_\alpha,\ell_\alpha}}{(\braket{\mathsf{k}_\alpha,\ell_\alpha}{\mathsf{k}_\alpha,\ell_\alpha})^{1/2}}, 
   \qquad \mathsf{k}_\alpha\in\mathbb{Z}/2\mathbb{Z},\quad \mathsf{\ell}_\alpha\in\mathbb{Z}/(L-r)\mathbb{Z},
\end{equation}
denote the normalized Bethe ground states, we consider the new set of normalized ground states $\ket{\phi_g^{(\epsilon,\mathsf{t})}}$,  $\epsilon\in\{0,1\}$, $\mathsf{t}\in \{0,1\ldots,L-r-1\}$, given as
\begin{equation}\label{change-basis}
  \ket{\phi_g^{(\epsilon,\mathsf{t})}}
  =\frac{1}{\sqrt{2(L-r)}}\sum_{\mathsf{k}_\alpha=0}^1\sum_{\mathsf{\ell}_\alpha=0}^{L-r-1} 
  (-1)^{\mathsf{k}_\alpha\epsilon}e^{-i\pi\frac{r\mathsf{k}_\alpha+2\mathsf{\ell}_\alpha}{L-r}(\mathsf{t}+s_0)}\, \ket{\psi_g^{(\mathsf{k}_\alpha,\ell_\alpha)}}.
\end{equation} 
The correspondence of the states \eqref{change-basis} with the flat ground states configurations identified in the low-temperature limit is the following: if we set $s_0=\frac{\tau}{2\eta}=-\frac{1}{2\tilde\eta}$, 
the state $\ket{\phi_g^{(\epsilon,\mathsf{t})}}$ with $\mathsf{t}=a-\lfloor \eta a\rfloor$ tends, in the low-temperature limit, to the elementary $\delta$-function state $\ket{s_0+a;+,-,+,-,\ldots}$
if $\epsilon-\lfloor \eta a\rfloor$ is even, and to $\ket{s_0+a+1;-,+,-,+,\ldots}$
if $\epsilon-\lfloor \eta a\rfloor$ is odd (here $\lfloor x \rfloor$ denotes the integer part of $x$).


\section{Local height probabilities in the ABA framework}
\label{sec-ABA-corr}

Let us consider $m$ given vertices of the lattice, labelled by their positions $(i_1,j_1), (i_2,j_2),\ldots,(i_m,j_m)$, $1\le i_k\le N$, $1\le j_k \le M$, starting from the reference vertex $(1,1)$ situated at the upper left corner of the lattice. We suppose that these vertices are ordered in such a way that $j_1\le j_2\le\ldots\le j_m$.
The probability that the heights at these sites have respective values $s_1,s_2,\ldots,s_m$ is given by
\begin{multline}\label{LHP-0}
  \mathbf{P}^{(M,N)}_{(i_1,j_1), (i_2,j_2),\ldots,(i_m,j_m)}(s_1,s_2,\ldots,s_m)
  =\left[ Z^{(M,N)}\right]^{-1}
  \\
  \times
  \sum_{\substack{\text{heights } \tilde{s}_{i,j}\\ 1\le i \le N,\ 1\le j\le M}}
  \prod_{k=1}^m\delta_{s_k}(\tilde{s}_{i_k,j_k}) \cdot \prod_{i=1}^N\prod_{j=1}^M
  W\binom{\tilde{s}_{i,j}\quad \tilde{s}_{i,j+1}}{\tilde{s}_{i+1,j}\ \tilde{s}_{i+1,j+1}},
\end{multline}
where we have  imposed periodic boundary conditions both along the vertical and horizontal directions of the lattice (i.e. $\tilde{s}_{N+1,j}\equiv\tilde{s}_{1,j}$, and $\tilde{s}_{i,M+1}\equiv\tilde{s}_{i,1}$). 
$Z^{(M,N)}$ denotes here the partition function of the model with periodic boundary conditions:
\begin{equation}\label{Z_MN}
   Z^{(M,N)}= \sum_{\substack{\text{heights } \tilde{s}_{i,j}\\ 1\le i \le N,\ 1\le j\le M}}
  \prod_{i=1}^N\prod_{j=1}^M
  W\binom{\tilde{s}_{i,j}\quad \tilde{s}_{i,j+1}}{\tilde{s}_{i+1,j}\ \tilde{s}_{i+1,j+1}}.
\end{equation}
As usual \cite{KauO49}, the multi-point local probability \eqref{LHP-0} can be expressed in terms of the (vertical) transfer matrices $\widehat{t}(w_j)\equiv\widehat{t}(w_j;\xi_1,\ldots,\xi_N)$ \eqref{transfer} as the following trace over the space of states $\mathbb{H}[0]$,
\begin{multline}\label{LHP-1}
  \mathbf{P}^{(M,N)}_{(i_1,j_1), (i_2,j_2),\ldots,(i_m,j_m)}(s_{1},s_{2},\ldots,s_{m})\\
  =\frac{\tr_{\mathbb{H}[0]}\left(  
             \prod\limits_{k=1}^{j_1-1}\widehat{t}(w_k)\cdot \widehat{\delta}_{s_{1}}^{(i_1)}\cdot 
             \prod\limits_{k=j_1}^{j_2-1}\widehat{t}(w_k)\cdot \widehat{\delta}_{s_{2}}^{(i_2)}\ldots 
             \prod\limits_{k=j_{m-1}}^{j_m-1}\!\!\! \widehat{t}(w_k)\cdot \widehat{\delta}_{s_{m}}^{(i_m)}\cdot
             \prod\limits_{k=j_m}^M\!\! \widehat{t}(w_k)\right)}
            {\tr_{\mathbb{H}[0]}\left(  \prod\limits_{k=1}^M \widehat{t}(w_k)\right)},
\end{multline}
and the trace in \eqref{LHP-1} can be expressed in terms of the Bethe eigenstates of the transfer matrix so that, in the limit $M\to\infty$, only the ground states characterized in Section~\ref{sec-gs-th} effectively contribute to the trace:
\begin{align}\label{LHP-2}
\mathbf{P}^{(N)}_{(i_1,j_1), (i_2,j_2),\ldots,(i_m,j_m)}(s_{1},s_{2},\ldots,s_{m})
&=
 \lim_{M\to\infty}\mathbf{P}^{(M,N)}_{(i_1,j_1), (i_2,j_2),\ldots,(i_m,j_m)}(s_{1},s_{2},\ldots,s_{m})
   \nonumber\\
&=\sum_{\mathsf{k},\ell}\,
       \bra{\psi_g^{(\mathsf{k},\ell)} }\, 
       \prod\limits_{k=1}^{j_1-1}\widehat{t}(w_k)\cdot \widehat{\delta}_{s_{1}}^{(i_1)}\cdot 
             \prod\limits_{k=j_1}^{j_2-1}\widehat{t}(w_k)\cdot \widehat{\delta}_{s_{2}}^{(i_2)}\ldots 
             \nonumber\\
 &\quad  \ldots         
             \prod\limits_{k=j_{m-1}}^{j_m-1}\!\!\! \widehat{t}(w_k)\cdot \widehat{\delta}_{s_{m}}^{(i_m)}\cdot
             \prod\limits_{k=1}^{j_m-1}\!\! \widehat{t}^{-1}(w_k) \,
       \ket{\psi_g^{(\mathsf{k},\ell)} }.
\end{align}
Here, $\ket{\psi_g^{(\mathsf{k},\ell)} }$ denotes the normalized Bethe ground state \eqref{basis-Bethe}, and the sum runs over all values of $\mathsf{k}\in\mathbb{Z}/2\mathbb{Z}$ and of $\ell\in\mathbb{Z}/(L-r)\mathbb{Z}$.
In \eqref{LHP-1} and \eqref{LHP-2}, $\widehat{\delta}_s^{(i)}$ denotes the local operator whose action on elementary $\delta$-function states \eqref{delta-state} is defined as:
\begin{equation}\label{delta_s^i}
   \widehat{\delta}_s^{(i)}\, \ket{s_1;\varepsilon_1,\varepsilon_2,\ldots,\varepsilon_N}
   =\delta_s(s_1+\varepsilon_1+\varepsilon_2+\cdots+\varepsilon_{i-1})\,
     \ket{s_1;\varepsilon_1,\varepsilon_2,\ldots,\varepsilon_N},
\end{equation}
i.e. it amounts to fixing to $s$ the value of the height at the $i$-th site of the considered vertical line.

To compute the quantity \eqref{LHP-2}, the problem is therefore to be able to act with such operators on the ground states (and more generally on the Bethe eigenstates \eqref{state}) so as to reduce the mean value \eqref{LHP-2} to a sum of scalar products of Bethe states \eqref{scalarproduct}, or at least of partial scalar products \eqref{Sn}. As in the case, considered in \cite{LevT13a}, of elementary operators $E_i^{\alpha \beta}$ acting on vertical {\em bonds} of the lattice (i.e. acting on the corresponding spin variable), the idea is to solve the inverse problem \cite{KitMT99,MaiT00,GohK00}.
We recall that, in \cite{LevT13a}, a local operator  $E_i^{\alpha \beta}$ acting on the $i$-th  bond of a given vertical line was reconstructed in terms of the entries of the (vertical) monodromy matrix \eqref{mon-op} in the following way:
\begin{thm} \cite{LevT13a}
  Let $E_i^{\alpha\beta}$ be the elementary matrix,  acting on the $i$-th space of the tensor product $\mathcal{H}=(\mathbb{C}^2)^{\otimes N}$, with elements $(E_i^{\alpha\beta})_{jk}=\delta_j^\alpha \delta_k^\beta$, where $\alpha$ and $\beta$ are equal to $\pm 1$. It can be expressed in terms of the entries of the monodromy matrix \eqref{mon-op} in the following way:
  \begin{equation}\label{inv-pb1}
    E_i^{\alpha\beta}    
    = \prod_{k=1}^{i-1} \widehat{t}(\xi_k) \cdot \widehat{T}_{\beta \alpha} (\xi_i)
        \cdot \prod_{k=1}^i \big[ \widehat{t}(\xi_k) \big]^{-1} \cdot\, \widehat{\tau}_s^{\beta-\alpha}.
  \end{equation}
  %
  %
\end{thm}
The idea is here to express similarly the local operators $\widehat{\delta}_s^{(i)}$ in terms of elements of the monodromy matrix and, in addition, of simple functions of the operators $\widehat{s},\widehat{\tau}_s$.

In fact, since the action of $\widehat{\delta}_s^{(1)}$ on any elementary $\delta$-function state \eqref{delta-state} amounts to fixing the reference value (height at site 1) of the dynamical parameter, it means that this action on any element $\ket{\psi}\in\mathbb{H}$ corresponds to the multiplication by the numerical function $\delta_s$ \eqref{delta-s}:
\begin{equation}\label{act-delta1}
   \Big(\widehat{\delta}_s^{(1)}\,\ket{\psi}\Big)(\tilde{s})
   =\Big(\delta_s(\widehat{s})\,\ket{\psi}\Big)(\tilde{s})
   =\delta_s(\tilde{s})\cdot\ket{\psi}(\tilde{s})=\ket{\psi}(s).
\end{equation} 

\begin{rem}\label{act-delta}
It follows from \eqref{act-delta} and from the decomposition
 \begin{equation}
    \delta_s(\tilde{s})=\frac{1}{L}\sum_{j=0}^{L-1} e^{-2\pi i \frac{j(s-\tilde{s})}{L}},
 \end{equation} 
 that the action of the operator  $\widehat{\delta}_s^{(1)}$ on a Bethe state $\ket{\{v\},\omega_v}$ \eqref{state} leads to a linear combination of $L$ Bethe-type states:
 \begin{equation}\label{delta-Bethe}
   \widehat{\delta}_s^{(1)}\, \ket{\{v\},\omega_v}
   =\delta_s(\widehat{s})\, \ket{\{v\},\omega_v}
   =\frac{1}{L}\sum_{j=0}^{L-1} e^{-2\pi i\frac{j s}{L}}\,
   \ket{\{v\},\omega_v\, e^{2\pi i \frac{j}{L}}}.
 \end{equation}
 Note that the resulting Bethe-type states $\ket{\{v\},\omega_v\, e^{2\pi i \frac{j}{L}}}$   are a priori no longer eigenstates of the transfer matrix (even  if $\ket{\{v\},\omega_v}$ was), since the combination of parameters $(\{v\},\omega_v\, e^{2\pi i \frac{j}{L}})$ does not in general satisfy the Bethe equations.
\end{rem}

\begin{rem}\label{ff-delta}
The form factor of the local   $\widehat{\delta}_s^{(1)}$ operator between two Bethe eigenstates is proportional to the partial scalar product \eqref{Sn}:
\begin{equation}\label{ff-delta-Sn}
   \bra{\{u\},\omega_u}\,  \widehat{\delta}_s^{(1)} \,  \ket{\{v\},\omega_v}
   = \bra{\{u\},\omega_u}\, {\delta}_s( \widehat{s})\,  \ket{\{v\},\omega_v}
   = \widetilde\varphi_{\omega_u}(s)\,\varphi_{\omega_v}(s) \, S_n ( \{u\}; \{v\} ; s).
\end{equation}
It can be expressed as sum over $L$ determinants, either by virtue of Remark~\ref{act-delta} and using the determinant representation for the scalar product of a Bethe state with a Bethe eigenstate, or simply via the formula~\eqref{result-spL}.
\end{rem}
More generally, the local operator $\widehat{\delta}_s^{(i)}$ can be reconstructed as follows.
\begin{thm} 
  The operator $\widehat{\delta}_s^{(i)}\in\mathrm{End}(\mathbb{H})$, defined by its action \eqref{delta_s^i} on the basis \eqref{delta-state} of $\mathbb{H}$, can be expressed as
  \begin{equation}\label{inv-pb2}
    \widehat{\delta}_s^{(i)}    
    = \prod_{k=1}^{i-1} \widehat{t}(\xi_k) \cdot \delta_s(\widehat{s})
        \cdot \prod_{k=1}^{i-1} \big[ \widehat{t}(\xi_k) \big]^{-1}.
  \end{equation}
\end{thm}
\Proof
The relation \eqref{inv-pb2} follows from a trivial recursion on $i$ using the fact that
\begin{equation}\label{rec-delta}
   \widehat{\delta}_s^{(i)} 
   = \widehat{\delta}_{s-1}^{(i-1)} \, E_{i-1}^{++} + \widehat{\delta}_{s+1}^{(i-1)} \, E_{i-1}^{--} ,
\end{equation}
and the solution of the inverse problem \eqref{inv-pb1} for $E_{i-1}^{++}$ and $E_{i-1}^{--} $.
The equality \eqref{rec-delta} can easily be proven by acting with both sides of the equation on the basis of elementary $\delta$-function states \eqref{delta-state}.
\qed

Hence, a completely arbitrary multi-point local height probability such as \eqref{LHP-2} can be expressed in this framework as
\begin{multline}\label{LHP-3}
\mathbf{P}^{(N)}_{(i_1,j_1), (i_2,j_2),\ldots,(i_m,j_m)}(s_{1},s_{2},\ldots,s_{m})
=\sum_{\mathsf{k},\ell}\,
       \bra{\psi_g^{(\mathsf{k},\ell)} }\, 
       \prod\limits_{k=1}^{j_1-1}\widehat{t}(w_k)\,
       \prod\limits_{l=1}^{i_1-1}\widehat{t}(\xi_l)
       \cdot\delta_{s_{1}}(\widehat{s})\cdot
             \prod\limits_{k=j_1}^{j_2-1}\widehat{t}(w_k)\\
             \times
             \prod\limits_{l=i_1}^{i_2-1}\big[ \widehat{t}(\xi_l)\big]^{H(i_2-i_1)}\,
             \prod\limits_{l=i_2}^{i_1-1}\big[ \widehat{t}(\xi_l)\big]^{-H(i_1-i_2)}\,
             \cdot \delta_{s_{2}}(\widehat{s})\ldots       
             \prod\limits_{k=j_{m-1}}^{j_m-1}\!\!\! \widehat{t}(w_k)\,
             \prod\limits_{l=i_{m-1}}^{i_m-1}\!\!\!\big[ \widehat{t}(\xi_l)\big]^{H(i_m-i_{m-1})}\\
             \times
             \prod\limits_{l=i_{m}}^{i_{m-1}-1}\!\!\!\big[ \widehat{t}(\xi_l)\big]^{-H(i_{m-1}-i_{m})}
             \cdot \delta_{s_{m}}(\widehat{s})\cdot
             \prod\limits_{k=1}^{j_m-1}\! \widehat{t}^{-1}(w_k) \,
             \prod\limits_{l=1}^{i_m-1}\! \widehat{t}^{-1}(\xi_l)\,
       \ket{\psi_g^{(\mathsf{k},\ell)} },
\end{multline}
in which $H$ denotes the Heaviside step function.
%
%

  The fact that we have access to {\em any} multi-point local height probability on the lattice, and not only to local probabilities at points on a same line, as it is usually the case by means of other approaches \cite{JimM95L,LukP96}, is due to the solution of the inverse problem \eqref{inv-pb2}.
  In the same way, one could also compute, using the solution of the inverse problem \eqref{inv-pb1}, the correlation function between spins sitting on any (vertical or horizontal) bonds of the lattice (or more generally the correlation between any combination of local heights and local spins). 
  More precisely, the contribution to the trace of a spin $\alpha$ sitting 
  on the vertical link $(i_1,j_1)_{\text{v}}$ between vertices $(i_1,j_1)$ and $(i_1+1,j_1)$
  corresponds to the insertion of the local operator $E_{i_1}^{\alpha \alpha}$, reconstructed as in \eqref{inv-pb1}, whereas the contribution  of a spin $\beta$ sitting 
  on the horizontal link $(i_2,j_2)_{\text{h}}$ between vertices $(i_2,j_2)$ and $(i_2,j_2+1)$  corresponds to the insertion of the product
  \begin{equation}
     \prod_{l=1}^{i_2-1}\widehat{t}(\xi_l) \cdot \widehat{T}_{\beta \beta} (w_{j_2})
        \cdot \prod_{l=1}^{i_2-1} \big[ \widehat{t}(\xi_l) \big]^{-1}.
  \end{equation}
%

Hence, in general, the probability to find any combination of given values $x_1,x_2,\ldots,x_m$ of heights or spins on positions $\mathbf{I}_1,\mathbf{I}_2,\ldots,\mathbf{I}_m$ of vertices or (vertical or horizontal) bonds on the lattice can be expressed as follows:
\begin{multline}\label{anyProb}
\mathbf{P}^{(N)}_{\mathbf{I}_1,\mathbf{I}_2,\ldots,\mathbf{I}_m}(x_1,x_2,\ldots,x_m)
=\sum_{\mathsf{k},\ell}\,
       \bra{\psi_g^{(\mathsf{k},\ell)} }\, 
       \prod\limits_{k=1}^{j_1-1}\widehat{t}(w_k)\,
              \prod\limits_{l=1}^{i_1-1}\widehat{t}(\xi_l)
       \cdot\widehat{X}_{\mathbf{I}_1}^{x_1}\cdot 
             \prod\limits_{k=j_1}^{j_2-1}\widehat{t}(w_k)\,\\
       \times
\prod\limits_{l=i_1}^{i_2-1}\big[ \widehat{t}(\xi_l)\big]^{H(i_2-i_1)}\,
             \prod\limits_{l=i_2}^{i_1-1}\big[ \widehat{t}(\xi_l)\big]^{-H(i_1-i_2)}\,
             \cdot \widehat{X}_{\mathbf{I}_2}^{x_2}\ldots 
              \prod\limits_{k=j_{m-1}}^{j_m-1}\!\!\! \widehat{t}(w_k)\,
             \prod\limits_{l=i_{m-1}}^{i_m-1}\!\!\!\big[ \widehat{t}(\xi_l)\big]^{H(i_m-i_{m-1})}\\
             \times
             \prod\limits_{l=i_{m}}^{i_{m-1}-1}\!\!\!\big[ \widehat{t}(\xi_l)\big]^{-H(i_{m-1}-i_{m})}
             \cdot \widehat{X}_{\mathbf{I}_m}^{x_m}\cdot
             \prod\limits_{k=1}^{j_m-1}\! \widehat{t}^{-1}(w_k) \,
             \prod\limits_{l=1}^{i_m-1}\! \widehat{t}^{-1}(\xi_l)\,
       \ket{\psi_g^{(\mathsf{k},\ell)} }.
\end{multline}
In \eqref{anyProb}, the prescription is as follows:
\begin{itemize}
 \item $\widehat{X}_{\mathbf{I}_k}^{x_k}=\delta_{s_k}(\widehat{s})\quad$ if $\mathbf{I}_k\equiv (i_k,j_k)$ labels the position of a vertex and $x_k\equiv s_k$ is the value of the height at this vertex;
 \item $\widehat{X}_{\mathbf{I}_k}^{x_k}=\widehat{T}_{\alpha_k \alpha_k}(\xi_{i_k})\cdot \widehat{t}^{-1}(\xi_{i_k})\quad$
 if $\mathbf{I}_k\equiv (i_k,j_k)_{\text{v}}$ labels the position of a vertical bond and $x_k\equiv \alpha_k$ is the value of the spin on this bond;
 \item $\widehat{X}_{\mathbf{I}_k}^{x_k}=\widehat{T}_{\beta_k \beta_k}(w_{j_k})\cdot \widehat{t}^{-1}(w_{j_k})\quad$
 if $\mathbf{I}_k\equiv (i_k,j_k)_{\text{h}}$ labels the position of a horizontal bond and $x_k\equiv \beta_k$ is the value of the spin on this bond
\end{itemize}

\begin{figure}[h]
   \begin{minipage}[c]{.46\linewidth}
      \includegraphics[width=7cm]{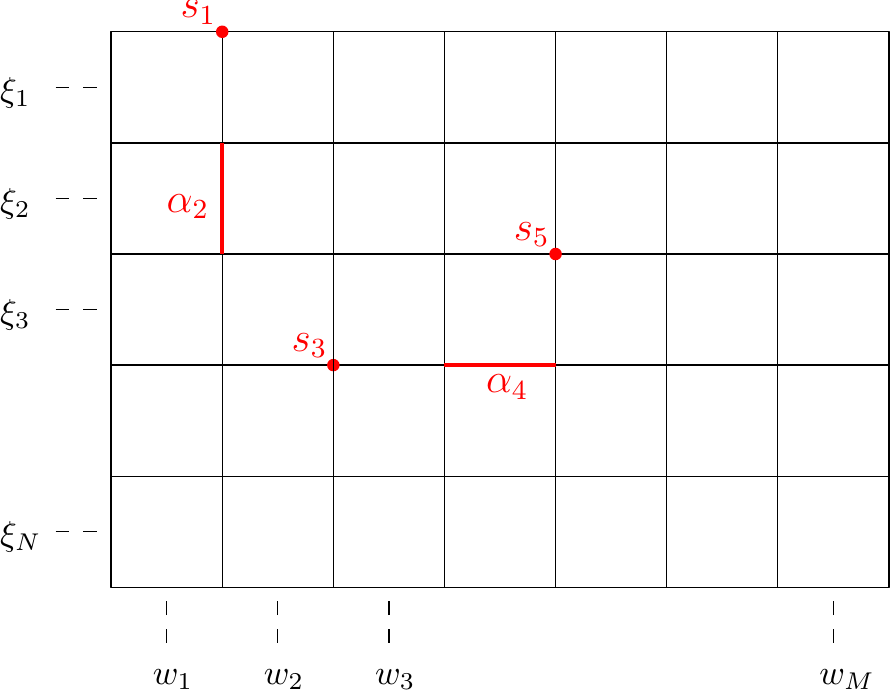}
   \end{minipage} \hfill
   \begin{minipage}[c]{.46\linewidth}
      \includegraphics[width=7cm]{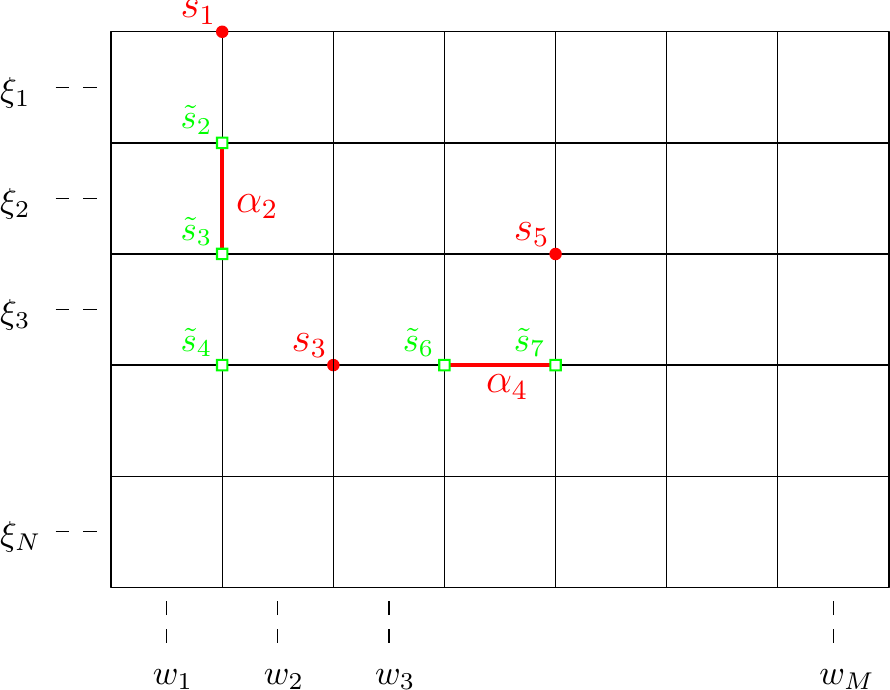}
   \end{minipage}
   \caption{\label{LHPAP} The probability associated to the configuration of spins/heights on the left figure is $\mathbf{P}_{(1,2),(2,2)_{\text{v}},(4,3),(4,4)_{\text{h}},(3,5)}(s_1,\alpha_2,s_3,\alpha_4,s_5)$.
   It can be computed as the sum over the LHPAP 
   $\mathbf{P}_{(1,2),(2,2),(3,2),(4,2),(4,3),(4,4),(4,5),(3,5)}({s}_1,\tilde{s}_2,\tilde{s}_3,\tilde{s}_4,{s}_3,\tilde{s}_6,\tilde{s}_7,{s}_8)$, associated to all possible height configurations at adjacent points presented on the right figure such that $\tilde{s}_3=\tilde{s}_2+\alpha_2$ and $\tilde{s}_7=\tilde{s}_6+\alpha_4$.}
\end{figure}

It is therefore possible to compute any of these quantities by summation over the corresponding form factors. We recall that the form factors of local spin operators can be expressed as a determinant of usual functions of the model (see \cite{LevT13a}), and that the form factors of local height operators \eqref{ff-delta-Sn} can be expressed as a sum of $L$ such determinants according to Remark~\ref{ff-delta}.
Alternatively, one can decompose \eqref{anyProb} as a sum over local height probabilities at adjacent points (see Fig.~\ref{LHPAP}). In the remaining part of this paper, we will explain how to compute these local height probabilities at adjacent points (LHPAP) in the case of the CSOS model at the thermodynamic limit.

Let us therefore consider $m+1$ vertices of the lattice at positions $\mathbf{I}_1=(i_1,j_1)$, $\mathbf{I}_2=(i_2,j_2),\ldots,$ $\mathbf{I}_{m+1}=(i_{m+1},j_{m+1})$, with $j_1\le j_2\le\ldots\le j_{m+1}$. We suppose that, for any $1\le k \le m$, $\mathbf{I}_k$ and $\mathbf{I}_{k+1}$ are nearest neighbors on the lattice, i.e. $(i_{k+1},j_{k+1})=(i_k+\varepsilon_k,j_k+\varepsilon'_k)$, with either $(\varepsilon_k,\varepsilon'_k)=(\pm 1,0)$ (if $\mathbf{I}_k$ and $\mathbf{I}_{k+1}$ are separated by a vertical bound) or $(\varepsilon_k,\varepsilon'_k)=(0, 1)$ (if $\mathbf{I}_k$ and $\mathbf{I}_{k+1}$ are separated by a horizontal bound).
The probability that the heights at these sites have respective values $s_1,s_2,\ldots,s_{m+1}$ is hence given by the expression
\begin{multline}\label{LHP-4}
\mathbf{P}^{(N)}_{\mathbf{I}_1, \mathbf{I}_2,\ldots,\mathbf{I}_{m+1}}(s_{1},s_{2},\ldots,s_{m+1})
=\sum_{\mathsf{k},\ell}\,
       \bra{\psi_g^{(\mathsf{k},\ell)} }\, 
       \prod\limits_{k=1}^{j_1-1}\widehat{t}(w_k)\,
       \prod\limits_{l=1}^{i_1-1}\widehat{t}(\xi_l)\\
       \times
       \delta_{s_{1}}(\widehat{s})\cdot 
             \widehat{t}^{\varepsilon'_1}(w_{j_1})\,
             \widehat{t}^{\,\varepsilon_1 H(\varepsilon_1)}(\xi_{i_1})\,
             \widehat{t}^{\,\varepsilon_1 H(-\varepsilon_1)}(\xi_{i_2})\,
             \cdot \delta_{s_{2}}(\widehat{s})\ldots 
             \widehat{t}^{\varepsilon'_m}(w_{j_{m}})\, 
             \widehat{t}^{\,\varepsilon_m H(\varepsilon_m)}(\xi_{i_m})\,
             \\
             \times
             \widehat{t}^{\,\varepsilon_m H(-\varepsilon_m)}(\xi_{i_{m+1}})\,
             \cdot \delta_{s_{m+1}}(\widehat{s})\cdot
             \prod\limits_{k=1}^{j_{m+1}-1}\! \widehat{t}^{-1}(w_k) \,
             \prod\limits_{l=1}^{i_{m+1}-1}\! \widehat{t}^{-1}(\xi_l)\,
       \ket{\psi_g^{(\mathsf{k},\ell)} },
\end{multline}
in which we recall that $H$ denotes the Heaviside step function.
Using moreover the inversion property of the transfer matrix evaluated at any of the inhomogeneity parameters,
\begin{equation}\label{invert-t}
   \frac{\widehat{t}(\xi_i)}{\mathsf{a}(\xi_i)}\cdot  \frac{\widehat{t}(\xi_i-1)}{\mathsf{d}(\xi_i-1)}
   =\frac{[\widehat{s}]}{[\widehat{s}+h_{1\ldots N}]},
\end{equation}
we can rewrite \eqref{LHP-4} in terms of a simple product of matrix elements of the monodromy matrix as
\begin{multline}\label{LHP-5}
\mathbf{P}^{(N)}_{\mathbf{I}_1, \mathbf{I}_2,\ldots,\mathbf{I}_{m+1}}(s_{1},s_{2},\ldots,s_{m+1})
=\sum_{\mathsf{k},\ell}\,
       \bra{\psi_g^{(\mathsf{k},\ell)} }\, 
       \prod\limits_{k=1}^{j_1-1}\widehat{t}(w_k)\,
       \prod\limits_{l=1}^{i_1-1}\widehat{t}(\xi_l)\cdot
       \delta_{s_{1}}(\widehat{s})\cdot
             \widehat{T}_{\alpha_1\alpha_1}(\zeta_1)\\
             \times
             \widehat{T}_{\alpha_2\alpha_2}(\zeta_2)\ldots 
             \widehat{T}_{\alpha_m\alpha_m}(\zeta_m)\, 
             \prod_{k=1}^m\widehat{t}^{-1}(\zeta_k)\,             \prod\limits_{k=1}^{j_{1}-1}\! \widehat{t}^{-1}(w_k) \,
             \prod\limits_{l=1}^{i_{1}-1}\! \widehat{t}^{-1}(\xi_l)\,
       \ket{\psi_g^{(\mathsf{k},\ell)} }.
\end{multline}
In \eqref{LHP-5}, we have set, for  $1\le k\le m$,
\begin{equation}\label{alpha}
\alpha_k= s_{k+1}-s_k ,
\end{equation}
and
\begin{equation}\label{zeta}
   \zeta_k=
   \begin{cases}
      w_{j_k} &\text{if}\quad (\varepsilon_k,\varepsilon'_k)=(0,1),\\
      \xi_{i_k} &\text{if}\quad (\varepsilon_k,\varepsilon'_k)=(1,0),\\
      \xi_{i_k-1}-1 &\text{if}\quad (\varepsilon_k,\varepsilon'_k)=(-1,0).
   \end{cases}   
\end{equation}

We will in fact consider more specific quantities, namely local height probabilities in one of the background ground states $\ket{\phi_g^{(\eps,\mathsf{t})}}$ tending to the flat ground state configurations identified in \cite{PeaS89} in the low-temperature limit:
\begin{multline}\label{LHP-6}
\mathbf{P}^{(\epsilon,\mathsf{t})}_{\mathbf{I}_1, \mathbf{I}_2,\ldots,\mathbf{I}_{m+1}}(s_{1},s_{2},\ldots,s_{m+1})
\equiv  \bra{\phi_g^{(\epsilon,\mathsf{t})} }\, 
       \prod\limits_{k=1}^{j_1-1}\widehat{t}(w_k)\,
       \prod\limits_{l=1}^{i_1-1}\widehat{t}(\xi_l)\cdot
       \delta_{s_{1}}(\widehat{s})\cdot
             \widehat{T}_{\alpha_1\alpha_1}(\zeta_1)\\
             \times
             \widehat{T}_{\alpha_2\alpha_2}(\zeta_2)\ldots 
             \widehat{T}_{\alpha_m\alpha_m}(\zeta_m)\, 
             \prod_{k=1}^m\widehat{t}^{-1}(\zeta_k)\,             \prod\limits_{k=1}^{j_{1}-1}\! \widehat{t}^{-1}(w_k) \,
             \prod\limits_{l=1}^{i_{1}-1}\! \widehat{t}^{-1}(\xi_l)\,
        \ket{\phi_g^{(\epsilon,\mathsf{t})} },
\end{multline}
so that, according to the change of basis  \eqref{change-basis}, we will need to compute not only means values in the Bethe ground states \eqref{basis-Bethe}, as in \eqref{LHP-5}, but also more general matrix elements of the form
\begin{multline}\label{LHP-7}
\mathbb{P}^{(\mathsf{k}_1,\ell_1;\mathsf{k}_2,\ell_2)}_{\mathbf{I}_1, \mathbf{I}_2,\ldots,\mathbf{I}_{m+1}}(s_{1},s_{2},\ldots,s_{m+1})
\equiv   \bra{\psi_g^{(\mathsf{k}_1,\ell_1)} }\, 
       \prod\limits_{k=1}^{j_1-1}\widehat{t}(w_k)\,
       \prod\limits_{l=1}^{i_1-1}\widehat{t}(\xi_l)\cdot
       \delta_{s_{1}}(\widehat{s})\cdot
             \widehat{T}_{\alpha_1\alpha_1}(\zeta_1)\\
             \times
             \widehat{T}_{\alpha_2\alpha_2}(\zeta_2)\ldots 
             \widehat{T}_{\alpha_m\alpha_m}(\zeta_m)\, 
             \prod_{k=1}^m\widehat{t}^{-1}(\zeta_k)\,             \prod\limits_{k=1}^{j_{1}-1}\! \widehat{t}^{-1}(w_k) \,
             \prod\limits_{l=1}^{i_{1}-1}\! \widehat{t}^{-1}(\xi_l)\,
         \ket{\psi_g^{(\mathsf{k}_2,\ell_2)} },
\end{multline}
where the parameters $\alpha_k$, $\zeta_k$, $1\le k\le m$, are defined according to the prescriptions \eqref{alpha}, \eqref{zeta}.

For simplicity, we will suppose in the following that the column inhomogeneity parameters $w_{j_k}$ involved in the definition \eqref{zeta} of the parameter $\zeta_k$ belong to the set of (possibly shifted) line inhomogeneity parameters\footnote{This choice is compatible with the homogeneous limit. It is possible to consider more general inhomogeneity parameters, but the resulting representations for the local height probabilities could be slightly more complicated.}, i.e. that $w_{j_k}\in\{\xi_1,\ldots,\xi_N\}\cup\{\xi_1-1,\ldots,\xi_N-1\}$. For the ease of computations, we will also suppose that the parameters $\zeta_k$ are all distinct (modulo $\frac{1}{\eta}\mathbb{Z}+\frac{\tau}{\eta}\mathbb{Z}$): if two or more of these parameters are equal, the corresponding homogeneous limit should be taken in our final formulas.

In the next section, we explain how to reduce the computation of the quantity \eqref{LHP-7} to a sum over partial scalar products \eqref{Sn}, and hence to a sum over ratios of determinants of the form \eqref{def-OmegaL} and \eqref{mat-Phi}. Then, in Section~\ref{sec-therm}, we explain how to take the thermodynamic limit of the resulting expression. Finally, in Section~\ref{sec-MPLHP}, we give the result for the quantity \eqref{LHP-6} at the thermodynamic limit.

\section{Finite-size multi-point matrix elements}
\label{sec-MPME}

So as to compute the local  height probabilities at adjacent points \eqref{LHP-6}, we consider the normalized multi-point matrix elements of the form
\begin{align}\label{MPME1}
   &\mathbb{P}^{(\{u\},\omega_u;\{v\},\omega_v)}_{\{\zeta\} }(s;\alpha_1,\ldots,\alpha_m) 
   \nonumber\\
   &\hspace{3cm}=  \frac{\bra{\{u\},\omega_u}\, \delta_s(\widehat{s})\, \widehat{T}_{\alpha_1\alpha_1}(\zeta_1)\ldots 
             \widehat{T}_{\alpha_m\alpha_m}(\zeta_m)\, 
             \prod_{k=1}^m\widehat{t}^{-1}(\zeta_k)\,  \ket{\{v\},\omega_v} }{ \big( \braket{ \{u\},\omega_u}{ \{u\},\omega_u } \braket{ \{v\},\omega_v}{ \{v\},\omega_v } \big)^{1/2}} ,
            \nonumber \\
  &\hspace{3cm} =  \frac{\omega_v^{s+\alpha_{1,\ldots,m}}}{L\,\omega_u^s}\prod_{j=1}^n\frac{[s+j-1]}{[s+\alpha_{1,\ldots,m}-j]}\
\prod_{j=1}^m \tau^{-1}(\zeta_j;\{v\},\omega_v)\nonumber\\
 &\hspace{1.5cm} \times
      \frac{\bra{\mathbb{0}}\, \widehat{C}(u_1)\ldots\widehat{C}(u_n)\, \delta_s(\widehat{s})\, \widehat{T}_{\alpha_1\alpha_1}(\zeta_1)\ldots \widehat{T}_{\alpha_m\alpha_m}(\zeta_m) \, \widehat{B}(v_n)\ldots\widehat{B}(v_1)\, \ket{\mathbb{0}} }{  \big( \braket{ \{u\},\omega_u}{ \{u\},\omega_u } \braket{ \{v\},\omega_v}{ \{v\},\omega_v } \big)^{1/2}} ,     
\end{align}
where $\ket{\{u\},\omega_u}$ and $\ket{\{v\},\omega_v}$ are two (possibly different) Bethe eigenstates of the model,
and $\{\zeta_1,\ldots,\zeta_m\}\equiv\{\zeta\}$ is a set of $m$ arbitrary complex parameters.
Here and in the following, we use the simplified notation: $\alpha_{i,\ldots,j}=\sum_{l=i}^j\alpha_l$ for $i<j$.

The matrix elements \eqref{MPME1} can be evaluated, as usual \cite{KitMT99,KitMT00}, by acting with the operators $\widehat{T}_{\alpha_i\alpha_i}(\zeta_i)$ on one of the (left or right) Bethe eigenstates, and  by computing the scalar product of the resulting state with the remaining Bethe eigenstate. The multiple action of the operator entries $\widehat{T}_{\alpha_i\alpha_i}(\zeta_i)$ on the right state $\widehat{B}(v_n)\ldots\widehat{B}(v_1)\, \ket{\mathbb{0}}$ in \eqref{MPME1} can be computed similarly as in \cite{KitMT00} from the quadratic commutation relations given by the $R$-matrix.
Using in particular that
\begin{align}
   &\widehat{A}(v_{n+1})\ \prod_{j=1}^n \widehat{B}(v_j) \, \ket{\mathbb{0}}
     = 
     \sum_{j=1}^{n+1} \mathsf{a}(v_j)\, \frac{ [\widehat{s}+v_j-v_{n+1}] }{ [\widehat{s}-n] }  \,
     \frac{ \pl_{l=1}^n [v_l-v_j+1] }{ \pl_{\substack{l=1\\ l\not= j}}^{n+1} [v_l-v_j] }\
     \prod_{\substack{l=1\\ l\not= j}}^{n+1} \widehat{B}(v_l)  \ket{\mathbb{0}}  \label{ABB},
\\
    & \widehat{D}(v_{n+1})\ \prod_{j=1}^n \widehat{B}(v_j) \, \ket{\mathbb{0}}
     = \frac{[\widehat{s}-n-1]}{[\widehat{s}-1]}
     \sum_{j=1}^{n+1} \mathsf{d}(v_j)\, \frac{ [\widehat{s}+v_j-v_{n+1}] }{ [\widehat{s}] }  \nonumber\\
    &\hspace{7.7cm}\times 
     \frac{ \pl_{l=1}^n [v_l-v_j-1] }{ \pl_{\substack{l=1\\ l\not= j}}^{n+1} [v_l-v_j] }\
     \prod_{\substack{l=1\\ l\not= j}}^{n+1} \widehat{B}(v_l)  \ket{\mathbb{0}}  \label{DBB},
 \end{align}
and defining the following sets of indices,
\begin{xalignat}{2}
&\boldsymbol{\alpha}_- = \{ j: 1\le j\le m, \alpha_j = -1 \}
   =\{i_p\}_{p\in\{1,\ldots,|\boldsymbol{\alpha}_-| \} }
    & &\text{with}\ \ i_k<i_l \ \ \text{if}\ \ k<l\le |\boldsymbol{\alpha}_-|,\label{alpha-}
 \\
&\boldsymbol{\alpha}_+ = \{ j: 1\le j\le m, \alpha_j = 1 \}=\{i_p\}_{p\in\{|\boldsymbol{\alpha}_-|+1,\ldots,m\} }
   & &\text{with}\ \ i_k>i_l \ \ \text{if}\ \ |\boldsymbol{\alpha}_-|<k<l, \label{alpha+}
\end{xalignat}
where $|\boldsymbol{\alpha}_-|$ denotes the cardinality of $\boldsymbol{\alpha}_-$, one obtains that
\begin{equation}\label{TTTBBB}
   \widehat{T}_{\alpha_1\alpha_1}(\zeta_1)\ldots \widehat{T}_{\alpha_m\alpha_m}(\zeta_m) \
   \prod_{j=1}^n \widehat{B}(v_j) \, \ket{\mathbb{0}}
   =\sum_{ \mathbf{b} } F_{\mathbf{b}}(\widehat{s};\{v\},\{\zeta\}) \
      \prod_{k=1}^n  \widehat{B}(v_{b_{m+k}}) \, \ket{\mathbb{0}}.
\end{equation}
In \eqref{TTTBBB} the summation runs over all $m$-tuples of indices $\mathbf{b}=(b_1,\ldots, b_m)$ such that
\begin{equation}\label{sum-beta}
\begin{cases}
b_{p}\in \{1,\ldots,n+m+1-i_{p}\}, \\
b_j\not= b_k \quad \text{if} \quad j\not= k,
\end{cases}
\end{equation}
and we have set $v_{n+j}=\zeta_{m+1-j}$, as well as
\begin{equation}
   \{1,\ldots, n+m\}\setminus \{b_1,\ldots, b_m\} =\{b_{m+1},\ldots,b_{m+n}\}.
\end{equation}
%
With these conventions, the coefficient $F_{\mathbf{b} }(\widehat{s};\{v\},\{\zeta\}) \equiv F_{(b_1,\ldots,b_m) }(\widehat{s};\{v_1,\ldots, v_n\},\{\zeta_1,\ldots,\zeta_m\}) $ is given as
\begin{multline}
  F_{\mathbf{b} }(\widehat{s};\{v\},\{\zeta\})
  = f_{\alpha_1,\ldots,\alpha_m}(\widehat{s})\
     \prod_{p=1}^{ |\boldsymbol{\alpha}_-| }
   \mathsf{d}(v_{b_p})
     \prod_{p=|\boldsymbol{\alpha}_-|+1}^{m} \hspace{-3mm}
   \mathsf{a}(v_{b_p})\
     \prod_{1\leq i<j\leq m}\!\frac{[v_{b_i}-v_{b_j}]}{[v_{b_i}-v_{b_j}+1]}\
     \\
     \times
     \prod_{p=1}^{m}\left\{ 
     \frac{\big[\widehat{s}+\alpha_{1,\ldots,i_p-1}+v_{b_p}-\zeta_{i_p} \big]}{\big[\widehat{s}+\alpha_{1,\ldots,i_p-1} \big]}\
    \frac{\pl_{k=1}^{n} [v_k-v_{b_p}+\alpha_{i_p}]}
    {\pl_{\substack{ k=1 \\ k\neq b_p}}^{n} [v_k-v_{b_p}] }\
    \frac{\prod\limits_{k=i_{p}+1}^{m} [\zeta_{k}- v_{b_{p}}+\alpha_{i_p}]}
           {\prod\limits_{\substack{ k=i_{p} \\ k\neq n+m+1-b_{p}}}^{m}
            [\zeta_{k}-v_{b_{p}}]}
     \right\},
\end{multline}
with
\begin{equation*}
    f_{\alpha_1,\ldots,\alpha_m}(\widehat{s})=
       \prod_{j\in\boldsymbol{\alpha}_- }\!\!
        \frac{  \big[\widehat{s}+\alpha_{1,\ldots,j-1}-n-1\big] }{  \big[\widehat{s}+\alpha_{1,\ldots,j-1} -1\big] } 
        \prod_{j\in\boldsymbol{\alpha}_+}\!
        \frac{\big[\widehat{s}+\alpha_{1,\ldots,j-1} \big]}{  \big[\widehat{s}+\alpha_{1,\ldots,j-1} -n\big] } 
        =\prod_{j=1}^n\!\frac{\big[\widehat{s}+\alpha_{1,\ldots,m}-j\big]}{\big[\widehat{s}-j\big]}.
\end{equation*}
Hence the quantity \eqref{MPME1} can be rewritten as
\begin{multline}\label{MPME4}
 \mathbb{P}^{(\{u\},\omega_u;\{v\},\omega_v)}_{\{\zeta\} }(s;\alpha_1,\ldots,\alpha_m) 
   =  \frac{1}{L}\frac{\omega_v^{s+\alpha_{1,\ldots,m}}}{\omega_u^s}\prod_{j=1}^n\frac{[s+j-1]}{[s+\alpha_{1,\ldots,m}-j]}\
\prod_{j=1}^m \tau^{-1}(\zeta_j;\{v\},\omega_v)\\
   \times \sum_{ \mathbf{b} } F_{\mathbf{b} }(s;\{v\},\{\zeta\}) \
 \frac{S_n(\{u\},\{v_{b_{m+k}}\};s)}{  \big( \braket{ \{u\},\omega_u}{ \{u\},\omega_u } \braket{ \{v\},\omega_v}{ \{v\},\omega_v } \big)^{1/2} } ,
\end{multline}
in terms of the partial scalar products $S_n(\{u\},\{v_{b_{m+k}} \} ; s)$ \eqref{Sn} associated to the sets of variables $\{u\}\equiv\{u_j\}_{1\le j\le n}$ and $\{v_{b_{m+k}}\}\equiv\{v_{b_{m+k}}\}_{1\le k\le n}$. 
These partial scalar products can be expressed as sums of $L$ determinants as in \eqref{result-spL}-\eqref{def-OmegaL}, whereas the normalization factors $\braket{ \{u\},\omega_u}{ \{u\},\omega_u }$ and $\braket{ \{v\},\omega_v}{ \{v\},\omega_v }$ can themselves be expressed as a unique determinant as in \eqref{gaudin}-\eqref{mat-Phi}.

We ultimately want to compute the thermodynamic limit of the quantity \eqref{MPME1} when the two states  $\ket{\{u\},\omega_u}$ and $\ket{\{v\},\omega_v}$ tend to ground states of the infinite-size model.
We have seen in our previous paper \cite{LevT13b} that the large-size behavior of the determinant of the matrix $\Phi(\{u\})$ \eqref{mat-Phi}, appearing in the denominator of \eqref{MPME4} throughout the normalization factor \eqref{gaudin}, is given in terms of a Fredholm determinant which can be explicitly computed.
However, it is more difficult to directly determine the large-size behavior of the determinants of the matrices $\Omega_\gamma^{(\nu)}(\{u\},\omega_u;\{v_{b_{m+k}}\})$ appearing in the numerator of \eqref{MPME4} through the expression of the partial scalar product \eqref{result-spL}, especially when the two considered ground states are different.
In fact, as in the simpler case of the form factor considered in \cite{LevT13b}, one should modify these determinants so as to obtain more convenient representations for taking the thermodynamic limit.

Let us set, for $\varepsilon=\pm 1$,
\begin{equation}\label{Lambda-pm}
   \Lambda_\varepsilon(\zeta;\{v\},\omega_v)=\varepsilon\,\omega_v^{\varepsilon-1}
   \prod_{j=1}^N\big[\zeta-\xi_j+\frac{1+\varepsilon}{2}\big]\cdot 
   \prod_{\ell=1}^n\big[v_\ell-\zeta+\varepsilon\big].
\end{equation}
%
Using the Bethe equations for $(\{v\},\omega_v)$, we can rewrite, when $\{u\}$ and $\{v\}$ are pairwise distinct, the determinant of the matrix $\Omega_\gamma^{(\nu)}(\{u\},\omega_u;\{v_{b_{m+k}}\})$ as
\begin{multline}\label{id-det1}
  \det_n \big[  \Omega_\gamma^{(\nu)}(\{u\},\omega_u;\{v_{b_{m+k}}\}) \big] 
  =\frac{\pl_{k=1}^n\left\{\mathsf{a}(v_{b_{m+k}})\pl_{l=1}^n [v_l-v_{b_{m+k}}+1]\right\} }
            {\pl_{\substack{k=1\\ b_{m+k}>n}}^n\Lambda_+(v_{b_{m+k}};\{v\},\omega_v) }
  \\
  \times
    \det_n \big[  H_{\gamma;\mathbf{b}}^{(\nu)}(\{u\},\omega_u;\{v\},\omega_v|\{v_{b_{m+k}}\})\big], 
\end{multline}
in which
\begin{multline}\label{H-ell}
   \big[  H_{\gamma;\mathbf{b}}^{(\nu)}(\{u\},\omega_u;\{v\},\omega_v|\{v_{b_{m+k}}\})\big] _{jk} \\
   =
   \begin{cases}
   \big[ H_{\gamma}^{(\nu)} (\{u\}, \omega_u;\{v\}, \omega_v) \big ]_{j b_{m+k}}  &\text { if }\ b_{m+k}\le n, \\
   \big [ Q_\gamma^{(\nu)} (\{u\},\omega_u;\{v\} | \{\zeta\}) \big]_{j, n+m+1-b_{m+k}} 
   &\text{ if }\ b_{m+k}> n,
   \end{cases}
\end{multline}
with
\begin{multline}\label{H-ell-0}
   \big[ H_{\gamma}^{(\nu)} (\{u\}, \omega_u;\{v\}, \omega_v) \big ]_{jk}
   = 
   \sum_{\varepsilon=\pm}
    \frac{\varepsilon}{[\gamma]} \bigg\{ \frac{[u_{j}-v_{k}+\gamma]}{[u_{j}-v_{k}]} 
    - q^{-\varepsilon\,\nu} \  \frac{[u_{j}-v_{k}+\gamma+\varepsilon]}{[u_{j}-v_{k}+\varepsilon]} \bigg\} 
    \\
    \times
    \Big(\frac{\omega_v}{\omega_u}\Big)^{1-\varepsilon}
    \prod_{t=1}^n \frac{[u_t - v_k +\varepsilon]}{[v_t - v_k +\varepsilon]} ,
\end{multline}
\begin{multline}
\big [ Q_\gamma^{(\nu)} (\{u\},\omega_u;\{v\} | \{\zeta\}) \big]_{jk} 
=
 \sum_{\varepsilon=\pm}
    \frac{\varepsilon}{[\gamma]} \bigg\{ \frac{[u_{j}-\zeta_{k}+\gamma]}{[u_{j}-\zeta_{k}]} 
    - q^{-\varepsilon\,\nu} \  \frac{[u_{j}-\zeta_{k}+\gamma+\varepsilon]}{[u_{j}-\zeta_{k}+\varepsilon]} \bigg\} 
    \\
    \times
    \Lambda_\varepsilon(\zeta_k;\{v\},\omega_v)\,
    \Big(\frac{\omega_v}{\omega_u}\Big)^{1-\varepsilon}
    \prod_{t=1}^n \frac{[u_t - \zeta_k +\varepsilon]}{[v_t - \zeta_k +\varepsilon]} .
\end{multline}
The determinant of the matrix \eqref{H-ell} can then be transformed by means of the identity of Appendix~\ref{app-sp}. It gives
\begin{multline}
   \det_n \big[ H^{(\nu)}_{\gamma;\mathbf{b}}(\{u\},\omega_u;\{v\},\omega_v|\{v_{b_{m+k}}\} )\big]
   =\frac{\big[|u|-|v|+\gamma\big]}{(-[0]')^{n}\,[\gamma]}
                 \prod_{j<k} \frac{[u_j - u_k]}{[v_j - v_k]}\\
                 \times
        \det_n \big[ \mathcal{H}^{(\nu)}_{\gamma;\mathbf{b}}(\{u\},\omega_u;\{v\},\omega_v|\{v_{b_{m+k}}\}) \big]   ,     
\end{multline}
in which
\begin{multline}\label{calH-ell}
   \big[  \mathcal{H}_{\gamma;\mathbf{b}}^{(\nu)}(\{u\},\omega_u;\{v\},\omega_v|\{v_{b_{m+k}}\})\big] _{jk} 
   \\
   =
   \begin{cases}
   \big[ \mathcal{H}_{\gamma}^{(\nu)} (\{u\}, \omega_u;\{v\}, \omega_v) \big ]_{j b_{m+k}}  &\text { if }\ b_{m+k}\le n, \\
   \big [ \mathcal{Q}_\gamma^{(\nu)} (\{u\},\omega_u;\{v\}| \{\zeta\} )\big]_{j, n+m+1-b_{m+k}}    &\text{ if }\ b_{m+k}> n,
   \end{cases}
\end{multline}
with
\begin{multline}\label{calH-ell0}
\big[ \mathcal{H}_\gamma^{(\nu)} (\{u\}, \omega_u;\{v\}, \omega_v) \big]_{jk}
 = \delta_{jk} \, [0]' \,\frac{\pl_{l\neq j} [v_j - v_l]}{\pl_{l=1}^n [v_j - u_l]} 
    \Bigg\{ \prod_{l=1}^n \frac{[u_l - v_k +1]}{[v_l - v_k +1]} 
  - \Big( \frac{\omega_v}{\omega_u} \Big)^{\! 2} \prod_{l=1}^n\frac{[u_l - v_k - 1]}{[v_l - v_k - 1]} \Bigg\} 
  \\
 +\frac{[0]'}{[|u| - |v| + \gamma]} 
   \Bigg\{ q^{-\nu} \frac{[v_j - v_k + |u| - |v| + \gamma + 1]}{[v_j - v_k +1]}
         - q^{\nu} \Big( \frac{\omega_v}{\omega_u} \Big)^{\! 2} \frac{[v_j - v_k + |u| - |v| +\gamma- 1]}{[v_j - v_k - 1]} \Bigg\},
\end{multline}
\begin{multline}\label{calQ-ell}
\big[ \mathcal{Q}_{\gamma}^{(\nu)}   (\{u\},\omega_u;\{v\}| \{\zeta\} ) \big]_{jk} 
=  \sum_{\varepsilon=\pm} \frac{\varepsilon\, \Lambda_\varepsilon(\zeta_k;\{v\},\omega_v)\, [0]'}{[|u| - |v| + \gamma]} \left(\frac{\omega_v}{\omega_u}\right)^{1-\varepsilon} \\
\times
\Bigg\{q^{-\varepsilon\nu}  \frac{[v_j - \zeta_k + |u| - |v| +\gamma +\varepsilon]}{[v_j - \zeta_k +\varepsilon]}  
  -  \frac{[v_j - \zeta_k + |u| -|v| + \gamma]}{[v_j - \zeta_k]}
     \prod_{l=1}^n \frac{[v_l - \zeta_k] [u_l - \zeta_k +\varepsilon]}{[u_l - \zeta_k] [v_l - \zeta_k +\varepsilon]} \Bigg\} .
\end{multline}

Hence, combining all these expressions together with the representation \eqref{gaudin} of the normalization factor, one obtains that
\begin{multline}\label{MPME5}
  \mathbb{P}^{(\{u\},\omega_u;\{v\},\omega_v)}_{\{\zeta\} }(s;\alpha_1,\ldots,\alpha_m)
  =
  \left(\frac{ \braket{ \{v\},\omega_v}{ \{v\},\omega_v } }{\braket{ \{u\},\omega_u}{ \{u\},\omega_u } } \right)^{\! 1/2}\
        \sum_{ \mathbf{b} } G_{\mathbf{b} }(s;\{v\},\{\zeta\})
      \\
   \times
   \frac{[s] \, [|u|-|v|+\gamma]}{[0]'\, [|u|-|v_{b_{m+k}}|+\gamma+s]}  \
      \frac{1}{L}
   \sum_{ \nu=0  }^{L-1} q^{\nu s} \, a_\gamma^{(\nu)}(s_0)\,
   \frac{ \det_n \big[ \mathcal{H}_{\gamma;\mathbf{b}}^{(\nu)}(\{u\},\omega_u;\{v\},\omega_v|\{v_{b_{m+k}}\}) \big]  }{\det_n [\Phi(\{v\})] },
\end{multline}
with
\begin{equation}\label{ratio2}
  \frac{ \braket{ \{v\},\omega_v}{ \{v\},\omega_v } }{\braket{ \{u\},\omega_u}{ \{u\},\omega_u } }
  =
  \prod_{k=1}^n\! \frac{\mathsf{a}(v_k) \mathsf{d}(v_k)}{\mathsf{a}(u_k) \mathsf{d}(u_k) }
  \prod_{j,k=1}^n\! \frac{[v_j-v_k+1]}{[u_j-u_k+1]}\prod_{j\not= k}\! \frac{[u_j-u_k]}{[v_j-v_k]}
  \frac{\det_n [\Phi(\{v\})]}{\det_n [\Phi(\{u\})]}.
\end{equation}
In \eqref{MPME5}, $\mathcal{H}_{\gamma;\mathbf{b}}^{(\nu)}$ is given by \eqref{calH-ell}, $\Phi$ by \eqref{mat-Phi}, and $a_\gamma^{(\nu)}(s_0)$ by \eqref{a-nu}.
We recall that the sum is over all $m$-tuples $\mathbf{b}=(b_1,\ldots,b_m)$ satisfying the condition \eqref{sum-beta}.
The different sets of parameters involved in the expression \eqref{MPME5} should be understood as follows: $\{u\}\equiv\{u_j\}_{1\le j \le n}$,  $\{v\}\equiv\{v_j\}_{1\le j \le n}$, $\{v_{b_p}\}\equiv\{v_{b_p}\}_{1\le p \le m}$, $\{\zeta\}=\{\zeta_p\}_{1\le p \le m}$ and $\{v_{b_{m+k}}\}\equiv\{v_{b_{m+k}}\}_{1\le k\le n}=\{v\}\cup\{\zeta\}\setminus\{v_{b_p}\}$.
Also, we have set $|u|=u_1+\dots+ u_n$, $|v|=v_1+\dots+ v_n$, $|v_{b_{m+k}}|=v_{b_{m+1}}+\dots+v_{b_{m+n}}$, so that $|v_{b_{m+k}}|=|v|+|\zeta|-|v_{b_p}|$, with $|\zeta|=\zeta_1+\dots+\zeta_m$ and $|v_{b_p}|=v_{b_1}+\dots+v_{b_m}$.
Finally, the algebraic factor $G_{\mathbf{b} }(s;\{v\},\{\zeta\})$ is
\begin{multline}\label{G_b}
   G_{\mathbf{b} }(s;\{v\},\{\zeta\})
   =
   (-1)^{mn+\epsilon(\mathbf{b})+|\boldsymbol{\alpha}_-|} \ \left(\frac{\omega_v}{\omega_u}\right)^{\! s}\
   \prod_{k=1}^n\frac{\mathsf{d}(u_k)}{\mathsf{d}(v_k)}\\
   \times
   \frac{\pl_{\substack{k=1\\ b_k>n}}^{|\boldsymbol{\alpha}_-| }   \Lambda_-(v_{b_k};\{v\},\omega_v) \cdot
            \pl_{\substack{k=|\boldsymbol{\alpha}_-| +1\\ b_k>n}}^{m }    \Lambda_+(v_{b_k};\{v\},\omega_v) }
           {\pl_{j=1}^m \Big(  \Lambda_+(\zeta_j;\{v\},\omega_v)- \Lambda_-(\zeta_j;\{v\},\omega_v) \Big) } \
       \prod_{j<k}^m\frac{1}{[\zeta_j-\zeta_k]}\ \prod_{\substack{i,j=1\\ i<j}}^m\frac{1}{[v_{b_i}-v_{b_j}+1]} 
       \\
  \times                                                      
      \prod_{k=1}^m\left\{ \frac{\big[s+\alpha_{1,\ldots,i_k-1}+v_{b_k}-\zeta_{i_k} \big]}
                                                  {\big[s+\alpha_{1,\ldots,i_k-1}    \big]}\
      \prod\limits_{l=1}^{i_k-1} [\zeta _{l}-v_{b_{k}}]
          \prod\limits_{l=i_{k}+1}^{m} [\zeta_{l}- v_{b_{k}}+\alpha_{i_k}]      \right\}    ,                                                
\end{multline}
where $\epsilon(\mathbf{b})$ denotes the number of inversions of the permutation $j\mapsto b_j$, $1\le j \le n+m$.

\begin{rem}\label{rem-mean}
   In the case $\{u\}=\{v\}$, the expression \eqref{MPME5} remains valid provided we replace the matrix $\mathcal{H}_\gamma^{(\nu)} (\{u\}, \omega_u;\{v\}, \omega_v)$ \eqref{calH-ell0} by the matrix $\Phi_\gamma^{(\nu)}(\{v\})$ with matrix elements
 \begin{multline}\label{Phi-ell}
 \big[ \Phi^{(\nu)}_\gamma(\{v\} ) \big]_{jk} 
 =  \delta_{jk} \Bigg\{ \log' \frac{\mathsf{a}}{\mathsf{d}}(v_j) + \sum_{t=1}^n \Bigg( \frac{[v_j - v_k -1]'}{[v_j - v_k -1]} - \frac{[v_j - v_k +1]'}{[v_j - v_k +1]} \Bigg) \Bigg\}  \\
 - \frac{[0]'}{[\gamma]} \Bigg\{ q^\nu \frac{[v_j - v_k +\gamma-1]}{[v_j - v_k -1]} 
 - q^{-\nu} \frac{[v_j - v_k+\gamma -1]}{[v_j - v_k -1]} \Bigg\}.
 \end{multline}  
\end{rem}

At this stage, let us make some comments about the expression \eqref{MPME5}.
This expression is quite similar to the corresponding one obtained in \cite{KitMT00} when considering $m$-point elementary building blocks for correlation functions in the XXZ model, with however three main differences.

The first one is of course that we have here an extra sum (over the index $\nu$ running from $0$ to $L-1$) coming from the corresponding sum in the expression \eqref{result-spL} for the partial scalar product \eqref{Sn}. Although the resulting expression looks slightly more complicated than in the XXZ case, the fact that we have an extra sum here is not really problematic for taking the thermodynamic limit since the number of terms remains finite in this limit.

The second difference comes from the nature of the determinants involved in the expression \eqref{MPME5}. In the XXZ case, the ratios of two determinants obtained at this level of the computation for the renormalized mean values considered in \cite{KitMT00} can be reduced to a single determinant of size $m$ due to the fact that the corresponding matrices coincide up to a change of only $m$ columns.
Here this is no longer the case. This is due to the fact that we consider more general matrix elements (and not only mean values as in \cite{KitMT00}), but also to the fact that, even in the case of the mean value (see Remark~\ref{rem-mean}), all the matrix elements of the determinant in the numerator are modified with respect to the corresponding ones in the denominator, notably due to the presence of a `twist' by $q^{\pm \nu}$.
Hence the corresponding ratio of determinants cannot be reduced to a single determinant of size $m$ as in the case considered in \cite{KitMT00}. This point is of course a priori more problematic for taking the thermodynamic limit since the size of these determinants diverges in the thermodynamic limit. 
Although it is easy to expressed the determinant of the denominator in terms of a Fredholm determinant that can be explicitely computed in the thermodynamic limit (see \cite{LevT13b}), this is not the case for the determinant in the numerator. The idea to solve this problem is however quite simple: it is enough to multiply and divide by the determinant of the matrix \eqref{calH-ell0}:
\begin{align}
   \det_n \big[ \mathcal{H}^{(\nu)}_{\gamma;\mathbf{b}} \big] 
   &= \det_n \big[ \mathcal{H}_\gamma^{(\nu)}  \big] \cdot
         \det_n \big[ \big( \mathcal{H}^{(\nu)}_{\gamma} \big)^{-1} \cdot \mathcal{H}^{(\nu)}_{\gamma;\mathbf{b}} \big] \nonumber\\
   &= (-1)^{m(n+1)+\frac{m(m-1)}2+\epsilon(\mathbf{b})}      \det_n \big[ \mathcal{H}_\gamma^{(\nu)}  \big] \cdot 
         \det_m \big[\mathcal{S}^{(\nu)}_{\gamma;\mathbf{b}}\big].\label{det-nm}
\end{align}
Here $\mathcal{S}^{(\nu)}_{\gamma;\mathbf{b} }$ is the $m\times m$ matrix with elements
\begin{equation}
  \big[\mathcal{S}^{(\nu)}_{\gamma;\mathbf{b}}\big]_{jk}=
  \begin{cases}
    \big[ \big( \mathcal{H}^{(\nu)}_{\gamma} \big)^{-1} \cdot \mathcal{Q}^{(\nu)}_{\gamma} \big]_{b_j k}
     &\text{if } b_j \le n, \\
   -\delta_{n+m+1-b_j, k} &\text{if } b_j >n.
  \end{cases}   
   \label{det-m1}
\end{equation}
In its turn, the determinant of $\mathcal{H}^{(\nu)}_{\gamma}$ can be computed in terms of a Fredholm determinant when $n$ becomes large.
In the case $\{u\}=\{v\}$, one has of course the same kind of identity with $\mathcal{H}_\gamma^{(\nu)} $ simply replaced by $\Phi_\gamma^{(\nu)}$ in \eqref{det-nm}-\eqref{det-m1} by virtue of Remark~\ref{rem-mean}.

Finally, the last difference comes from the fact that we have considered here generic parameters $\zeta_k$ (which do not obligatory coincide with inhomogeneity parameters $\xi_{j_k}$ as in \cite{KitMT00}). Hence, the expression of the $m$ modified lines in the determinant is slightly more complicated, as well as the expression of the algebraic factor \eqref{G_b}. Note however that, when $\zeta_k$ coincides either with an inhomogeneity parameter $\xi_{j_k}$ or with a shifted inhomogeneity parameter $\xi_{j_k}-1$, only one of the two terms survives in the expression \eqref{calQ-ell}.


\section{Multi-point matrix elements in the thermodynamic limit}
\label{sec-therm}

Based on the finite-size computation performed in the previous section, we now want to evaluate the large size behavior of a matrix element of the form $\mathbb{P}^{(\mathsf{k}_x,\ell_x;\mathsf{k}_y,\ell_y)}_{\mathbf{I}_1, \mathbf{I}_2,\ldots,\mathbf{I}_{m+1}}(s_{1},s_{2},\ldots,s_{m+1})$ \eqref{LHP-7} between two Bethe ground states $\ket{\mathsf{k}_x,\ell_x}$ and $\ket{\mathsf{k}_y,\ell_y}$, associated to a particular configuration of heights $s_1,s_2,\ldots,s_{m+1}$ on $m+1$ adjacent sites of the lattice at respective positions $\mathbf{I}_1, \mathbf{I}_2,\ldots,\mathbf{I}_{m+1}$. For simplicity, we set $\mathbf{I}_1=(1,1)$.
The matrix element \eqref{LHP-7} is hence given by the renormalized matrix element  \eqref{MPME1}, with the identifications $\ket{\{u\},\omega_u}\equiv\ket{\mathsf{k}_x,\ell_x}$, $\ket{\{v\},\omega_v}\equiv\ket{\mathsf{k}_y,\ell_y}$, $s\equiv s_1$, and with $\alpha_k$, $\zeta_k$, defined by the prescriptions \eqref{alpha}, \eqref{zeta}, according to the respective positions of the neighboring vertices $\mathbf{I}_k$ and $\mathbf{I}_{k+1}$ ($1\le k\le m$).

Rewriting the expression \eqref{MPME5} in terms of real Bethe roots $x_j\equiv\tilde\eta u_j$, $y_j\equiv \tilde\eta v_j$, $j=1,\ldots,n,$ by means of Jacobi's imaginary transformation \eqref{jacobi}, we obtain
\begin{multline}\label{MPME6}
\mathbb{P}^{(\mathsf{k}_x,\ell_x;\mathsf{k}_y,\ell_y)}_{\mathbf{I}_1,\ldots,\mathbf{I}_{m+1}}(s_{1},\ldots,s_{m+1})
=\left(\frac{ \braket{ \mathsf{k}_y,\ell_y}{ \mathsf{k}_y,\ell_y} }{\braket{ \mathsf{k}_x,\ell_x}{ \mathsf{k}_x,\ell_x } } \right)^{\! 1/2}\
       \Big( \frac{\omega_y}{\omega_x}\Big)^{s_1 -2n}
       \\
       \times
        \sum_{ \mathbf{b} } 
         \frac{\pl_{\substack{k=1\\ b_k>n}}^{|\boldsymbol{\alpha}_-| }   \widetilde{\Lambda}_-(y_{b_k};\{y\},\omega_y) \cdot
            \pl_{\substack{k=|\boldsymbol{\alpha}_-| +1\\ b_k>n}}^{m }    \widetilde{\Lambda}_+(y_{b_k};\{y\},\omega_y) }
           {\pl_{j=1}^m \Big( \widetilde{ \Lambda}_+(\tilde\zeta_j;\{y\},\omega_y)- \widetilde{\Lambda}_-(\tilde\zeta_j;\{y\},\omega_y) \Big) }\
        \widetilde{G}_{\alpha_1,\ldots,\alpha_m }(s_1;\{y_{b_p}\},\{\tilde\zeta\})
      \\
   \times
   q^{- s_1 (|x|-|y|+\tilde\gamma)}
   \frac{\theta_1(\tilde\eta s_1)\, \theta_1(|x|-|y|+\tilde\gamma)}{\tilde\eta\,\theta_1'(0)\, \theta_1(|x|-|y_{b_{m+k}}|+\tilde\gamma+\tilde\eta s_1)} 
   \\
   \times
   \frac{1}{L}
   \sum_{ \nu=0  }^{L-1} q^{\nu s_1} \ a_\gamma^{(\nu)}(s_0)\
   \frac{\det_n\big[\widetilde{\mathcal{H}}_\gamma^{(\nu)}(\{x\},\omega_x;\{y\},\omega_y)\big]}
           {\det_n\big[\widetilde{\Phi}(\{y\})\big]}\,
   \det_m\big[\widetilde{\mathcal{S}}_{\gamma;\mathbf{b}} ^{(\nu)}\big]    .    
\end{multline}
Here we have set $\omega_x\equiv\omega_u, \ \omega_y\equiv\omega_v$, $\tilde\gamma=\tilde\eta \gamma$ and $\tilde\zeta_j=\tilde\eta\zeta_j$, $j=1,\ldots,m$.
We have also set
\begin{equation}\label{tLambda-pm}
   \widetilde{\Lambda}_\varepsilon(\tilde\zeta;\{y\},\omega_y)
   =\varepsilon\left(\omega_y\, e^{i\pi\eta(2|y|+\bar\xi)}\right)^{\varepsilon-1}
   \prod_{j=1}^N\theta_1\big(\tilde\zeta-\tilde\xi_j+\frac{1+\varepsilon}{2}\tilde\eta\big)\cdot 
   \prod_{\ell=1}^n\theta_1\big(y_\ell-\tilde\zeta+\varepsilon\tilde\eta\big),
\end{equation}
for $\varepsilon=\pm 1$.
The algebraic factor $\widetilde{G}_{\alpha_1,\ldots,\alpha_m }(s;\{\lambda\},\{\mu\})$ is
\begin{multline} \label{Gtilde}
   \widetilde{G}_{\alpha_1,\ldots,\alpha_m }(s;\{\lambda\},\{\mu\})
   =     (-1)^{|\boldsymbol{\alpha}_+|} \,
     \prod_{j=1}^m \left\{ 
     \frac{\theta_1\big(\tilde\eta ( s+\alpha_{1,\ldots,i_j-1})+\lambda_j-\mu_{i_j} \big)}{\theta_1\big(\tilde\eta (s+\alpha_{1,\ldots,i_j-1})  \big)}   \right\}   \,     \prod_{j<k}^m\frac{1}{\theta_1(\mu_k-\mu_j)}\
   \\
   \times
        \prod_{\substack{i,j=1\\ i<j}}^m\frac{1}{\theta_1(\lambda_i-\lambda_j+\tilde\eta)} \,
          \prod_{j=1}^{m} \Bigg\{
          \prod\limits_{k=1}^{i_j-1} \theta_1(\mu_{k}-\lambda_j)
          \prod\limits_{k=i_{j}+1}^{m} \theta_1(\mu_{k}- \lambda_j+\tilde\eta\alpha_{i_j})     \Bigg\}   .
\end{multline}

The matrix $\widetilde{\Phi}$ is given as
\begin{equation}
\big[ \widetilde\Phi\big]_{jk}
      = -2\pi i \tilde\eta N \delta_{jk}\bigg\{\frac{ {p_0}'_{\text{tot}}(y_j)}{2\pi}-\frac{1}{N} \sum_{l=1}^n K(y_j-y_l) \bigg\} - 2\pi i\tilde\eta K(y_j-y_k) +4\pi i \tilde\eta \eta,
      \label{mat-tPhi}  
\end{equation}  
and it has been shown in \cite{LevT13b} that its determinant could be written in terms of a Fredholm determinant for large $N$:
\begin{equation}\label{Fred-norm}
  \det_n\big[ \widetilde\Phi(\{y\})\big]
   = (-2\pi i \tilde\eta N)^n \prod_{l=1}^n\rho_{\text{tot}}(y_l)  \left\{ \det \big[ 1+ \widehat{K} -\widehat{V}_0\big]+O(N^{-\infty}) \right\}.
\end{equation}
Here $\widehat{K}$ and $\widehat{V}_0$ are integral operators acting on the interval $[-\frac12,\frac12]$, with respective kernels $K(y-z)$ given by \eqref{K}, and $V_0(y-z)=2\eta$.

In their turn, the elements of the matrix $\widetilde{\mathcal{H}}_\gamma^{(\nu)}$ are expressed as
\begin{align}
  \big[ \mathcal{\tilde{H}}_\gamma^{(\nu)} \big]_{jk} 
&=  \delta_{jk}  \, \tilde\eta\, \theta_1'(0)\,
      \frac{\prod_{l\not= j} \theta_1(y_j-y_l)}{\prod_{l=1}^n\theta_1(y_j-x_l)}
    \Bigg\{ q^{|x| - |y|} \, \prod_{l=1}^n \frac{\theta_1(x_l-y_k+\tilde\eta)}{\theta_1(y_l-y_k+\tilde\eta)} 
                   \nonumber\\
&\hspace{2cm}                   
   - \Big( \frac{\omega_y}{\omega_x} \Big)^{\! 2} q^{-|x| + |y|}\, \prod_{l=1}^n \frac{\theta_1(x_l-y_k-\tilde\eta)}{\theta_1(y_l-y_k-\tilde\eta)}
    \Bigg\}
   \nonumber\\
&\hspace{2cm}     + \frac{\tilde\eta\, \theta_1'(0)}{\theta_1(|x| - |y| + \tilde{\gamma})}
        \Bigg\{ q^{-\nu+|x| - |y| + \tilde{\gamma}}\,
        \frac{\theta_1(y_j - y_k + |x|- |y| + \tilde{\gamma} + \tilde{\eta})}{\theta_1(y_j - y_k + \tilde{\eta})} 
     \nonumber\\
 &\hspace{2cm}   - q^{\nu-|x| + |y|-\tilde\gamma}\, \Big(\frac{\omega_y}{\omega_x} \Big)^{\! 2}\,
      \frac{\theta_1(y_j - y_k + |x|- |y| + \tilde{\gamma} - \tilde{\eta})}{\theta_1(y_j - y_k - \tilde{\eta})} \Bigg\}
      \nonumber\\
   & = - 2 i \pi \tilde\eta N \, \rho_{\text{tot}}(y_k)\,  e^{2 i \pi (\eta-1)(|x|-|y|)}
        \Bigg\{ \delta_{jk} + \frac{1}{N \rho_{\text{tot}}(y_k) } \,
        K^{(\eta(\tilde\gamma-\nu)+|x|-|y|)}_{\tilde\gamma+|x|-|y|}(y_j-y_k) \Bigg\} 
        \nonumber\\
  &\hspace{2cm}       
 + O(N^{-\infty}),
\end{align}
in which we have used \eqref{id-om}, \eqref{phi-t}, \eqref{phi-zero}, and where the function $K_X^{(Y)}(z)$ is given by \eqref{K-X-Y}.
Hence the corresponding determinant can also be expressed in terms of a Fredholm determinant for large $N$:
\begin{multline}\label{det-H-Fred}
 \det_n\big[\widetilde{\mathcal{H}}_\gamma^{(\nu)}(\{x\},\omega_x;\{y\},\omega_y)\big]
   = \left(-2\pi i \tilde\eta N\,  e^{2 i \pi (\eta-1)(|x|-|y|)} \right)^{\! n}  \, \prod_{l=1}^n\rho_{\text{tot}}(y_l)
   \\
   \times
     \Big\{ \det \Big[ 1+ \widehat{K}^{(\eta(\tilde\gamma-\nu)+|x|-|y|)}_{\tilde\gamma+|x|-|y|}\Big]
     +O(N^{-\infty}) \Big\},
\end{multline}
where $\widehat{K}^{(Y)}_{X}$ is an integral operator acting on the interval $[-\frac12,\frac12]$, with kernel $K_X^{(Y)}$ \eqref{K-X-Y}.

Finally, the elements of the $m\times m$ matrix $\widetilde{\mathcal{S}}_{\gamma;\mathbf{b}}$ are given as
\begin{equation}
  \big[\widetilde{\mathcal{S}}^{(\nu)}_{\gamma;\mathbf{b} }\big]_{jk}=
  \begin{cases}
    \big[ \widetilde{\mathcal{S}}^{(\nu)}_\gamma(\{y\},\{\tilde\zeta\} )\big]_{b_j, k}
     &\text{if } b_j \le n, \\
  - \delta_{n+m-1-b_j , k} &\text{if } b_j >n,
  \end{cases}   
   \label{det-m2}
\end{equation}
in terms of the elements of the $n\times m$ matrix $\widetilde{\mathcal{S}}^{(\nu)}_\gamma(\{y\},\{\tilde\zeta\} )$ solution of the following equation:
\begin{equation}\label{eq-S-finite}
  \sum_{b=1}^n 
     \big[ \widetilde{\mathcal{H}}_\gamma^{(\nu)}(\{x\},\omega_x;\{y\},\omega_y)\big]_{jb} \cdot 
     \big[\widetilde{\mathcal{S}}^{(\nu)}_\gamma(\{y\},\{\tilde\zeta\} )\big]_{bk} 
 = \big[ \mathcal{\widetilde{Q}}_{\gamma}^{(\nu)} (\{x\},\{y\}|\{\tilde\zeta\})\big]_{jk}.
\end{equation}
%
Here
%
%
\begin{multline}\label{mat-tildeQ}
\big[ \mathcal{\widetilde{Q}}_{\gamma}^{(\nu)} (\{x\},\{y\}|\{\tilde\zeta\}) \big]_{jk} 
=
 \sum_{\varepsilon=\pm}
 \frac{\varepsilon\, \widetilde{\Lambda}_\varepsilon(\tilde\zeta_k;\{y\},\omega_y)\, \tilde\eta\,\theta_1'(0)}{\theta_1(|x| - |y| + \tilde{\gamma})}
 \left(\frac{\omega_y}{\omega_x}\right)^{1-\varepsilon} 
 q^{\varepsilon(|x|-|y|)}\\
\times
\Bigg\{q^{-\varepsilon(\nu-\tilde\gamma)}   \frac{\theta_1(y_j - \tilde\zeta_{k} + |x| - |y| + \tilde{\gamma} + \varepsilon\tilde{\eta})}{\theta_1(y_j - \tilde\zeta_{k} +\varepsilon \tilde{\eta})} 
\\
 - \frac{\theta_1(y_j -\tilde\zeta_{k} + |x| - |y| + \tilde{\gamma} )}{\theta_1(y_j - \tilde\zeta_{k})} 
      \prod_{l=1}^n \frac{\theta_1(y_l - \tilde\zeta_{k}) \theta_1(x_l - \tilde\zeta_{k} + \varepsilon\tilde{\eta}) }
                                       {\theta_1(x_l - \tilde\zeta_{k}) \theta_1(y_l - \tilde\zeta_{k} + \varepsilon\tilde{\eta} )}\Bigg\} .
\end{multline}

Let us now suppose that $\{\tilde\zeta\}\subset\{\tilde\xi_1,\ldots,\tilde\xi_N\}\cup\{\tilde\xi_1-\tilde\eta,\ldots,\tilde\xi_N-\tilde\eta\}$.
If $\tilde\zeta_k\in\{\tilde\xi_1,\ldots,\tilde\xi_N\}$, then $\widetilde{\Lambda}_-(\tilde\zeta_k;\{y\},\omega_y)=0$ and only the term $\varepsilon=+$ contributes to \eqref{mat-tildeQ}. In that case, it is easy to see, using \eqref{phi-t}, that
\begin{multline}
\big[ \mathcal{\widetilde{Q}}_{\gamma}^{(\nu)} (\{x\},\{y\}|\{\tilde\zeta\}) \big]_{jk} 
=-2\pi i\tilde\eta\,e^{2\pi i(\eta-1)(|x|-|y|)}\, \widetilde{\Lambda}_+(\tilde\zeta_k;\{y\},\omega_y)\\
\times
   t^{(\eta(\tilde\gamma-\nu)+|x|-|y|)}_{\tilde\gamma+|x|-|y|}(y_j,\tilde\zeta_k)
   +O(N^{-\infty}),                                     
\end{multline}
where the function $t_X^{(Y)}(y,\zeta)$ is given by \eqref{t-X-Y}.
A solution of the equation \eqref{eq-S-finite} can easily be obtained at the thermodynamic limit.
Indeed, it follows from the use of \eqref{sum-int2} that this equation turns into an integral equation of the form \eqref{eq-S-X-Y}, which can easily be solved by means of the Fourier series (see Appendix~\ref{app-fourier}). Hence
\begin{equation}\label{mat-S1}
    \big[\widetilde{\mathcal{S}}^{(\nu)}_\gamma(\{y\},\{\zeta\} ) \big]_{jk}
    = \frac{ \widetilde{\Lambda}_+(\tilde\zeta_k;\{y\},\omega_y)}{N \rho_{\text{tot}}(y_j)}\ S^{(\eta(\tilde\gamma-\nu)+|x|-|y|)}(y_j -\tilde\zeta_k)+O(N^{-\infty}),
\end{equation}
with
\begin{equation}\label{S-nu}
  S^{(\eta(\tilde\gamma-\nu)+|x|-|y|)}(z)
     =\frac{1}{2\pi i}\,\frac{\theta_1'\big(0;\tilde\eta\big)\, \theta_2\big(z+\eta(\tilde\gamma-\nu)+|x|-|y|;\tilde\eta\big)}{\theta_2\big(\eta(\tilde\gamma-\nu)+|x|-|y|;\tilde\eta\big)\,\theta_1\big(z;\tilde\eta\big)}.
\end{equation}
If now $\tilde\zeta_k\in\{\tilde\xi_1-\tilde\eta,\ldots,\tilde\xi_N-\tilde\eta\}$, then $\widetilde{\Lambda}_+(\tilde\zeta_k;\{y\},\omega_y)=0$ and only the term $\varepsilon=-$ contributes to \eqref{mat-tildeQ}, leading to
\begin{multline}
\big[ \mathcal{\widetilde{Q}}_{\gamma}^{(\nu)} (\{x\},\{y\}|\{\tilde\zeta\}) \big]_{jk} 
=-2\pi i\tilde\eta\,e^{2\pi i(\eta-2)(|x|-|y|)}\, q^{\nu-\tilde\gamma}\,\widetilde{\Lambda}_-(\tilde\zeta_k;\{y\},\omega_y)\\
\times
   t^{(\eta(\tilde\gamma-\nu)+|x|-|y|)}_{\tilde\gamma+|x|-|y|}(y_j,\tilde\zeta_k+\tilde\eta)
   +O(N^{-\infty}),                                     
\end{multline}
in which we have used \eqref{phi-t}, \eqref{id-om}. Hence
\begin{align}
    \big[\widetilde{\mathcal{S}}^{(\nu)}_\gamma(\{y\},\{\zeta\} ) \big]_{jk}
    &= e^{-2\pi i(|x|-|y|)}\, q^{\nu-\tilde\gamma}\,\frac{ \widetilde{\Lambda}_-(\tilde\zeta_k;\{y\},\omega_y)}{N \rho_{\text{tot}}(y_j)}\nonumber\\
    &\hspace{3cm}\times S^{(\eta(\tilde\gamma-\nu)+|x|-|y|)}(y_j -\tilde\zeta_k-\tilde\eta)+O(N^{-\infty}),
    \nonumber\\
    &=-\frac{ \widetilde{\Lambda}_-(\tilde\zeta_k;\{y\},\omega_y)}{N \rho_{\text{tot}}(y_j)}\ S^{(\eta(\tilde\gamma-\nu)+|x|-|y|)}(y_j -\tilde\zeta_k)+O(N^{-\infty}),\label{mat-S2}
\end{align}
in which we have used the $\tilde\eta$-quasi-periodicity of the function \eqref{S-nu}.
Finally, we can gather these two cases into a single formulation of the matrix $\widetilde{\mathcal{S}}^{(\nu)}_\gamma(\{y\},\{\zeta\} )$ in the case $\{\tilde\zeta\}\subset\{\tilde\xi_1,\ldots,\tilde\xi_N\}\cup\{\tilde\xi_1-\tilde\eta,\ldots,\tilde\xi_N-\tilde\eta\}$, namely
\begin{multline}\label{mat-S3}
    \big[\widetilde{\mathcal{S}}^{(\nu)}_\gamma(\{y\},\{\zeta\} ) \big]_{jk}
    = \frac{ \widetilde{\Lambda}_+(\tilde\zeta_k;\{y\},\omega_y)-\widetilde{\Lambda}_-(\tilde\zeta_k;\{y\},\omega_y)}{N \rho_{\text{tot}}(y_j)}\\
    \times S^{(\eta(\tilde\gamma-\nu)+|x|-|y|)}(y_j -\tilde\zeta_k)+O(N^{-\infty}).
\end{multline}

\begin{rem}\label{rem-mean2}
  By considering the large $N$ behavior of the elements of the matrix $\Phi_\gamma^{(\nu)}$ \eqref{Phi-ell}, one obtains that the previous study is also valid in the case $\{x\}=\{y\}$ (see Remark~\ref{rem-mean}).
  It is enough in that case to set $|x|=|y|$ in the above formulas.
\end{rem}

Hence, gathering all these results and using also that the ratio of the two normalization factors is simply given by a phase factor (see \cite{LevT13b}),
\begin{equation}
\frac{ \moy{\mathsf{k}_y,\ell_y\mid \mathsf{k}_y,\ell_y}}
       { \moy{\mathsf{k}_x,\ell_x\mid \mathsf{k}_x,\ell_x}}
  =\left(\frac{\omega_y}{\omega_x}\right)^{\! 2n}+O(N^{-\infty}),
\end{equation} 
we obtain that the properly normalized multi-point matrix element \eqref{MPME1} is given as the following multiple sum
\begin{multline}\label{MPME7}
\mathbb{P}^{(\mathsf{k}_x,\ell_x;\mathsf{k}_y,\ell_y)}_{\mathbf{I}_1,\ldots,\mathbf{I}_{m+1}}(s_{1},\ldots,s_{m+1})
   = 
          \sum_{ \mathbf{b} } 
         \frac{\pl_{\substack{k=1\\ b_k>n}}^{|\boldsymbol{\alpha}_-| }   \widetilde{\Lambda}_-(y_{b_k};\{y\},\omega_y) \cdot
            \pl_{\substack{k=|\boldsymbol{\alpha}_-| +1\\ b_k>n}}^{m }    \widetilde{\Lambda}_+(y_{b_k};\{y\},\omega_y) }
           {\pl_{j=1}^m \Big( \widetilde{ \Lambda}_+(\tilde\zeta_j;\{y\},\omega_y)- \widetilde{\Lambda}_-(\tilde\zeta_j;\{y\},\omega_y) \Big) }
      \\
   \times
      \widetilde{G}_{\alpha_1,\ldots,\alpha_m }(s_1;\{y_{b_p}\},\{\tilde\zeta\})\
   \left( q^{- |x|+|y|-\tilde\gamma}\,  \frac{\omega_y}{\omega_x}\right)^{\! s_1 }
   \frac{\theta_1(\tilde\eta s_1)\, \theta_1(|x|-|y|+\tilde\gamma)}{\tilde\eta\,\theta_1'(0)\, \theta_1(|x|-|y_{b_{m+k}}|+\tilde\gamma+\tilde\eta s_1)} 
   \\
      \times 
   \frac{1}{L}   
   \sum_{ \nu=0  }^{L-1} q^{\nu s_1} \ a_\gamma^{(\nu)}(s_0)\
   \frac{\det \Big[ 1+ \widehat{K}^{(\eta(\tilde\gamma-\nu)+|x|-|y|)}_{\tilde\gamma+|x|-|y|}\Big]}
           {\det \big[ 1+ \widehat{K} -\widehat{V}_0\big]}\,
   \det_m \big[\widetilde{\mathcal{S}}^{(\nu)}_{\gamma;\mathbf{b}}\big]
   +O(N^{-\infty})   ,
\end{multline}
with $\widetilde{\mathcal{S}}^{(\nu)}_{\gamma;\mathbf{b}}$ given by \eqref{det-m2}, \eqref{mat-S3}.

Let us now suppose that the set of parameters $\{\tilde\zeta\}\subset\{\tilde\xi_1,\ldots,\tilde\xi_N\}\cup\{\tilde\xi_1-\tilde\eta,\ldots,\tilde\xi_N-\tilde\eta\}$ is such that $\tilde\zeta_j-\tilde\zeta_k\not= \tilde\eta$, for all $j,k\in\{1,\ldots,m\}$, i.e. that it does not contain any pair of the type $\{\tilde\xi_l,\tilde\xi_l-\tilde\eta\}$ associated to a same inhomogeneity parameter $\tilde\xi_l$. In that case, one can extend the summation  over all $m$-tuples of indices $\mathbf{b}$ submitted to the condition \eqref{sum-beta} to sums over $m$ independent indices $b_j$, $1\le j\le m$, taking values in the set $\{1, \ldots,n+m\}$, due to the vanishing of terms corresponding to configurations of indices not satisfying \eqref{sum-beta}.
Hence, at this stage, similarly as in \cite{KitMT00}, the multiple sums over the indices $b_j$ from $1$ to $n$ become multiple integrals due to Proposition~\ref{prop-sum-int}, whereas the sums over the indices $b_j>n$ for $1\le j \le |\boldsymbol{\alpha}_-|$ (respectively for $|\boldsymbol{\alpha}_-|+1\le j\le m$) become contour integrals around the shifted inhomogeneity parameters $\{\tilde\xi-\tilde\eta\}$ (respectively around the inhomogeneity parameters $\{\tilde\xi\}$) due to the fact that
\begin{equation}
   2 i \pi \,\mathrm{Res}  \Big[ S^{(\eta(\tilde\gamma-\nu)+|x|-|y|)}(z)\Big]_{z=0} = 1,
\end{equation}
and to the vanishing of $\widetilde{ \Lambda}_\varepsilon(\tilde\xi_l-\frac{1+\varepsilon}{2}\tilde\eta;\{y\},\omega_y)=0$.
We therefore obtain that
\begin{multline}\label{MPME8}
\mathbb{P}^{(\mathsf{k}_x,\ell_x;\mathsf{k}_y,\ell_y)}_{\mathbf{I}_1,\ldots,\mathbf{I}_{m+1}}(s_{1},\ldots,s_{m+1})
   = 
      \int\limits_{\mathcal{C}_-} \prod_{j=1}^{|\boldsymbol{\alpha}_-|}\! d \lambda_j \
      \int\limits_{\mathcal{C}_+} \prod_{j=|\boldsymbol{\alpha}_-|+1}^m\hspace{-4mm} d \lambda_j \
      \widetilde{G}_{\alpha_1,\ldots,\alpha_m }(s_1;\{\lambda\},\{\tilde\zeta\})
      \\
   \times
   \Big( \frac{\omega_y}{\omega_x}\Big)^{\! s_1 }
   e^{-2\pi i\eta s_1 (|x|-|y|+\tilde\gamma)}
   \frac{\theta_1(\tilde\eta s_1)\, \theta_1(|x|-|y|+\tilde\gamma)}{\tilde\eta\,\theta_1'(0)\, \theta_1(|x|-|y|+|\lambda|-|\tilde\zeta|+\tilde\gamma+\tilde\eta s_1)} \,
   \frac{1}{L}
     \sum_{ \nu=0  }^{L-1} q^{\nu s_1} \ a_\gamma^{(\nu)}(s_0)\
   \\
   \times
   \frac{\det \Big[ 1+ \widehat{K}^{(\eta(\tilde\gamma-\nu)+|x|-|y|)}_{\tilde\gamma+|x|-|y|}\Big]}
           {\det \big[ 1+ \widehat{K} -\widehat{V}_0\big]}\,
   \det_{1\le j,k \le m} \big[ S^{(\eta(\tilde\gamma-\nu)+|x|-|y|)}(\lambda_j-\tilde\zeta_k) \big]
   +O(N^{-\infty})   .
\end{multline}
In \eqref{MPME8},  the integration contours are
\begin{align}
   &\mathcal{C}_-=[-1/2,1/2]\cup \Gamma_+(\{\tilde\xi-\tilde\eta\}), \label{C-}\\
   &\mathcal{C}_+=[-1/2,1/2]\cup \Gamma_-(\{\tilde\xi\}), \label{C+}
\end{align}
where $ \Gamma_+(\{\tilde\xi-\tilde\eta\})$ (respectively  $\Gamma_-(\{\tilde\xi\})$) is such that it surrounds the points $\tilde\xi_1-\tilde\eta,\ldots,\tilde\xi_m-\tilde\eta$ with index $+1$ (respectively the points $\tilde\xi_1,\ldots,\tilde\xi_m$ with index $-1$), all other poles of the integrand being outside. 

However, when at least one pair of the type $\{\tilde\xi_l,\tilde\xi_l-\tilde\eta\}$ associated to a same inhomogeneity parameter $\tilde\xi_l$ occurs within the set $\{\tilde\zeta\}$ (which may happen when considering general matrix elements of the form \eqref{LHP-7}, \eqref{zeta}),
the previous procedure, and in particular the reconstruction of the $m\times m$ matrix of elements $S^{(\eta(\tilde\gamma-\nu)+|x|-|y|)}(\lambda_j-\tilde\zeta_k)$ by means of contour integrals selecting the appropriate residues, cannot be performed directly due to the $\tilde\eta$-quasi-periodicity of the function \eqref{S-nu}. In that case, one can nevertheless raise the degeneracy by  introducing auxiliary contour integrals around the parameters $\zeta$ at the level of \eqref{MPME7}, so as to replace the $\tilde\zeta$'s by some auxiliary integration variables $\mu$'s on which the previous procedure can be applied. This enables us to formally  write a generalization of \eqref{MPME8} as
\begin{multline}\label{MPME8bis}
\mathbb{P}^{(\mathsf{k}_x,\ell_x;\mathsf{k}_y,\ell_y)}_{\mathbf{I}_1,\ldots,\mathbf{I}_{m+1}}(s_{1},\ldots,s_{m+1})
   = 
     \prod_{j=1}^m \left( \ \oint\limits_{\Gamma_+(\tilde\zeta_j)} \frac{d\mu_j}{2\pi i}\frac{\theta_1'(0)}{\theta_1(\mu_j-\tilde\zeta_j)}\right)
     \\
     \times
            \int\limits_{\mathcal{C}_-(\{\mu\})} \prod_{j=1}^{|\boldsymbol{\alpha}_-|}\! d \lambda_j \
      \int\limits_{\mathcal{C}_+(\{\mu\})} \prod_{j=|\boldsymbol{\alpha}_-|+1}^m\hspace{-4mm} d \lambda_j \quad
      \widetilde{G}_{\alpha_1,\ldots,\alpha_m }(s_1;\{\lambda\},\{\mu\})\
      \\
   \times 
   \Big( \frac{\omega_y}{\omega_x}\Big)^{\! s_1 }
   e^{-2\pi i\eta s_1 (|x|-|y|+\tilde\gamma)}\
   \frac{\theta_1(\tilde\eta s_1)\, \theta_1(|x|-|y|+\tilde\gamma)}{\tilde\eta\,\theta_1'(0)\, \theta_1(|x|-|y|+|\lambda|-|\mu|+\tilde\gamma+\tilde\eta s_1)} \,
   \frac{1}{L}
     \sum_{ \nu=0  }^{L-1} q^{\nu s_1} \ a_\gamma^{(\nu)}(s_0)\
   \\
   \times
   \frac{\det \Big[ 1+ \widehat{K}^{(\eta(\tilde\gamma-\nu)+|x|-|y|)}_{\tilde\gamma+|x|-|y|}\Big]}
           {\det \big[ 1+ \widehat{K} -\widehat{V}_0\big]}\,
   \det_{1\le j,k \le m} \big[ S^{(\eta(\tilde\gamma-\nu)+|x|-|y|)}(\lambda_j-\mu_k) \big]
   +O(N^{-\infty})   .
\end{multline}
Here the contour $\Gamma_+(\tilde\zeta_j)$ surrounds the point $\tilde\zeta_j$ with index 1, whereas the contours $\mathcal{C}_\pm(\{\mu\})$ are defined as
\begin{align}
   &\mathcal{C}_-(\{\mu\})=[-1/2,1/2]\cup \Gamma_+(\{\mu\}_{_-}), \label{C-mu}\\
   &\mathcal{C}_+(\{\mu\})=[-1/2,1/2]\cup \Gamma_-(\{\mu\}_{_+}), \label{C+mu}
\end{align}
where $\{\mu\}_{_-}$ (respectively $\{\mu\}_{_+}$) corresponds to the set of integration variables $\mu_j$ integrated around the $\tilde\xi_l-\tilde\eta$, i.e. such that $0<-\Im\mu_j<\Im\tilde\eta$ (respectively integrated around the $\tilde\xi_l$, i.e. such that $0<\Im\mu_j<\Im\tilde\eta$).

The determinants appearing in the expressions \eqref{MPME8} or \eqref{MPME8bis} can be explicitly computed.
The two Fredholm determinants can be computed from the Fourier coefficients obtained in Appendix~\ref{app-fourier}.
Indeed, as the kernel of the integral operator $\widehat{K}-\widehat{V}_0$ (respectively $\widehat{K}^{(\eta(\tilde\gamma-\nu)+|x|-|y|)}_{\tilde\gamma+|x|-|y|}$) depends only on the difference of two variables, its eigenvalues correspond to the Fourier coefficients of the function $K-V_0$ (respectively ${K}^{(\eta(\tilde\gamma-\nu)+|x|-|y|)}_{\tilde\gamma+|x|-|y|}$). We obtain that
\begin{multline}\label{ratio-Fred}
     \frac{\det \Big[ 1+ \widehat{K}^{(\eta(\tilde\gamma-\nu)+|x|-|y|)}_{\tilde\gamma+|x|-|y|}\Big]}
             {\det \big[  1 + \widehat{K}-\widehat{V}_0 \big] }
      =\frac{1}{1-\eta}
        \frac{\theta_1\big( (1-\eta)\tilde\gamma+\eta\nu;\tilde\tau-\tilde\eta)}{\theta'_1(0;\tilde\tau-\tilde\eta)}
        \frac{\theta'_1(0;\tilde\tau)}{\theta_1(|x|-|y|+\tilde\gamma;\tilde\tau)}
        \\
        \times
         \frac{\theta_2\big(|x|-|y|+\eta(\tilde\gamma-\nu);\tilde\eta)}{\theta_2(0;\tilde\eta)}  .
\end{multline}
In its turn, the $m\times m$ determinant is simply given as
\begin{equation*}
  \det_{1\le j,k \le m} \big[ S^{(\eta(\tilde\gamma-\nu)+|x|-|y|)}(\lambda_j-\mu_k) \big]
  = \frac{\theta_2(|\lambda|-|\mu|+|x|-|y|+\eta(\tilde\gamma-\nu);\tilde\eta)}
             {\theta_2(|x|-|y|+\eta(\tilde\gamma-\nu);\tilde\eta)}\
               \bar{\mathcal{S}}_m(\{\lambda\};\{\mu\}),            
\end{equation*}
in terms of a common part independent from $\tilde\gamma$, $\nu$, and from the states $\{x\},\omega_x$ and $\{y\},\omega_y$:
\begin{equation}\label{S-bar}
  \bar{\mathcal{S}}_m(\{\lambda\};\{\mu\})=\left(\frac{\theta_1'(0;\tilde\eta)}{2\pi i}\right)^{\! m}\
    \frac{\prod_{i<j}\theta_1(\lambda_i-\lambda_j;\tilde\eta)\, \theta_1(\mu_j-\mu_i;\tilde\eta)}
            {\prod_{i,j=1}^m\theta_1(\lambda_i-\mu_j;\tilde\eta)}.
\end{equation}
Hence
\begin{multline}\label{MPME9}
\mathbb{P}^{(\mathsf{k}_x,\ell_x;\mathsf{k}_y,\ell_y)}_{\mathbf{I}_1,\ldots,\mathbf{I}_{m+1}}(s_{1},\ldots,s_{m+1})
   = 
     \prod_{j=1}^m \! \left( \ \oint\limits_{\Gamma_+(\tilde\zeta_j)} \hspace{-2mm}\frac{d\mu_j}{2\pi i}\frac{\theta_1'(0)}{\theta_1(\mu_j-\tilde\zeta_j)}\right)\ 
     \int\limits_{\mathcal{C}_-(\{\mu\})}\hspace{-1mm} \prod_{j=1}^{|\boldsymbol{\alpha}_-|}\! d \lambda_j 
     \int\limits_{\mathcal{C}_+(\{\mu\})}\! \prod_{j=|\boldsymbol{\alpha}_-|+1}^m\hspace{-4mm} d \lambda_j 
     \\
     \times
      \widetilde{G}_{\alpha_1,\ldots,\alpha_m }(s_1;\{\lambda\},\{\mu\})\
     \bar{\mathcal{S}}_m(\{\lambda\};\{\mu\})\
    \bar{\mathbb{P}}(s_1, |\lambda|-|\mu|;\mathsf{k},\ell)   +O(N^{-\infty})   ,
\end{multline}
in which we have separated the purely algebraic part, issued from the commutation relations of the Yang-Baxter algebra, and a purely analytic one, encoding all the information about the states we consider.
The latter is given as
\begin{align}
    \bar{\mathbb{P}}(s, Z ;\mathsf{k},\ell)
   & =
    e^{-i\pi s \big(\mathsf{k}-\frac{L\mathsf{k}+2\ell}{L-r}+2\eta\tilde\gamma\big)}
  \frac{\theta_1(\tilde\eta s)\, \theta_1\big(-\frac{L\mathsf{k}+2\ell}{2(L-r)}+\tilde\gamma\big)}
         {\tilde\eta\,\theta_1'(0)\, \theta_1\big(Z-\frac{L\mathsf{k}+2\ell}{2(L-r)}+\tilde\gamma+\tilde\eta s\big)}
   \nonumber\\
   &\hspace{-1.5cm}\times
    \frac{1}{L}
    \sum_{ \nu=0  }^{L-1} q^{\nu s} \ a_\gamma^{(\nu)}(s_0)\   
    \frac{\det \bigg[ 1+ \widehat{K}^{\big(\eta(\tilde\gamma-\nu)-\frac{L\mathsf{k}+2\ell}{2(L-r)}\big)}_{\tilde\gamma-\frac{L\mathsf{k}+2\ell}{2(L-r)}}\bigg]}
           {\det \big[ 1+ \widehat{K} -\widehat{V}_0\big]}
           \frac{\theta_2\big(Z-\frac{L\mathsf{k}+2\ell}{2(L-r)}+\eta(\tilde\gamma-\nu);\tilde\eta\big)}
             {\theta_2\big(-\frac{L\mathsf{k}+2\ell}{2(L-r)}\eta(\tilde\gamma-\nu);\tilde\eta\big)}
             \label{1PME1}\\
     &=e^{-i\pi s \big(-\frac{r\mathsf{k}+2\ell}{L-r}+2\eta\tilde\gamma\big)}
      \frac{\theta_1(\tilde\eta s;\tilde\tau)}
             {\tilde\eta\, \theta_1\big(Z-\frac{L\mathsf{k}+2\ell}{2(L-r)}+\tilde\gamma+\tilde\eta s;\tilde\tau\big)} 
     \nonumber\\
  &\hspace{-1.5cm}\times
    \frac{1}{L-r}
    \sum_{ \nu=0  }^{L-1} q^{\nu s} \ a_\gamma^{(\nu)}(s_0)\    
        \frac{\theta_1\big( (1-\eta)\tilde\gamma+\eta\nu;\tilde\tau-\tilde\eta)}{\theta'_1(0;\tilde\tau-\tilde\eta)}  \,         
        \frac{\theta_2\big(Z-\frac{L\mathsf{k}+2\ell}{2(L-r)}+\eta(\tilde\gamma-\nu);\tilde\eta\big)}
               {\theta_2(0;\tilde\eta)}     .    \label{1PME2}
\end{align}
where we have set $\mathsf{k}=\mathsf{k}_y-\mathsf{k}_x$, $\ell=\ell_y-\ell_x$.
We have also used \eqref{dif-sum} to express the quantity $|x|-|y|$ in terms of $\mathsf{k}$ and $\ell$.
In the case where the set $\{\tilde\zeta\}$ does not contain any pair of the type $\{\tilde\xi_l,\tilde\xi_l-\tilde\eta\}$, the expression \eqref{MPME9} simplifies into
\begin{multline}\label{MPME9bis}
\mathbb{P}^{(\mathsf{k}_x,\ell_x;\mathsf{k}_y,\ell_y)}_{\mathbf{I}_1,\ldots,\mathbf{I}_{m+1}}(s_{1},\ldots,s_{m+1})
   = 
     \int\limits_{\mathcal{C}_-} \prod_{j=1}^{|\boldsymbol{\alpha}_-|}\! d \lambda_j 
     \int\limits_{\mathcal{C}_+} \prod_{j=|\boldsymbol{\alpha}_-|+1}^m\hspace{-4mm} d \lambda_j 
     \\
     \times
      \widetilde{G}_{\alpha_1,\ldots,\alpha_m }(s_1;\{\lambda\},\{\tilde\zeta\})\
     \bar{\mathcal{S}}_m(\{\lambda\};\{\tilde\zeta\})\
    \bar{\mathbb{P}}(s_1, |\lambda|-|\tilde\zeta|;\mathsf{k},\ell)   +O(N^{-\infty})   .
\end{multline}

Note that, apart from this quantity $\bar{\mathbb{P}}(s, |z|-|\tilde\zeta|;\mathsf{k},\ell)$, the expression \eqref{MPME9bis} has a very similar form to the representation for the elementary building blocks for the correlation functions of the XXZ chain obtained in \cite{KitMT00}.
Hence, the whole complexity due to the presence of the dynamical parameter is contained into the dressing factor $\bar{\mathbb{P}}(s, |\lambda|-|\tilde\zeta|;\mathsf{k},\ell)$, which also encodes the information about the particular states we consider.
In fact, this dressing factor corresponds simply to the one-point local height matrix element, slightly deformed by the quantity $Z\equiv |\lambda|-|\tilde\zeta|$. Indeed, the  one-point local height matrix element is given as
\begin{align}
\mathbb{P}^{(\mathsf{k}_x,\ell_x;\mathsf{k}_y,\ell_y)}_{\mathbf{I}_1}(s)
   = \bar{\mathbb{P}}(s, 0 ;\mathsf{k},\ell)+O(N^{-\infty}).
\end{align}
Hence, the problem is now to obtain a simpler expression of this quantity. It is convenient for this to perform a change of basis in the subspace of the space of states generated by all degenerated ground states, so as to express the local height probabilities in a basis in which they are diagonal. This is done in the next section.

\section{Multi-point local height probabilities}
\label{sec-MPLHP}

We now compute the multi-point local height probabilities at adjacent sites \eqref{LHP-6} by performing the change of basis \eqref{change-basis}.
It is easy to see that, similarly as what happens for the local operator $\sigma_m^z$ (see \cite{LevT13b}),  all combinations of local operators are diagonal in the new basis \eqref{change-basis}. Hence,
\begin{multline}\label{MPLHP1}
 \bra{\phi_g^{(\epsilon_1,\mathsf{t}_1)}}\, \delta_{s_1}(\widehat{s})\, \widehat{T}_{\alpha_1\alpha_1}(\zeta_1)\ldots 
             \widehat{T}_{\alpha_m\alpha_m}(\zeta_m)\, 
             \prod_{k=1}^m\widehat{t}^{-1}(\zeta_k) \, \ket{\phi_g^{(\epsilon_2,\mathsf{t}_2)}}\\
 = \delta_{\epsilon_1,\epsilon_2}\, \delta_{\mathsf{t}_1,\mathsf{t}_2} \,
     \bar{\mathbf{P}}^{(\epsilon_1,\mathsf{t}_1)}_{\mathbf{I}_1, \ldots,\mathbf{I}_{m+1}}(s_{1},\ldots,s_{m+1})
 +O(N^{-\infty}),
\end{multline}
where $\bar{\mathbf{P}}^{(\epsilon,\mathsf{t})}_{\mathbf{I}_1, \ldots,\mathbf{I}_{m+1}}(s_{1},\ldots,s_{m+1})$ represents the limiting value,  at the thermodynamic limit, of the multi-point  local height probability $\mathbf{P}^{(\epsilon,\mathsf{t})}_{\mathbf{I}_1, \ldots,\mathbf{I}_{m+1}}(s_{1},\ldots,s_{m+1})$ \eqref{LHP-6} at adjacent sites $\mathbf{I}_1, \ldots,\mathbf{I}_{m+1}$ (we still suppose for simplicity that $\mathbf{I}_1=(1,1)$). We recall that, as previously, the `spin' variables $\alpha_k$ and the parameters $\zeta_k$ are respectively defined by the prescription \eqref{alpha} and \eqref{zeta}, according to the respective positions of the neighboring vertices $\mathbf{I}_k$ and $\mathbf{I}_{k+1}$ ($1\le k\le m$).
It follows from \eqref{MPME9} that these multi-point local height probabilities are given as
\begin{multline}\label{MPLHP2}
 \bar{\mathbf{P}}^{(\epsilon,\mathsf{t})}_{\mathbf{I}_1, \ldots,\mathbf{I}_{m+1}}(s_{1},\ldots,s_{m+1})
   = 
   \prod_{j=1}^m \! \left( \ \oint\limits_{\Gamma_+(\tilde\zeta_j)} \hspace{-2mm}\frac{d\mu_j}{2\pi i}\frac{\theta_1'(0)}{\theta_1(\mu_j-\tilde\zeta_j)}\right)\ 
     \int\limits_{\mathcal{C}_-(\{\mu\})}\hspace{-1mm} \prod_{j=1}^{|\boldsymbol{\alpha}_-|}\! d \lambda_j 
     \int\limits_{\mathcal{C}_+(\{\mu\})}\! \prod_{j=|\boldsymbol{\alpha}_-|+1}^m\hspace{-4mm} d \lambda_j 
     \\
     \times
      \widetilde{G}_{\alpha_1,\ldots,\alpha_m }(s_1;\{\lambda\},\{\mu\})\
     \bar{\mathcal{S}}_m(\{\lambda\};\{\mu\})\
    \bar{\mathbf{P}}(s_1,  |\lambda|-|\mu|;\epsilon,\mathsf{t}).  
\end{multline}
We recall that, in \eqref{MPLHP2}, we have set $\tilde\zeta_j=\tilde\eta\zeta_j$, and that we have supposed that $\{\tilde\zeta\}\subset\{\tilde\xi_1,\ldots,\tilde\xi_N\}\cup\{\tilde\xi_1-\tilde\eta,\ldots,\tilde\xi_N-\tilde\eta\}$, i.e. that the column inhomogeneity parameters $w_{j_k}$ involved in the set $\{\zeta\}$ through \eqref{zeta} belong to $\{\xi_1,\ldots,\xi_N\}\cup\{\xi_1-1,\ldots,\xi_N-1\}$. The algebraic part is unchanged with respect to the corresponding multiple-point matrix elements \eqref{MPME9} in the Bethe basis. In particular, the purely algebraic factor $\widetilde{G}_{\alpha_1,\ldots,\alpha_m}$ is given by \eqref{Gtilde} in terms of the sets \eqref{alpha-}-\eqref{alpha+},  and the common part $\bar{\mathcal{S}}_m$ coming from the computation of the scalar product determinants is given by \eqref{S-bar}. We also recall that $\Gamma_\pm(X)$ stands for a little contour encircling the (set of) point(s) $X$ with index $\pm1$ (all other poles of the integrand being outside), and that the integration contours $\mathcal{C}_\pm(\{\mu\})$ are defined in \eqref{C-mu}-\eqref{C+mu}.

Hence, the whole problem is now reduced to the computation of a compact expression for the (modified) one-point local height probabilities:
\begin{equation}\label{LHP1}
  \bar{\mathbf{P}}(s, Z;\epsilon,\mathsf{t}) = \sum_{\mathsf{k}=0}^1 \sum_{\ell=0}^{L-r-1}
   (-1)^{\mathsf{k}\epsilon}\, e^{-i\pi\frac{r\mathsf{k}+2\ell}{L-r}(\mathsf{t}+s_0)}\ 
   \bar{\mathbb{P}}(s, Z;\mathsf{k},\ell) .
\end{equation}
This quantity is computed in Appendix~\ref{app-LHP}, using several summation formulas of Appendix~\ref{app-theta}. The result is given by formulas \eqref{LHP-even-odd}, \eqref{LHP-ev-ev-phys} and \eqref{LHP-odd}. In terms of the original theta functions with imaginary period $\tau$, these expressions slightly simplify and we obtain
\begin{equation}\label{LHP-ev-odd}
\bar{\mathbf{P}}(s, Z;\epsilon,\mathsf{t}) \Big|_{\begin{subarray}{l} 
\text{$L$ even}\\ \text{$\epsilon\!+\!\mathsf{t}\!+\!s_0\!-\!s$ odd} \end{subarray}}  \hspace{-3mm}
  = 0,
\end{equation}
\begin{equation}\label{LHP-ev-ev}
  \bar{\mathbf{P}}(s, Z;\epsilon,\mathsf{t}) \Big|_{\begin{subarray}{l} 
\text{$L$ even}\\ \text{$\epsilon\!+\!\mathsf{t}\!+\!s_0\!-\!s$ even} \end{subarray}}
 \hspace{-4mm}
  = 2e^{i\pi\left(2\frac{r}{L}\tilde{s}Z+\frac{L-r}{r}Z^2\tau\right)}\,
       \frac{\theta_4\!\left(\frac{r\tilde{s}}{L};\tau\right) \,
               \theta_3\!\left(\frac{\tilde{s}_0+\mathsf{t}}{L-r}-\frac{\tilde{s}}{L}+\frac{Z\tau}{r} ; \frac{\tau}{r(L-r)}\right)}
              {L\, \theta_4\!\left(0;\frac{L}{r}\tau \right)\, 
               \theta_4\!\left(\frac{r(\tilde{s}_0+\mathsf{t})}{L-r};\frac{L}{L-r}\tau\right)}   ,
\end{equation}
\begin{multline}\label{LHP-odd-bis}
  \bar{\mathbf{P}}(s, Z;\epsilon,\mathsf{t}) \Big|_{\text{$L$ odd} }
  =  e^{i\pi\left(2\frac{r}{L}\tilde{s}Z+\frac{L-r}{r}Z^2\tau\right)}\\
  \times
     \frac{\theta_4\!\left(\frac{r\tilde{s}}{L};\tau\right) \,
               \theta_3\!\left(\big(\frac{1}{2}-\frac{1}{2L}\big)\tilde{s}
               -\big(\frac{1}{2}-\frac{1}{2(L-r)}\big)(\tilde{s}_0+\mathsf{t})-\frac{\epsilon}{2}+\frac{Z}{2r}\tau ; \frac{\tau}{4r(L-r)}\right)}
              {L\, \theta_4\!\left(0;\frac{L}{r}\tau \right)\, 
               \theta_4\!\left(\frac{r(\tilde{s}_0+\mathsf{t})}{L-r};\frac{L}{L-r}\tau\right)}   ,
\end{multline}
in which we have set $\tilde{s}=s-\frac{\tau}{2\eta}=s+\frac{1}{2\tilde\eta}$, i.e. $\tilde{s}_0=s_0+\frac{1}{2\tilde\eta}$ with $\tilde{s}_0\in\mathbb{R}$ so as to be in agreement with the physical model considered in \cite{PeaS89,PeaB90}.

\begin{rem}\label{rem-LHP}
The one-point local height probabilities are given by \eqref{MPLHP2} in the particular case $m=0$. Hence,  they correspond to the quantities $\bar{\mathbf{P}}(s, 0;\epsilon,\mathsf{t})$. It is in fact easy to see that the expressions \eqref{LHP-ev-odd}-\eqref{LHP-odd-bis} for $Z=0$ coincide with the explicit formulas for the one-point local height probabilities obtained in \cite{PeaS89}.
\end{rem}

Finally, if the set of parameters $\{\tilde\zeta\}$ is such that $\tilde\zeta_j-\tilde\zeta_k\not=\tilde\eta$, $\forall j,k$, which is in particular the case when one considers, as it is usually done in the literature (see for instance \cite{JimM95L,LukP96}), multi-point local height probabilities on adjacent sites on a {\em same} vertical line, then the expression \eqref{MPLHP2} simplifies into
\begin{multline}\label{MPLHP3}
 \bar{\mathbf{P}}^{(\epsilon,\mathsf{t})}_{\mathbf{I}_1, \ldots,\mathbf{I}_{m+1}}(s_{1},\ldots,s_{m+1})
   = 
   \int\limits_{\mathcal{C}_-} \prod_{j=1}^{|\boldsymbol{\alpha}_-|}\! d \lambda_j 
     \int\limits_{\mathcal{C}_+} \prod_{j=|\boldsymbol{\alpha}_-|+1}^m\hspace{-4mm} d \lambda_j \
     \widetilde{G}_{\alpha_1,\ldots,\alpha_m }(s_1;\{\lambda\},\{\tilde\zeta\})
     \\
     \times
     \bar{\mathcal{S}}_m(\{\lambda\};\{\tilde\zeta\})\
    \bar{\mathbf{P}}(s_1,  |\lambda|-|\tilde\zeta|;\epsilon,\mathsf{t}).  
\end{multline}
This expression is very similar, in its structure, to what has been obtained in \cite{JimM95L,KitMT99} for the elementary building blocks of the XXZ chain, or in \cite{LukP96} for the multi-point local height probabilities (on adjacent sites of a same vertical line) of the RSOS model. The analogy is especially obvious in the last case, since \eqref{MPLHP3} exhibits the same algebraic part as in Eq. (5.11) of \cite{LukP96} (the CSOS and RSOS model sharing the same dynamical Yang-Baxter algebra), and since the analytic part of both results involves the one-point local height probability of the model deformed in a similar way.

\section{Conclusion}
\label{sec-concl}

In this paper we have 
shown that the ABA approach to correlation functions developed in our previous paper \cite{LevT13a} in the case of the CSOS model enables us to compute not only height-independent quantities, such as the spontaneous staggered polarizations of the model \cite{DatJKM90,LevT13b}, but also more general local height probabilities. As an example, we have obtained multiple integral representations for the multi-point local height probabilities at adjacent sites, which are the building blocks of any arbitrary correlation function on the face lattice. We would like to stress that the solution of the inverse problem enables us to consider completely general correlation functions, and not only correlation functions of heights (or spins) on a same line of the lattice, as it is usually the case in the literature. Nevertheless, in the particular case where the considered sites are all aligned, the structure of our result is very similar in its form to what has been obtained for other models \cite{JimM95L,KitMT00,LukP96}. 

The obtention of these results shows that, although intermediate formulas happen to be slightly more complicated than in the 6-vertex case, the difficulties related to the presence of the dynamical parameter are not an obstruction to the implementation of the ABA approach to correlation functions.
In fact, in the case of the CSOS model, which is the simplest representative of the class of so-called face models, i.e. of integrable models associated to a Yang-Baxter algebra of dynamical type, the setting is now quite complete.
Although we have here more specifically focused on the computation of local height probabilities at adjacent sites, it is basically possible, as explained in Section~\ref{sec-ABA-corr}, to compute any kind of correlation function. More complicated quantities, such as two-point (or multi-point) correlation functions are a priori accessible through a summation over the corresponding elementary form factors (of local spin or height operators), a method which happens to be quite efficient.

Finally, we would like to mention that it is probably possible to adapt our method to the study of the unrestricted SOS model as well. The only difficulty in this case is that we have to deal with series instead of finite sums, which means that we have to pay special attention to the convergence of the expressions we manipulate. 

\section*{Acknowledgements}

V. T. is supported by CNRS.
We also acknowledge  the support from the ANR grant DIADEMS 10 BLAN 012004.
V. T. would  like to thank LPTHE (Paris VI University) for hospitality.

\appendix

\section{Theta functions and useful identities}
\label{app-theta}

In this paper, $\theta_1(z;\tau)$ denotes the usual theta function with quasi-periods $1$ and $\tau$ ($\Im\tau>0$),
\begin{equation}\label{theta1}
  \theta_1(z;\tau)=-i\sum_{k=-\infty}^{\infty} (-1)^k e^{i\pi\tau (k+\frac12)^2} e^{2i\pi (k+\frac12)z},
  \qquad \Im\tau>0,
\end{equation}
which satisfies
\begin{equation}\label{periods}
   \theta_1(z+1;\tau)=-\theta_1(z;\tau), \qquad
   \theta_1(z+\tau;\tau)= -e^{-i\pi\tau}\, e^{-2\pi i z}\, \theta_1(z;\tau).
\end{equation}
We also denote
\begin{align}
  &\theta_2(z;\tau)=\theta_1\Big(z+\frac12;\tau\Big)
    =\sum_{k=-\infty}^{\infty} e^{i\pi\tau (k+\frac12)^2} e^{2i\pi (k+\frac12)z},\label{theta2}
  \\
  &\theta_4(z;\tau)=-i \, e^{\frac{i\pi\tau}{4}}\, e^{i\pi  z}\,\theta_1\Big(z+\frac{\tau}{2};\tau\Big),
  \\
  &\theta_3(z;\tau)= \theta_4\Big(z+\frac12;\tau\Big)
                               =e^{\frac{i\pi\tau}{4}}\, e^{i\pi  z}\,\theta_1\Big(z+\frac{1}{2}+\frac{\tau}{2};\tau\Big)
                               =\sum_{k=-\infty}^{\infty} e^{i\pi\tau k^2} e^{2i\pi k z}.\label{theta3}
\end{align}

These theta functions satisfy several useful identities that we use in the course of the paper and that we recall here.
\begin{itemize}
\item Jacobi's imaginary transformation:
\begin{align}\label{jacobi}
   &\theta_1(z;\tau)
   = -i\, (-i\tau)^{-\frac12}\, e^{-i\pi \frac{z^2}{\tau}}\ \theta_1\Big(-\frac{z}{\tau}\, ;\, -\frac1\tau\,\Big),
   \\
   &\theta_2(z;\tau)
   =  (-i\tau)^{-\frac12}\, e^{-i\pi \frac{z^2}{\tau}}\ \theta_4\Big(-\frac{z}{\tau}\, ;\, -\frac1\tau\,\Big),
   \\
   &\theta_3(z;\tau)
   =  (-i\tau)^{-\frac12}\, e^{-i\pi \frac{z^2}{\tau}}\ \theta_3\Big(-\frac{z}{\tau}\, ;\, -\frac1\tau\,\Big),
   \\
   &\theta_4(z;\tau)
   =  (-i\tau)^{-\frac12}\, e^{-i\pi \frac{z^2}{\tau}}\ \theta_2\Big(-\frac{z}{\tau}\, ;\, -\frac1\tau\,\Big).
\end{align}

\item Schr\"oter's Formula:
\begin{multline}\label{Schroter}
  \theta_3\Big(x;\frac{r}{L}\tau\Big)\, \theta_3\Big(y;\frac{L-r}{L}\tau\Big)
  =\sum_{k=0}^{L-1} e^{i\pi\frac{r}{L}\tau k^2}\, e^{2\pi i k x}\\
  \times
    \theta_3\Big(x-y+\frac{rk}{L}\tau;\tau\Big)\, \theta_3\Big((L-r)x+ry+\frac{r(L-r)k}{L}\tau;r(L-r)\tau\Big).
\end{multline}

\item two other useful summation identities (see for instance \cite{Liu12}):
\begin{align}
&\frac{1}{n}\sum_{\nu=0}^{n-1} e^{-2\pi i k \frac{\nu}{n} }\, 
  \frac{\theta_1\big(x+y+\frac{\nu}{n};\tau\big)\, \theta'_1\big(0;\tau\big)}
         {\theta_1\big(x ;\tau\big)\, \theta_1\big(y+\frac{\nu}{n};\tau\big)}
   =
    e^{2\pi i k y}\,
    \frac{\theta_1\big(x+ny+k\tau;n\tau\big)\, \theta'_1\big(0;n\tau\big)}
         {\theta_1\big(x+k\tau ; n\tau\big)\, \theta_1\big(ny;n\tau\big)},\label{id-sum1}
\\
  &\sum_{\nu=0}^{n-1} e^{2\pi i \frac{\nu}{n} x}\, 
  \frac{\theta_1\big(x+y+\frac{\nu}{n}\tau;\tau\big)\, \theta'_1\big(0;\tau\big)}
         {\theta_1\big(x ;\tau\big)\, \theta_1\big(y+\frac{\nu}{n}\tau;\tau\big)}
   =
    \frac{\theta_1\big(\frac{x}{n}+y;\frac{\tau}{n}\big)\, \theta'_1\big(0;\frac{\tau}{n}\big)}
         {\theta_1\big(\frac{x}{n} ;\frac{\tau}{n}\big)\, \theta_1\big(y;\frac{\tau}{n}\big)},
     \label{id-sum2}
  \end{align}
with $k\in\mathbb{Z}$.
These two identities are equivalent through Jacobi's imaginary transformation \eqref{jacobi} and quasi-periodicity property \eqref{theta1}.

\item Frobenius determinant formula:
for $2n$ complex variables $x_1,\ldots,x_n,y_1,\ldots,y_n$ and any arbitrary parameter $t$, 
\begin{multline}\label{Frob-det}
   \det_{1\le i,j\le n}\left[\frac{\theta_1(x_i-y_j+t;\tau)}{\theta_1(x_i-y_j;\tau)\,\theta_1(t;\tau)}\right]
   =\frac{\theta_1\big(\sum_{j=1}^n(x_j-y_j)+t;\tau\big)}{\theta_1(t;\tau)}
   \\
   \times 
    \frac{\prod_{1\le i<j\le n} \theta_1(x_i-x_j;\tau)\,\theta_1(y_j-y_i;\tau)}
                                              {\prod_{i,j=1}^n \theta_1(x_i-y_j;\tau)} .
\end{multline}

\end{itemize}

\section{A determinant identity}\label{app-sp}

In this appendix, we explain how to transform the determinant of the matrix \eqref{H-ell} in a more suitable form for taking the thermodynamic limit. The procedure is similar to what was explained in Appendix~B of \cite{LevT13b}, the main difference being that now $\gamma$ is an arbitrary parameter.

Let us therefore consider, for two different sets of $n$ pairwise distinct complex variables $\{ u\}$ and $\{v\}$, a set of $m$ complex variables $\{\zeta\}$ pairwise distinct from $\{u\}$, and two arbitrary 4-tuples of vectors $\boldsymbol{\alpha}\equiv(\boldsymbol{\alpha}_1,\boldsymbol{\alpha}_2,\boldsymbol{\alpha_3},\boldsymbol{\alpha_4})$ and $\boldsymbol{\beta}\equiv(\boldsymbol{\beta}_1,\boldsymbol{\beta}_2,\boldsymbol{\beta}_3,\boldsymbol{\beta}_4)$, the $n\times n$ matrix ${H}_{\gamma,\boldsymbol{\alpha}}(\{u\},\{v\})$ and the $n\times m$ matrix ${Q}_{\gamma,\boldsymbol{\beta}}(\{u\},\{v\}|\{\zeta\}) $ with respective elements
\begin{multline}\label{mat-Halpha}
   \big[ {H}_{\gamma,\boldsymbol{\alpha}}   \big]_{ij}
   =\frac{1}{[\gamma]}\left\{
         \alpha_{1;j} \frac{[u_i - v_j + \gamma]}{[u_i-v_j]}
        -\alpha_{2;j} \frac{[u_i-v_j+\gamma+1]}{[u_i - v_j + 1]}
        \right\}
     \prod_{l=1}^n\frac{[u_l-v_j+1]}{[v_l-v_j+1]} 
     \\
     -
     \frac{1}{[\gamma]}\left\{
         \alpha_{3;j} \frac{[u_i-v_j+\gamma]}{[u_i-v_j]}
        -\alpha_{4;j} \frac{[u_i-v_j+\gamma-1]}{[u_i-v_j-1]}
        \right\} 
     \prod_{l=1}^n\frac{[u_l-v_j-1]}{[v_l-v_j-1]}  , 
\end{multline}
\begin{multline}\label{mat-Qalpha}
 \big[ {Q}_{\gamma,\boldsymbol{\beta}} \big]_{ij} 
  = \frac{1}{[\gamma]} \Bigg\{ \beta_{1;j} \frac{[u_i - \zeta_j + \gamma]}{[u_i - \zeta_j]} 
        - \beta_{2;j} \frac{[u_i - \zeta_j + \gamma+1]}{[u_i - \zeta_j+1]} \Bigg\}
        \prod_{l=1}^n \frac{[u_l - \zeta_j + 1]}{[v_l - \zeta_j +1]}
     \\
     -
     \frac{1}{[\gamma]}\left\{
         \beta_{3;j} \frac{[u_i-\zeta_j+\gamma]}{[u_i-\zeta_j]}
        -\beta_{4;j} \frac{[u_i-\zeta_j+\gamma-1]}{[u_i-\zeta_j-1]}
        \right\} 
     \prod_{l=1}^n\frac{[u_l-\zeta_j-1]}{[v_l-\zeta_j-1]} ,
\end{multline}
where $\alpha_{i;j}$ (respectively $\beta_{i;j}$) corresponds to the $j$-th coordinate of the vector $\boldsymbol{\alpha}_i$ (respectively $\boldsymbol{\beta}_i$).
The idea is, an in Appendix~B of \cite{LevT13b}, to multiply and divide the determinant by the determinant of a conveniently chosen matrix $\mathcal{X}_t(\{u\},\{v\})$.
In the present case the latter is defined as
\begin{equation}\label{mat-X}
  \big[ \mathcal{X}_t(\{u\},\{v\})  \big]_{jk} 
  = \frac{[0]'}{[t]} \, \frac{\prod_{l=1}^n [u_k - v_l]}{\prod_{l\not= k} [u_k - u_l]} \,
     \frac{[v_j - u_k + t]}{[v_j - u_k]},
     \qquad\text{with}\quad
    t=\sum_{l=1}^n(u_l-v_l)+\gamma. 
\end{equation}
Its determinant is equal to
\begin{equation}\label{det-X}
  \det_n \big[\mathcal{X}_t (\{u\},\{v\}) \big] = (-[0]')^n\,\frac{[\gamma]}{[t]}
                 \prod_{j<k} \frac{[v_j - v_k]}{[u_j - u_k]}.
\end{equation}

To compute the matrix elements of 
$\mathcal{H}_{\gamma,\boldsymbol{\alpha}}(\{u\},\{v\})\equiv   \mathcal{X}_t(\{u\},\{v\}) \,H_{\gamma,\boldsymbol{\alpha}}(\{u\},\{v\}) $,
one considers the functions
\begin{equation}
    g_\epsilon^{(j,k)}(z) 
    = \frac{[z - v_k + \gamma + \epsilon]}{[z - v_k + \epsilon]} 
       \prod_{l=1}^n \frac{[z - v_l]}{[z - u_l]} \frac{[v_j - z + t]}{[v_j - z]}
\end{equation}
for $\epsilon\in\{0,+1,-1\}$ and $j,k=1,\ldots,n$.
These functions are elliptic functions of periods $1/\eta$ and $\tau/\eta$ and the sum of their residues inside an elementary cell cancels, which leads to the identities
\begin{multline}
  \sum_{b=1}^n \frac{\prod_{l=1}^n [u_b - v_l]}{\prod_{l \neq b} [u_b - u_l]}
    \frac{[v_j - u_b +t]}{[v_j - u_b]} \frac{[u_b - v_k + \gamma + \epsilon]}{[u_b - v_k + \epsilon]}
     = \delta_{j,k} \,\delta_{\epsilon,0}\, [\gamma]\, [t] \,\frac{\prod_{l \neq j}[v_j - v_l]}{\prod_{l=1}^n [v_j - u_l]} 
       \\
   - (1 - \delta_{\epsilon,0}) \, [\gamma] \, \prod_{l=1}^n  \frac{[v_k - \epsilon - v_l]}{[v_k - \epsilon - u_l]} \,\frac{[v_j - v_k + \epsilon + t]}{[v_j - v_k + \epsilon]}.
\end{multline}
It follows that
\begin{multline}
  \big[ \mathcal{H}_{\gamma,\boldsymbol{\alpha}} \big]_{jk} 
  = \delta_{j,k}\, [0]'  \, \frac{\prod_{l\neq j} [v_j - v_l]}{\prod_{l=1}^n [v_j - u_l]} 
     \Bigg\{ \alpha_{1;k} \prod_{l=1}^n \frac{[u_l - v_k +1]}{[v_l - v_k +1]} - \alpha_{3;k}  \prod_{l=1}^n\frac{[u_l - v_k - 1]}{[v_l - v_k - 1]} \Bigg\} 
      \\
 +\frac{[0]'}{[t]} \Bigg\{ \alpha_{2;k} \frac{[v_j - v_k + t + 1]}{[v_j - v_k +1]} - \alpha_{4;k} \prod_{l=1}^n \frac{[v_j - v_k + t - 1]}{[v_k - v_k - 1]} \Bigg\}.
\end{multline}
Similarly, the product of matrices $\mathcal{Q}_{\gamma,\boldsymbol{\beta}}(\{u\},\{v\}| \{\zeta\})\equiv   \mathcal{X}_t(\{u\},\{v\}) \, {Q}_{\gamma,\boldsymbol{\beta}}(\{u\},\{v\}|\{\zeta\})  $ can be computed by considering the elliptic functions
\begin{equation}
\tilde{g}_\epsilon^{(j,k)}(z) = \frac{[z - \zeta_k + \gamma + \epsilon]}{[z - \zeta_k + \epsilon]} \prod_{l=1}^n \frac{[z-v_l]}{[z - u_l]} \frac{[v_j - z + t]}{[v_j - z]}
\end{equation}
for $\epsilon\in\{0,+1,-1\}$ and $j=1,\ldots,n,\ k=1,\ldots,m$.
It gives
\begin{multline}
   \big[ \mathcal{Q}_{\gamma,\boldsymbol{\beta}} \big]_{jk} 
   = \frac{[0]'}{[t]} \Bigg\{ \beta_{2;k} \frac{[v_j - \zeta_k + t +1]}{[v_j - \zeta_k +1]} 
     - \beta_{1;k}  \frac{[v_j - \zeta_k + t]}{[v_j - \zeta_k]}
       \prod_{l=1}^n \frac{[v_l - \zeta_k] [u_l - \zeta_k +1]}{[u_l - \zeta_k] [v_l - \zeta_k +1]}
       \\
       -\beta_{4;k} \frac{[v_j - \zeta_k + t -1]}{[v_j - \zeta_k -1]}
       +  \beta_{3;k}  \frac{[v_j - \zeta_k + t]}{[v_j - \zeta_k]}
       \prod_{l=1}^n \frac{[v_l - \zeta_k] [u_l - \zeta_k -1]}{[u_l - \zeta_k] [v_l - \zeta_k -1]}
       \Bigg\}.
\end{multline}

In particular, using these identities, we get that the determinant of the matrix \eqref{H-ell} is given by
%
\begin{multline}
   \det_n \big[ H_{\gamma;\mathbf{b}}^{(\nu)}(\{u\},\omega_u;\{v\},\omega_v|\{v_{b_{m+k}}\}) \big]
   =\frac{\big[\gamma+\sum_k(u_k-v_k)\big]}{(-[0]')^{n}\,[\gamma]}
                 \prod_{j<k} \frac{[u_j - u_k]}{[v_j - v_k]}\\
                 \times
        \det_n \big[ \mathcal{H}_{\gamma;\mathbf{b}}^{(\nu)}(\{u\},\omega_u;\{v\},\omega_v|\{v_{b_{m+k}}\}) \big]         
\end{multline}
with $\mathcal{H}_{\gamma;\mathbf{b}}^{(\nu)}(\{u\},\omega_u;\{v\},\omega_v|\{v_{b_{m+k}}\})$ given by \eqref{calH-ell}.

\section{Some functions with their Fourier coefficients and applications}\label{app-fourier}

We gather in this appendix the definitions of some useful (1-periodic) functions together with the explicit expression of their Fourier coefficients. This enables us to solve integral equations and compute Fredholm determinants appearing in the course of the paper.

Let us first consider, for two complex numbers $t$ and $\tau$ such that $0<\Im t<\Im \tau$, the following 1-periodic functions:
\begin{equation}\label{Zprimet}
   \Theta_{\pm t}^{(0)}(z;\tau)
   =\frac{i}{2\pi}\,\frac{\theta_1'(z\pm t;\tau)}{\theta_1(z\pm t;\tau)}.
\end{equation}
Their Fourier coefficients are given as
\begin{align}\label{Z0-fourier}
   \big(\Theta_{\pm t}^{(0)}\big)_m
   &=\int_{-1/2}^{1/2} \Theta_{\pm t}^{(0)}(z;\tau)\, e^{-2\pi i mz}\, dz
         \nonumber\\
   &=\begin{cases} \pm \frac12 &\text{if } m=0,\\
                                  \pm  \frac{e^{\pm 2\pi i m t} }{1- e^{\pm 2\pi i m\tau}} &\text{otherwise}.
        \end{cases}                          
\end{align}
From \eqref{Z0-fourier} one obtains the Fourier coefficients of the 1-periodic functions $p'_0$ \eqref{p-prime} and $K$ \eqref{K}, which can be written as $p'_0(z)=2\pi\Big(\Theta_{\tilde\eta/2}^{(0)}(z;\tilde\tau)-\Theta_{-\tilde\eta/2}^{(0)}(z;\tilde\tau)\Big)$ and $K(z)=\Theta_{\tilde\eta}^{(0)}(z;\tilde\tau)-\Theta_{-\tilde\eta}^{(0)}(z;\tilde\tau)$:
\begin{align}
   &p_m'=\int_{-1/2}^{1/2} p'_0(z)\, e^{-2\pi i m z}\, dz
             = \begin{cases}
                 2\pi & \text{for } m=0,\\
                 2\pi\, e^{i\pi |m|\tilde\eta} \,\frac{1- e^{2\pi i |m|(\tilde\tau-\tilde\eta)}}{1-e^{2\pi i |m| \tilde\tau} } & \text{otherwise,}
  \end{cases}\\
  & K_m = \int_{-1/2}^{1/2} K(z)\, e^{-2\pi i m z}\, dz
              = \begin{cases}
                 1 & \text{for } m=0,\\
                  e^{2\pi i |m| \tilde\eta}\,\frac{1- e^{2\pi i |m|(\tilde\tau-2 \tilde\eta)}}{1- e^{2\pi i |m|\tilde\tau}}  & \text{otherwise.}
                 \end{cases} \label{km}
\end{align}
This enables us to obtain the solution \eqref{rho} of the integral equation \eqref{liebeq}. This also enables us to explicitly compute the Fredholm determinant $\det\big[1+\widehat{K}-\widehat{V}_0\big]$, where $\widehat{K}$ and $\widehat{V}_0$ are integral operators acting on the interval $[-\frac12,\frac12]$ with respective kernels $K(y-z)$ \eqref{K} and $V_0(y-z)=2\eta$. 
This Fredholm determinant is given as the infinite product of eigenvalues $2-2\eta$ and $1+K_m$ for $m\not=0$:
%
\begin{equation}\label{det-Fred0}
      \det \big[  1 + \widehat{K}-\widehat{V}_0 \big] 
       =2(1-\eta)\prod_{m=1}^{+\infty}
            \frac{\big(1+e^{2\pi i m\tilde\eta}\big)^2\, \big(1-e^{2\pi i m(\tilde\tau-\tilde\eta)}\big)^2}
                   {\big(1-e^{2\pi i m \tilde\tau}\big)^2}.
\end{equation}

\bigskip

Let us now consider, for two complex numbers $t$ and $\tau$ such that $0<\Im t<\Im \tau$, and an arbitrary parameter $X$, the following ratios of theta-functions:
\begin{equation}\label{Zt}
   \Theta_{X;\pm t}(z;\tau)
   =\frac{i}{2\pi}\,\frac{\theta_1'(0;\tau)\, \theta_1(z+X\pm t;\tau)}{\theta_1(X;\tau)\,\theta_1(z\pm t;\tau)}.
\end{equation}
They satisfy the following quasi-periodicity properties:
\begin{equation}
   \Theta_{X;\pm t}(z+1;\tau)=\Theta_{X;\pm t}(z;\tau),
   \qquad
   \Theta_{X;\pm t}(z+\tau;\tau)= e^{-2\pi i X}\, \Theta_{X;\pm t}(z;\tau).
\end{equation}
Their Fourier coefficients are given as
\begin{align}\label{Z-fourier}
   \big(\Theta_{X;\pm t}\big)_m
   &=\int_{-1/2}^{1/2} \Theta_{X;\pm t}(z;\tau)\, e^{-2\pi i mz}\, dz
         \nonumber\\
   &=\pm  \frac{e^{\pm 2\pi i m t} }{1- e^{\pm 2\pi i (X+m\tau)}}.
\end{align}

Let us then consider, for two arbitrary parameters $X$ and $Y$ and a complex number $\zeta$ such that $0<\Im \zeta<\Im\tilde\eta<\Im\tilde\tau$, the following functions, defined in terms of theta functions with imaginary quasi-period $\tilde\tau$:
\begin{align}
  K^{(Y)}_{X}(z)
    &=\frac{i}{2\pi}\frac{\theta_1'(0)}{\theta_1(X)}
           \left\{   e^{2i\pi Y}\, \frac{\theta_1(z+X+\tilde\eta)}{\theta_1(z+\tilde\eta)}
                    -  e^{-2i\pi Y}\, \frac{\theta_1(z+X-\tilde\eta)}{\theta_1(z-\tilde\eta)}   \right\}
                    \label{K-X-Y}\\
     &= e^{2\pi i Y} \,   \Theta_{X;\tilde\eta}(z;\tilde\tau)   -      e^{-2i\pi Y}\,     \Theta_{X;-\tilde\eta}(z;\tilde\tau),
\end{align}
\begin{align}     
  t^{(Y)}_{X}(z,\zeta)
     &=\frac{i}{2\pi}\frac{\theta_1'(0)}{\theta_1(X)}
           \left\{   e^{2i\pi Y}\, \frac{\theta_1(z-\zeta+X+\tilde\eta)}{\theta_1(z-\zeta+\tilde\eta)}
                    -  \frac{\theta_1(z-\zeta+X)}{\theta_1(z-\zeta)}   \right\}
                    \label{t-X-Y}\\
     &= e^{2\pi i Y} \,   \Theta_{X;\tilde\eta-\zeta}(z;\tilde\tau)   -      \Theta_{X;-\zeta}(z;\tilde\tau).
\end{align}   
It follows from \eqref{Z-fourier} that
\begin{align}
   \big(K_{X}^{(Y)}\big)_m
      &=\int_{-1/2}^{1/2} K_X^{(Y)}(z)\, e^{-2\pi i mz}\, dz \nonumber\\
      &= e^{2\pi i Y} \, \frac{e^{2\pi i m\tilde\eta}}{1-e^{2\pi i (X+m\tilde\tau)}}
        + e^{-2\pi i Y} \, \frac{e^{-2\pi i m\tilde\eta}}{1-e^{-2\pi i (X+m\tilde\tau)}},\label{K-X-Y_m}
\end{align}
\begin{align}
   \big(t_{X}^{(Y)}(\zeta)\big)_m
      &=\int_{-1/2}^{1/2} t_X^{(Y)}(z,\zeta)\, e^{-2\pi i mz}\, dz \nonumber\\
      &= e^{-2\pi i m\zeta} \left\{ e^{2\pi i Y} \, \frac{e^{2\pi i m\tilde\eta}}{1-e^{2\pi i (X+m\tilde\tau)}}
        + \frac{1}{1-e^{-2\pi i (X+m\tilde\tau)}} \right\},
\end{align}

The knowledge of the Fourier series of \eqref{K-X-Y} and \eqref{t-X-Y} enables us to solve the following integral equation:
\begin{equation}  \label{eq-S-X-Y}
  S_X^{(Y)}(y,\zeta)+\int_{-1/2}^{1/2} K^{(Y)}_{X}(y-z)\, S_X^{(Y)}(z,\zeta)\, dz
  = t^{(Y)}_{X}(y,\zeta),
\end{equation}
where $S_X^{(Y)}(y;\zeta)$ is a 1-periodic function of $y$ to be determined. We find that the latter has Fourier coefficients
\begin{equation}
   \big( S_X^{(Y)}(\zeta) \big)_m= \frac{ \big(t_{X}^{(Y)}(\zeta)\big)_m }{1+ \big(K_{X}^{(Y)}\big)_m }
          = \frac{ e^{-2\pi i m\zeta} }{1+ e^{-2\pi i (Y+m\tilde\eta)}},
\end{equation}
so that
\begin{equation}\label{S-X-Ybis}
  S_X^{(Y)}(y,\zeta)= - \Theta_{Y+\frac{1}{2};-\zeta}(y;\tilde\eta)
     =\frac{1}{2\pi i}\,\frac{\theta_1'(0;\tilde\eta)\, \theta_2(y-\zeta+Y;\tilde\eta)}{\theta_2(Y;\tilde\eta)\,\theta_1(y-\zeta;\tilde\eta)}
     \equiv S^{(Y)}(y-\zeta) .
\end{equation}
Note that it neither depends on $X$ nor on $\tilde\tau$.

The knowledge of the Fourier series of \eqref{K-X-Y} also enables us to explicitly compute the Fredholm determinant $\det\big[ 1+\widehat{K}^{(Y)}_{X}\big]$, where $\widehat{K}^{(Y)}_{X}$ is an integral operator acting on the interval $[-\frac12,\frac12]$ with kernel ${K}^{(Y)}_{X}(y-z)$ given by \eqref{K-X-Y}. This Fredholm determinant is given as the infinite product of the eigenvalues $1+\big(K_{X}^{(Y)}\big)_m$ for $m\in\mathbb{Z}$. We obtain:
\begin{equation}\label{det-Fred-X-Y}
      \det\big[ 1+\widehat{K}^{(Y)}_{X}\big]
       =\frac{\theta_1(X-Y;\tilde\tau-\tilde\eta)\, \theta_2(Y;\tilde\eta)}{\theta_1(X;\tilde\tau)}\,
         \prod_{m=1}^{+\infty}
            \frac{\big(1-e^{2\pi i m \tilde\tau}\big)}
                    {\big(1-e^{2\pi i m\tilde\eta}\big) \big(1-e^{2\pi i m(\tilde\tau-\tilde\eta)}\big)}  .
\end{equation}

\section{Computation of the modified one-point local height probability}
\label{app-LHP}

In this appendix, we explain how to obtain a more compact formula for the (modified) one-point local height probabilities \eqref{LHP1}.

One can first remark that it is possible to obtain an alternative formula for the modified one-point matrix element \eqref{1PME2}. Indeed, one can use the series expansions \eqref{theta1}, \eqref{theta2}, to rewrite the product of theta functions in the last line of \eqref{1PME2} as
\begin{multline}\label{1PME3}
   \theta_1\Big( (1-\eta)\tilde\gamma+\eta\nu;\tilde\tau-\tilde\eta\Big) \,         
   \theta_2\!\left(Z-\frac{L\mathsf{k}+2\ell}{2(L-r)}+\eta(\tilde\gamma-\nu);\tilde\eta\right)
   \\
   =
   \sum_{j=-\infty}^{+\infty} e^{i\pi\tilde\eta j^2}\,
   e^{2\pi i j \left(Z-\frac{L\mathsf{k}+2\ell}{2(L-r)}+\eta(\tilde\gamma-\nu)\right)}
   \,
   \theta_1\!\left(\tilde\gamma-\frac{L\mathsf{k}+2\ell}{2(L-r)}+Z+\tilde\eta j;\tilde\tau\right).
\end{multline}
Since the resulting theta functions in \eqref{1PME3} do not depend any more on $\nu$, one can then exchange the two summation symbols so as to explicitly compute the sum over $\nu$ by means of the summation formula \eqref{id-sum2}.
One obtains
\begin{multline}\label{1PME4}
  \bar{\mathbb{P}}(s, Z ;\mathsf{k},\ell)
  = \frac{e^{i\pi s \frac{r\mathsf{k}+2\ell}{L-r}}}{L-r}\,
  \sum_{j=-\infty}^{+\infty} e^{i\pi\tilde\eta j^2}\,
   e^{2\pi i j \left(Z-\frac{L\mathsf{k}+2\ell}{2(L-r)}\right)}
   \\
   \times
   \frac{\theta_1\!\left(\tilde\eta s;\tilde\tau\right)\,
            \theta_1\!\left(\tilde\gamma-\frac{L\mathsf{k}+2\ell}{2(L-r)}+Z+\tilde\eta j;\tilde\tau\right)}
          {\theta_1\!\left(Z-\frac{L\mathsf{k}+2\ell}{2(L-r)}+\tilde\gamma+\tilde\eta s;\tilde\tau\right) \, 
            \theta_1'\!\left(0;\tilde\tau-\tilde\eta\right)\,\theta_2\!\left(0;\tilde\eta\right)}\,
    \frac{\theta_1\!\left(\tilde\gamma+\tilde\eta(s-j);\tilde\tau\right)\, \theta_1'\!\left(0;\tilde\tau\right)}
          {\theta_1\!\left(\tilde\eta(s-j);\tilde\tau\right)\,
            \theta_1\!\left(\tilde\gamma;\tilde\tau\right)}.
\end{multline}

Let us now consider the (modified) one-point local height probabilities \eqref{LHP1}.
By means of \eqref{1PME4}, it is given as
\begin{align}
  \bar{\mathbf{P}}(s, Z;\epsilon,\mathsf{t}) 
  &= \sum_{\mathsf{k}=0}^1 \sum_{\ell=0}^{L-r-1}
    (-1)^{\mathsf{k}\epsilon}\, e^{-i\pi\frac{r\mathsf{k}+2\ell}{L-r}(\mathsf{t}+s_0)}\  \bar{\mathbb{P}}(s, Z;\mathsf{k},\ell) 
    \nonumber\\
  &= \sum_{j=-\infty}^{+\infty} e^{i\pi\tilde\eta j^2}\, e^{2\pi i j Z}\,
       \frac{\theta_1\!\left(\tilde\eta s;\tilde\tau\right)\, \theta_1\!\left(\tilde\gamma+\tilde\eta(s-j);\tilde\tau\right)}
              {\theta_1'\!\left(0;\tilde\tau-\tilde\eta\right)\,\theta_2\!\left(0;\tilde\eta\right)\, \theta_1\!\left(\tilde\gamma;\tilde\tau\right)}\
           \sum_{\mathsf{k}=0}^1 
           e^{i\pi\mathsf{k}\big(\epsilon+\frac{r(s-s_0-\mathsf{t})-Lj}{L-r}\big)}   
             \nonumber \\
  &\times
   \frac{1}{L-r} \sum_{\ell=0}^{L-r-1} e^{2\pi i \frac{\ell}{L-r}(s-s_0-\mathsf{t}-j)}\,
      \frac{\theta_1\!\left(\tilde\gamma-\frac{L\mathsf{k}+2\ell}{2(L-r)}+Z+\tilde\eta j;\tilde\tau\right)\, \theta_1'\!\left(0;\tilde\tau\right)}
          {\theta_1\!\left(Z-\frac{L\mathsf{k}+2\ell}{2(L-r)}+\tilde\gamma+\tilde\eta s;\tilde\tau\right)\, \theta_1\!\left(\tilde\eta(s-j);\tilde\tau\right) } .
          \label{LHP2}
\end{align}
The sum over $\ell$ can then be computed by means of \eqref{id-sum1}, which gives
\begin{multline}
  \bar{\mathbf{P}}(s, Z;\epsilon,\mathsf{t}) 
  = \sum_{j=-\infty}^{+\infty} e^{i\pi\tilde\eta j^2}\, e^{2\pi i j Z}\,
       \frac{\theta_1\!\left(\tilde\eta s;\tilde\tau\right)\, \theta_1\!\left(\tilde\gamma+\tilde\eta(s-j);\tilde\tau\right)}
              {\theta_1'\!\left(0;\tilde\tau-\tilde\eta\right)\,\theta_2\!\left(0;\tilde\eta\right)\, \theta_1\!\left(\tilde\gamma;\tilde\tau\right)}\
           \sum_{\mathsf{k}=0}^1 
           e^{i\pi\mathsf{k}(\epsilon-s+s_0+\mathsf{t})}   \\
   \times   
   e^{2\pi i (s-s_0-\mathsf{t}-j)(Z+\tilde\gamma+\tilde\eta s)}   \,
   \frac{\theta_1'\!\left(0;(L-r)\tilde\tau\right)}{\theta_1\!\left((s-j)\tilde\eta+(  \mathsf{t}+j+s_0-s)\tilde\tau;(L-r)\tilde\tau\right)}\\
   \times
   \frac{\theta_1\!\left(\frac{L\mathsf{k}}{2}+\tilde\eta(s-j)-(L-r)(Z+\tilde\gamma+\tilde\eta s) +(\mathsf{t}+j+s_0-s)\tilde\tau;(L-r)\tilde\tau\right)}{\theta_1\!\left(\frac{L\mathsf{k}}{2}-(L-r)(Z+\tilde\gamma+\tilde\eta s);(L-r)\tilde\tau\right)}.
\end{multline}
Before going further, it is in fact convenient to re-expand the last ratios of theta functions, using once agin \eqref{id-sum1}, as a sum over $L$ terms in terms of theta functions with imaginary period $\frac{L-r}L \tilde\tau$:
\begin{multline}
  \hspace{-2mm}\frac{\theta_1\big(\tilde\eta(s_0+\mathsf{t})+\frac{L\mathsf{k}}{2}-(L-r)(Z+\tilde\gamma+\tilde\eta s) +(\mathsf{t}+j+s_0-s)\frac{L-r}{L}\tilde\tau;(L-r)\tilde\tau\big)\, 
           \theta_1'\big(0;(L-r)\tilde\tau\big)}
        {\theta_1\big( \tilde\eta(s_0+\mathsf{t})+(  \mathsf{t}+j+s_0-s)\frac{L-r}{L}\tilde\tau;(L-r)\tilde\tau\big)\,
         \theta_1\big(\frac{L\mathsf{k}}{2}-(L-r)(Z+\tilde\gamma+\tilde\eta s);(L-r)\tilde\tau\big)}
         \\
    =\frac{1}{L}\sum_{l=0}^{L-1}    
    e^{-2\pi i (\mathsf{t}+j+s_0-s)\big(\frac{l}{L}+\frac{\mathsf{k}}{2}-\frac{L-r}{L}(Z+\tilde\gamma+\tilde\eta s)\big)} 
    \\
    \times
    \frac{\theta_1\Big(\tilde\eta(s_0+\mathsf{t})+\frac{\mathsf{k}}{2}-\frac{L-r}{L}(Z+\tilde\gamma+\tilde\eta s)+\frac{l}{L};\frac{L-r}{L}\tilde\tau\Big)\,
             \theta_1'\Big(0;\frac{L-r}{L}\tilde\tau\Big)}
           {\theta_1\Big(\tilde\eta(s_0+\mathsf{t});\frac{L-r}{L}\tilde\tau\Big)\,
            \theta_1\Big(\frac{\mathsf{k}}{2}-\frac{L-r}{L}(Z+\tilde\gamma+\tilde\eta s)+\frac{l}{L};\frac{L-r}{L}\tilde\tau\Big)},
\end{multline}
so that
\begin{multline}
  \bar{\mathbf{P}}(s, Z;\epsilon,\mathsf{t}) 
  = \sum_{j=-\infty}^{+\infty} e^{i\pi\tilde\eta j^2}\, e^{2\pi i j Z}\,
       \frac{\theta_1\!\left(\tilde\eta s;\tilde\tau\right)\, \theta_1\!\left(\tilde\gamma+\tilde\eta(s-j);\tilde\tau\right)}
              {\theta_2\!\left(0;\tilde\eta\right)\, \theta_1\!\left(\tilde\gamma;\tilde\tau\right)\,
               \theta_1\!\left(\tilde\eta(s_0+\mathsf{t});\tilde\tau-\tilde\eta\right)}\
           \sum_{\mathsf{k}=0}^1 
           e^{i\pi\mathsf{k}(\epsilon-j)}   \\
   \times   
   \frac{1}{L}\sum_{l=0}^{L-1}    
    e^{-2\pi i (\mathsf{t}+j+s_0-s)\big(\frac{l}{L}+\eta(Z+\tilde\gamma+\tilde\eta s)\big)}
     \frac{\theta_1\Big(\tilde\eta(s_0+\mathsf{t})+\frac{\mathsf{k}}{2}-\frac{L-r}{L}(Z+\tilde\gamma+\tilde\eta s)+\frac{l}{L};\frac{L-r}{L}\tilde\tau\Big)}
           {\theta_1\Big(\frac{\mathsf{k}}{2}-\frac{L-r}{L}(Z+\tilde\gamma+\tilde\eta s)+\frac{l}{L};\frac{L-r}{L}\tilde\tau\Big)}.
\end{multline}

One now wants to compute the sum over $j$. Performing a change of indices of the form $j=j_1+Lj_2$, with $0\le j_1 \le L-1$, and using the series expansion \eqref{theta3} of the theta function, one obtains that
\begin{align*}
  &\sum_{j=-\infty}^{+\infty} e^{i\pi\tilde\eta j^2}\, 
      e^{2\pi i j \big( (1-\eta)Z-\eta(\tilde\gamma+\tilde\eta s)-\frac{\mathsf{k}}{2}-\frac{l}{L}\big)}\,
      \theta_1\!\left(\tilde\gamma+\tilde\eta(s-j);\tilde\tau\right)
   \nonumber\\
  &\qquad = 
     \sum_{j_1=0}^{L-1} e^{i\pi\tilde\eta j_1^2}\,
       e^{2\pi i j_1 \big( (1-\eta)Z-\eta(\tilde\gamma+\tilde\eta s)-\frac{\mathsf{k}}{2}-\frac{l}{L}\big)}\,
       \theta_1\big(\tilde\gamma+\tilde\eta(s-j_1);\tilde\tau\big) \nonumber\\
   &\qquad \hspace{5cm} \times    
      \theta_3\Big((L-r)Z+\frac{r-L\mathsf{k}}{2}+rj_1(\tilde\tau-\tilde\eta); r(L-r)\tilde\tau\Big)
      \\
   &\qquad =
   \theta_2\Big( (1-\eta) Z-\eta (\tilde\gamma+\tilde\eta s) -\frac{\mathsf{k}}{2}-\frac{l}{L}   ;\tilde\eta\Big)\,
   \theta_1\Big( (1-\eta)( Z+\tilde\gamma+\tilde\eta s) -\frac{\mathsf{k}}{2}-\frac{l}{L}   ;\tilde\tau-\tilde\eta\Big),
\end{align*} 
in which we have also used Schr\"oter's formula \eqref{Schroter}.
Hence,
\begin{multline}
  \bar{\mathbf{P}}(s, Z;\epsilon,\mathsf{t}) 
  = 
       \frac{\theta_1\!\left(\tilde\eta s;\tilde\tau\right)\, 
                e^{-2\pi i\eta(\mathsf{t}+s_0-s)(Z+\tilde\gamma+\tilde\eta s)}}
              {\theta_2\!\left(0;\tilde\eta\right)\, \theta_1\!\left(\tilde\gamma;\tilde\tau\right)\,
               \theta_1\!\left(\tilde\eta(s_0+\mathsf{t});\tilde\tau-\tilde\eta\right)}\
           \sum_{\mathsf{k}=0}^1 
           e^{i\pi\mathsf{k}\epsilon}\
            \frac{1}{L}\sum_{l=0}^{L-1}    
    e^{-2\pi i (\mathsf{t}+s_0-s) \frac{l}{L}}  \\
   \times     
    \theta_1\Big((1-\eta)(Z+\tilde\gamma+\tilde\eta s)-\tilde\eta(s_0+\mathsf{t})-\frac{\mathsf{k}}{2}-\frac{l}{L};\tilde\tau-\tilde\eta\Big)\,
    \theta_2\Big((1-\eta)Z-\eta(\tilde\gamma+\tilde\eta s)-\frac{\mathsf{k}}{2}-\frac{l}{L};\tilde\eta\Big).
\end{multline}

Let us now compute the sum over $l$. To this aim, we expand the last product of theta functions by means of Schr\"oter's formula \eqref{Schroter} as
\begin{multline}
   \theta_1\Big((1-\eta)(Z+\tilde\gamma+\tilde\eta s)-\tilde\eta(s_0+\mathsf{t})-\frac{\mathsf{k}}{2}-\frac{l}{L};\tilde\tau-\tilde\eta\Big)\,
    \theta_2\Big((1-\eta)Z-\eta(\tilde\gamma+\tilde\eta s)-\frac{\mathsf{k}}{2}-\frac{l}{L};\tilde\eta\Big)
    \\
    = \sum_{j=0}^{L-1} e^{i\pi\tilde\eta j^2}\, e^{-2\pi i j\big( (1-\eta)Z-\eta(\tilde\gamma+\tilde\eta s)-\frac{\mathsf{k}}{2}-\frac{l}{L}\big)}\,
    \theta_1\big(\tilde\eta(s-s_0-\mathsf{t})+\tilde\gamma+j\tilde\eta;\tilde\tau\big)
    \\
    \times
     \theta_3\Big(r\tilde\eta(s_0+\mathsf{t})-(L-r)Z+\frac{L\mathsf{k}+r}{2}+\frac{r(L-r)}{L}j\tilde\tau;r(L-r)\tilde\tau\Big),
\end{multline}
so that, once we have exchanged the summation symbols, the sum over $l$ simply becomes
\begin{equation}
   \frac{1}{L}\sum_{l=0}^{L-1} e^{-2\pi i (\mathsf{t}+s_0-s-j)\frac{l}{L}}=\delta_{j,\mathsf{t}+s_0-s}.
\end{equation}
It follows that the expression for $\bar{\mathbf{P}}(s, Z;\epsilon,\mathsf{t}) $ is effectively independent from the value of $\tilde\gamma$ (as it should be), and we obtain
\begin{multline}\label{LHP-before-k}
  \bar{\mathbf{P}}(s, Z;\epsilon,\mathsf{t}) 
  = 
       \frac{\theta_1\!\left(\tilde\eta s;\tilde\tau\right) \,
        e^{i\pi\tilde\eta (\mathsf{t}+s_0-s)^2}\, e^{-2\pi i (\mathsf{t}+s_0-s)Z}}
              {\theta_2\!\left(0;\tilde\eta\right)\, 
               \theta_1\!\left(\tilde\eta(s_0+\mathsf{t});\tilde\tau-\tilde\eta\right)}\
           \sum_{\mathsf{k}=0}^1 
           e^{i\pi\mathsf{k}(\epsilon+\mathsf{t}+s_0-s)}\
              \\
   \times     
    \theta_3\Big(r\tilde\eta(s_0+\mathsf{t})-(L-r)Z+\frac{L\mathsf{k}+r}{2}+\frac{r(L-r)}{L}(\mathsf{t}+s_0-s)\tilde\tau;r(L-r)\tilde\tau\Big).
\end{multline}

It only remains now to compute the sum over $\mathsf{k}$. It is convenient to distinguish two cases according to the parity of $L$:
\begin{itemize}

\item If $L$ is even, then $r$ is odd and the theta function in the last line of \eqref{LHP-before-k} does not depend on $\mathsf{k}$. Hence the sum over $\mathsf{k}$ reduces to
\begin{equation}
   \sum_{\mathsf{k}=0}^1 
           e^{i\pi\mathsf{k}(\epsilon+\mathsf{t}+s_0-s)}
           =
           \begin{cases}
             0 & \text{if} \ \, \epsilon+\mathsf{t}+s_0-s\ \, \text{is odd},\\
             2& \text{if} \ \, \epsilon+\mathsf{t}+s_0-s\ \,  \text{is even}.
           \end{cases}
\end{equation}
It means that, 
\begin{equation}\label{LHP-even-odd}
  \bar{\mathbf{P}}(s, Z;\epsilon,\mathsf{t}) \Big|_{\begin{subarray}{l} 
\text{$L$ even}\\ \text{$\epsilon\!+\!\mathsf{t}\!+\!s_0\!-\!s$ odd} \end{subarray}}
  = 0,
\end{equation}
whereas
\begin{multline}\label{LHP-even-even}
  \bar{\mathbf{P}}(s, Z;\epsilon,\mathsf{t}) \Big|_{\begin{subarray}{l} 
\text{$L$ even}\\ \text{$\epsilon\!+\!\mathsf{t}\!+\!s_0\!-\!s$ even} \end{subarray}}
  = 
     2 \, \frac{\theta_1\!\left(\tilde\eta s;\tilde\tau\right) \,
        e^{i\pi\tilde\eta (\mathsf{t}+s_0-s)^2}\, e^{-2\pi i (\mathsf{t}+s_0-s)Z}}
              {\theta_2\!\left(0;\tilde\eta\right)\, 
               \theta_1\!\left(\tilde\eta(s_0+\mathsf{t});\tilde\tau-\tilde\eta\right)}              \\
   \times     
    \theta_4\big(r\tilde\eta s -(L-r)Z+r(\mathsf{t}+s_0-s)\tilde\tau ;r(L-r)\tilde\tau\big).
\end{multline}
Setting $\tilde{s}=s-\frac{\tau}{2\eta}=s+\frac{1}{2\tilde\eta}$, i.e. $\tilde{s}_0=s_0+\frac{1}{2\tilde\eta}$ with $\tilde{s}_0\in\mathbb{R}$ so as to be in agreement with the physical model considered in \cite{PeaS89,PeaB90}, we obtain that
\begin{multline}\label{LHP-ev-ev-phys}
  \bar{\mathbf{P}}(s, Z;\epsilon,\mathsf{t}) \Big|_{\begin{subarray}{l} 
\text{$L$ even}\\ \text{$\epsilon\!+\!\mathsf{t}\!+\!s_0\!-\!s$ even} \end{subarray}}
  = 
     2 \, \frac{\theta_2\!\left(\tilde\eta \tilde{s};\tilde\tau\right) \,
        e^{i\pi\tilde\eta (\mathsf{t}+\tilde{s}_0-\tilde{s})^2}\, e^{-2\pi i (\mathsf{t}+\tilde{s}_0-\tilde{s})Z}}
              {\theta_2\!\left(0;\tilde\eta\right)\, 
               \theta_2\!\left(\tilde\eta(\tilde{s}_0+\mathsf{t});\tilde\tau-\tilde\eta\right)}              \\
   \times     
    \theta_3\big(r\tilde\eta \tilde{s} -(L-r)Z+r(\mathsf{t}+\tilde{s}_0-\tilde{s})\tilde\tau ;r(L-r)\tilde\tau\big).
\end{multline}

\item If $L$ is odd, the computation of the sum over $\mathsf{k}$ leads to
\begin{multline}\label{LHP-odd}
  \bar{\mathbf{P}}(s, Z;\epsilon,\mathsf{t}) \Big|_{\text{$L$ odd} }
  = 
     2 \, \frac{\theta_2\!\left(\tilde\eta \tilde{s};\tilde\tau\right) \,
        e^{i\pi\tilde\eta (\mathsf{t}+\tilde{s}_0-\tilde{s})^2}\, 
        e^{-2\pi i (\mathsf{t}+\tilde{s}_0-\tilde{s})Z}}
              {\theta_2\!\left(0;\tilde\eta\right)\, 
               \theta_2\!\left(\tilde\eta(\tilde{s}_0+\mathsf{t});\tilde\tau-\tilde\eta\right)}              \\
   \times  
    \theta_3\big(2r\tilde\eta \tilde{s} -2(L-r)Z+2r(\mathsf{t}+\tilde{s}_0-\tilde{s})\tilde\tau-2r(\epsilon+\mathsf{t}+\tilde{s}_0-\tilde{s})(L-r)\tilde\tau ; 4r(L-r)\tilde\tau\big)
    \\
    \times
   e^{i\pi r(L-r)\tilde\tau(\epsilon+\mathsf{t}+\tilde{s}_0-\tilde{s})^2   }\,
   e^{-2\pi i(\epsilon+\mathsf{t}+\tilde{s}_0-\tilde{s})(r\tilde\eta \tilde{s} +r(\mathsf{t}+\tilde{s}_0-\tilde{s})\tilde\tau-(L-r)Z)},
\end{multline}
in which we have set, as in \eqref{LHP-ev-ev-phys}, $\tilde{s}=s+\frac{1}{2\tilde\eta}$ and $\tilde{s}_0=s_0+\frac{1}{2\tilde\eta}$.

\end{itemize}

\bibliographystyle{amsplain}
\bibliography{../biblio}

\end{document}